\documentclass[11pt]{article}
\usepackage{geometry}                
\usepackage{graphicx}
\usepackage{rawfonts}
\usepackage{amssymb}
\usepackage{epstopdf}
\usepackage{amsmath,amsthm,amssymb}
\usepackage{mathtools}
\usepackage{fullpage}
\usepackage{color}
\usepackage[usenames,dvipsnames,svgnames]{xcolor}
\usepackage[colorlinks=true,citecolor=blue,linkcolor=red,urlcolor=blue]{hyperref}
\usepackage[font=footnotesize, labelfont=bf, margin=0.5cm]{caption}
\usepackage[labelformat=simple]{subcaption}
\usepackage{rawfonts}
\usepackage{etex}

\usepackage{ulem}

\input prepictex.tex
\input pictex.tex
\input postpictex.tex

\definecolor{blue}{rgb}{0,0.18,0.39}
\definecolor{RoyalBlue}{rgb}{0,0.2,0.7}

\definecolor{Maroon}{cmyk}{0,0.87,0.68,0.62}
\definecolor{Brown}{rgb}{0.7,0.3,0}
\definecolor{Navy}{rgb}{0.3,0.0,0.4}
\definecolor{Red}{cmyk}{0,1,1,0}
\definecolor{BrickRed}{cmyk}{0.16,0.89,0.61,0.02}
\definecolor{DarkRed}{cmyk}{0,1,1,0.5}
\definecolor{DarkBlue}{cmyk}{1,1,0,0.2}
\definecolor{DarkGreen}{cmyk}{1,0,1,0.4}
\definecolor{Green}{cmyk}{1,0,1,0}
\definecolor{DarkBrown}{cmyk}{0,0.81,1,0.6}
\definecolor{OrangeRed}{cmyk}{0,1,0.87,0}
\definecolor{RedOrange}{cmyk}{0,0.77,0.87,0}
\definecolor{Orange}{cmyk}{0,0.61,0.87,0}
\definecolor{Offwhite}{rgb}{.8,0.9,.8}
\definecolor{Offwhite2}{cmyk}{.04,.02,.01,0}
\definecolor{Tan}{rgb}{0.82,0.70,0.55}
\definecolor{Blue}{rgb}{0,0,1}
\definecolor{RoyalBlue}{rgb}{0.25,0.41,0.88}
\definecolor{Sepia}{rgb}{0.37,0.14,0.07}
\definecolor{myblue}{cmyk}{0.5,0.3,0,0}
\definecolor{Mahogany}{cmyk}{0.18,0.87,1,0.08}

\definecolor{green1}{cmyk}{0.25,0,0.76,0}
\definecolor{green2}{cmyk}{0.25,0,0.76,0.07}
\definecolor{green3}{cmyk}{0.25,0,0.76,0.20}
\definecolor{green4}{cmyk}{0.25,0,0.75,0.30}
\definecolor{green5}{cmyk}{0.25,0,0.75,0.40}
\definecolor{green6}{cmyk}{0.25,0,0.75,0.50}

\definecolor{B02}{cmyk}{0,0.14,0.22,0.12}
\definecolor{B03}{cmyk}{0,0.16,0.26,0.16}
\definecolor{B04}{cmyk}{0,0.19,0.28,0.19}
\definecolor{B05}{cmyk}{0,0.25,0.32,0.25}
\definecolor{B06}{cmyk}{0,0.31,0.36,0.31}
\definecolor{B07}{cmyk}{0,0.37,0.40,0.37}
\definecolor{B08}{cmyk}{0,0.46,0.46,0.46}
\definecolor{B09}{cmyk}{0,0.55,0.52,0.54}
\definecolor{B10}{cmyk}{0,0.69,0.61,0.62}
\definecolor{B11}{cmyk}{0,0.78,0.70,0.68}
\definecolor{B12}{cmyk}{0,0.93,0.85,0.60}
\definecolor{B13}{cmyk}{0.5,1,0.6,0.30}
\definecolor{B14}{cmyk}{1,1,0,0.30}
\definecolor{B15}{cmyk}{1,1,0,0}

\newtheorem{theo}{Theorem}
\newtheorem{lemm}{Lemma}

\newtheorem{conj}{Conjecture}

\def\qed{$\Box$}

\def\sfrac#1#2{\hbox{\nor $\frac{#1}{#2}$}}
\def\Sfrac#1#2{\hbox{\large $\frac{#1}{#2}$}}

\def\shalf{{\sfrac{1}{2}}}

\def\Ref#1{(\ref{#1})}

\def\L{\left(} \def\R{\right)}
\def\LC{\left\{} \def\RC{\right\}}

\def\alev{\hbox{\textit{ae}}}

\usepackage{tikz}

\def\2;{\;\;}

\def\eps{\epsilon}
\def\IntZ{{\mathbb Z}}

\def\binom#1#2{{{#1}\choose{#2}}}

\def\W#1{\widehat{#1}}
\def\Ref#1{(\ref{#1})}
\def\Sfrac#1#2{\hbox{\Large $\frac{#1}{#2}$}}
\def\sfrac#1#2{\hbox{\normalsize $\frac{#1}{#2}$}}
\newcommand{\hqed}{\hbox{\hspace{3mm}$\square$}}

\title{Force-induced desorption of uniform block copolymers}
\author{E.J. Janse van Rensburg\thanks{\href{mailto:rensburg@yorku.ca}{rensburg@yorku.ca}}\\
\small Department of Mathematics and Statistics\\
\small York University, Toronto, Ontario M3J 1P3, Canada
\and
C.~E.~Soteros\thanks{\href{mailto:soteros@math.usask.ca}{soteros@math.usask.ca}}\\
\small Department of Mathematics and Statistics\\
\small The University of Saskatchewan, Saskatoon, Saskatchewan S7N 5E6, Canada
\and
S.~G.~Whittington\thanks{\href{mailto:swhittin@chem.utoronto.ca}{swhittin@chem.utoronto.ca}}\\
\small Department of Chemistry\\
\small University of Toronto, Toronto, Ontario M5S 3H6, Canada
}

\begin{document}

\maketitle

\begin{abstract}
We investigate self-avoiding walk models of linear block copolymers adsorbed at
a surface and desorbed by the action of a force.    We rigorously establish the 
dependence of the free energy on the adsorption and force parameters, and 
the form of the phase diagram for several cases, including 
$AB$-diblock copolymers and $ABA$-triblock copolymers, pulled from an
end vertex and from the central vertex.  Our interest in block
copolymers is partly motivated by the occurrence of a novel mixed
phase in a directed walk model of diblock copolymers \cite{Iliev}
and we believe that this paper is the first rigorous treatment of a self-avoiding 
walk model of the situation.
\end{abstract}


\section{Introduction}
\label{sec:Introduction}

An interesting question, both from the theoretical and from the practical point
of view, is how self-avoiding walks \cite{Rensburg2015,MadrasSlade} respond to 
tensile or compressive forces
\cite{Beaton2015,GuttmannLawler,IoffeVelenik,IoffeVelenik2010,Rensburg2009,Rensburg2016a}.
For a review see reference \cite{Orlandini}.  A particularly interesting case is 
a self-avoiding walk adsorbed at a surface and desorbed by the action of a force 
\cite{Guttmann2014,Rensburg2013,Rensburg2016b,Krawczyk2005,Krawczyk2004,Mishra2005}
as a model of the desorption of a linear polymer in an AFM experiment \cite{Haupt1999,Zhang2003}.

In this paper we address the question of copolymer adsorption and how the adsorbed copolymer
responds to a force.  Copolymers are polymers with more than one kind of monomer and we shall 
be concerned with the special case of two comonomers, $A$ and $B$.  In addition we only consider 
linear copolymers where the system is defined by the sequence of monomers along the linear chain.  
The sequence of comonomers $A$ and $B$ along the linear chain can be determined by a random 
process,  giving a random copolymer \cite{SoterosWhittington}, or the sequence can be deterministic.  
The sequences of $A$s and $B$s can be short as in an alternating copolymer, for instance, or we can 
have long blocks of $A$s followed by long blocks of $B$s.  These block copolymers are especially 
interesting since they are a very useful class of steric stabilizers of colloidal dispersions 
\cite{Fleer,Napper}.  In a diblock or triblock copolymer used as a steric stabilizer one type of 
monomer adsorbs strongly on the surface of the colloidal particle, to anchor the polymer, and the 
other extends into the dispersing medium and loses entropy when colloidal particles approach one 
another.

We investigate a cubic lattice self-avoiding walk model of a block copolymer in a good solvent. 
Specifically we consider a self-avoiding walk on the simple cubic lattice confined to a half-space,
with the confining plane acting as the adsorbing surface. The vertices of the walk are labelled 
$A$ or $B$ corresponding to the two comonomers.
We assume that the starting point of the walk is tethered to the adsorbing surface and show that 
the phase diagram depends on the relative strength of adsorption of the two types of comonomers, 
the number of blocks,  as well as on whether the walk is pulled from the endpoint or  the midpoint 
of the walk.  In contrast to the homopolymer phase diagram, we establish that, under certain 
conditions,  some of these copolymer models can exhibit a mixed adsorbed-ballistic phase.  These 
mixed phases are similar in nature to the mixed phase that exists for a square lattice
directed walk model of a copolymer \cite{Iliev}. 
The existence of mixed adsorbed-ballistic phases suggests that AFM experiments might 
be used to explore the blockiness of a linear polymer.

\begin{figure}[t!]
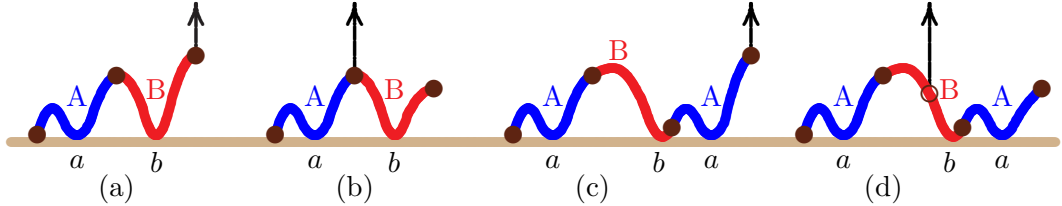

\beginpicture
\setcoordinatesystem units <0.5pt,0.5pt>
\setplotarea x from -60 to 300, y from -10 to 90
\setplotarea x from 0 to 300, y from 0 to 90

\setplotsymbol ({\large.})
\arrow <8pt>  [.2,.67] from 120 60 to 120 100

\setplotsymbol ({\footnotesize$\bullet$})

\color{Tan}
\plot -20 -5 770 -5 /

\setquadratic
\color{Blue}
\plot 0 0 10 20 20 10 30 0 40 10 50 35 60 45   /  \put {A} at 30 30
\color{Red}
\plot 60 45 70 40 80 20 90 0 100 20 110 50 120 60 /  \put {B} at 90 35

\color{Sepia}
\multiput {\LARGE$\bullet$} at 0 0 60 45 120 60  /

\color{black}
\normalcolor

\put {$a$} at 30 -20 
\put {$b$} at 90 -20 
\put {(a)} at 60 -40

\setcoordinatesystem units <0.5pt,0.5pt> point at -180 0 

\setplotsymbol ({\large.})
\arrow <8pt>  [.2,.67] from 60 50 to 60 100

\setplotsymbol ({\footnotesize$\bullet$})

\setquadratic
\color{Blue}
\plot 0 0 10 20 20 10 30 0 40 10 50 35 60 45   /  \put {A} at 30 30
\color{Red}
\plot 60 45 70 40 80 20 90 0 100 15 110 30 120 35 /  \put {B} at 90 35

\color{Sepia}
\multiput {\LARGE$\bullet$} at 0 0 60 45 120 35  /

\color{black}
\normalcolor

\put {$a$} at 30 -20 
\put {$b$} at 90 -20 
\put {(b)} at 60 -40

\setcoordinatesystem units <0.5pt,0.5pt> point at -360 0 

\setplotsymbol ({\large.})
\arrow <8pt>  [.2,.67] from 180 60 to 180 100

\setplotsymbol ({\footnotesize$\bullet$})

\setquadratic
\color{Blue}
\plot 0 0 10 20 20 10 30 0 40 10 50 35 60 45   /  \put {A} at 30 30
\color{Red}
\plot 60 45 70 50 80 50 90 40 100 20 110 0 120 5 /  \put {B} at 80 65
\color{Blue}
\plot 120 5 130 20 140 10 150 0 160 10 170 45 180 60    /  \put {A} at 150 30

\color{Sepia}
\multiput {\LARGE$\bullet$} at 0 0 60 45 120 5  180 60 /

\color{black}
\normalcolor

\put {$a$} at 30 -20 
\put {$b$} at 110 -20 
\put {$a$} at 150 -20 
\put {(c)} at 60 -40

\setcoordinatesystem units <0.5pt,0.5pt> point at -580 0 

\setplotsymbol ({\large.})
\arrow <8pt>  [.2,.67] from 95 31 to 95 100

\setplotsymbol ({\footnotesize$\bullet$})

\setquadratic
\color{Blue}
\plot 0 0 10 20 20 10 30 0 40 10 50 35 60 45   /  \put {A} at 30 30
\color{Red}
\plot 60 45 70 50 80 50 90 40 100 20 110 0 120 5 /  \put {B} at 110 35
\color{Blue}
\plot 120 5 130 20 140 10 150 0 160 10 170 25 180 35    /  \put {A} at 150 30

\color{Sepia}
\multiput {\LARGE$\bullet$} at 0 0 60 45 120 5  180 35 /
\multiput {\LARGE$\circ$} at 95 31 /

\color{black}
\normalcolor

\put {$a$} at 30 -20 
\put {$b$} at 110 -20 
\put {$a$} at 150 -20 
\put {(d)} at 60 -40

\endpicture
\caption{Models of adsorbing block copolymers pulled by a force: 
(a) An adsorbing diblock copolymer pulled at its endpoint.  The polymer is fixed 
in the adsorbing plane at its first vertex.  
(b) An adsorbing diblock copolymer pulled at its midpoint. 
(c) An adsorbing triblock copolymer of type $ABA$ pulled at its endpoint.
(d) An adsorbing triblock copolymer of type $ABA$ pulled at its midpoint.
Comonomers of type $A$ adsorb with activity $a$ in the adsorbing line, and 
comonomers of type $B$ with activity $b$.}
\label{figure1} 
\end{figure}

In figure \ref{figure1} we show the 
models that are considered in this paper, namely
diblock and triblock models of copolymers pulled either at an endpoint, or in the middle by an
external force.  The triblock copolymer models are of the type $ABA$, where blocks of comonomers
of types $A$ and $B$ are arranged in a sequence of $A$'s, then $B$'s, and then again $A$'s.
The models which exhibit a mixed adsorbed-ballistic phase are the models of figure 
\ref{figure1} (a), (c), (d) and the special case of (b) where both blocks have at 
least one vertex in the surface.  Note that although we will be working only
in the cubic lattice, our methods and results generalise, with minor changes, to
the $d$-dimensional hypercubic lattice (with the adsorbing surface being a $d{-}1$ dimensional
hyperplane, and the positive half-lattice defined so that its boundary is the adsorbing surface).

The plan of the paper is as follows.  In section \ref{sec:review} we give a brief 
review of some results about adsorbed and pulled self-avoiding walks and in 
section \ref{sec:lambda} we prove some results about the free energy of
pulled self-avoiding walks that will be useful later in the paper.  We examine in 
section \ref{sec:loops}  the behaviour of a self-avoiding walk model of a diblock 
copolymer where one end of the walk is attached to the surface and both blocks 
have at least one vertex in the surface (a special case of figure \ref{figure1} (b)).  
This is motivated by a directed walk model of this situation where a novel mixed 
phase was discovered \cite{Iliev}.  In section \ref{sec:diblock} we consider diblock 
copolymers either pulled at an end vertex (figure \ref{figure1} (a)) or at the 
central vertex (figure \ref{figure1} (b)), where we make use of the results derived in 
section \ref{sec:loops}.  Triblock $ABA$-copolymers are considered in 
section \ref{sec:triblock} with the force applied at the 
end (figure \ref{figure1} (c)) and central (figure \ref{figure1} (d)) vertices. 
For each model we derive expressions for the free energy and use
these to establish the form of the phase diagram.  The paper
ends with a brief discussion in section \ref{sec:discussion}.

\section{A brief review}
\label{sec:review}

In this section we give a brief review of some results about the adsorption 
of cubic lattice self-avoiding walks at a surface and the way that self-avoiding walks 
respond to applied tensile forces.  These results will be useful in the following 
sections.

Consider the simple cubic lattice $\IntZ^3$ and attach a coordinate 
system $(x_1,x_2,x_3)$ so that the vertices have integer coordinates.  
Let $c_n$ be the number of self-avoiding walks with $n$ edges, starting 
at the origin.  Hammersley \cite{Hammersley1957} showed that
\begin{equation}
\log 3 < \inf_{n>0} \sfrac{1}{n} \log c_n 
= \lim_{n\to\infty} \sfrac{1}{n} \log c_n = \log \mu_3 < \log 5
\end{equation}
where $\mu_3$ is the \textit{growth constant}.  Self-avoiding walks that 
start at the origin and where the $x_3$-coordinate of each vertex is 
non-negative are called \textit{positive walks}.  We write $c_n^+$ for 
the number of $n$-edge positive walks and we know that 
$\lim_{n\to\infty} \sfrac{1}{n} \log c_n^+ = \log \mu_3$ 
\cite{Whittington1975}.

Let $c_n(v,h)$ be the number of $n$-edge positive walks with $v{+}1$ 
vertices in  the surface $x_3=0$ and with the $x_3$-coordinate of the 
last vertex equal to $h$.  We say that the walk has $v$ \textit{visits} and 
the last vertex has \textit{height} equal to $h$.  Define the partition function
\begin{equation}
C_n(a,y) = \sum_{v,h} c_n(v,h)\, a^v y^h,
\label{Cnay}
\end{equation}
where $a=\exp(-{\epsilon}/{k_BT})$ and $y=\exp({F}/{k_BT})$ are 
the Boltzmann weights or activities associated with the monomer-surface 
interaction energy $\epsilon$ and the pulling force $F$ (in energy units),  respectively.
In this case, the pulling force $F$ is acting on the last vertex of the walk 
and we say the walk is being pulled from its endpoint. 

If the positive walk interacts with the surface but is not subject to 
a force then $y=1$.  The (reduced) free energy in this case is given by 
\begin{equation}
\label{walk-ads}
\kappa(a) = \lim_{n\to\infty} \sfrac{1}{n} \log C_n (a,1)
\end{equation}
and there exists a critical value of $a$, $a_c > 1$, such that 
$\kappa(a) = \log \mu_3$ when $a \le a_c$ and $\kappa(a) > \log \mu_3$ 
when $a > a_c$.  The free energy $\kappa(a)$ is singular at $a=a_c > 1$
\cite{HTW,Rensburg1998,Madras} and $\kappa(a)$ is a convex function of 
$\log a$ \cite{HTW}.

If the walk is subject to a force but does not interact with the (impenetrable) 
adsorbing surface then $a=1$ and the free energy is
\begin{equation}
\lambda(y) = \lim_{n\to\infty} \sfrac{1}{n} \log C_n (1,y).
\label{FE:lambda}
\end{equation}
The free energy
$\lambda(y)$ is a convex function of $\log y$ \cite{Rensburg2009} and 
is singular at $y=1$ \cite{Beaton2015,IoffeVelenik,IoffeVelenik2010}. 
When $y \le 1$, $\lambda(y)=\log\mu_3$ (it is a constant)  and $\lambda(y)$ 
is strictly increasing in $y$ for $y>1$. When $y > 1$ the walk is in 
a ballistic phase \cite{Beaton2015}.  

In the general situation where $a > 0$ and $y > 0$ the limit defining the 
free energy exists \cite{Rensburg2013} and the free energy is given by
\begin{equation}
\psi_e(a,y) = \lim_{n\to\infty} \sfrac{1}{n} \log C_n (a,y) = 
\max[\kappa(a), \lambda(y)],
\label{eqn:psicondition}
\end{equation}
where the subscript ``$e$" refers to the fact that the walk, in this case, 
is being pulled at its endpoint. When $a \le a_c$ and $y \le 1$, 
$\psi_e(a,y) = \log \mu_3$ and the walk 
is in a \textit{free phase}.  For $a > a_c$ and $y > 1$ there is a phase 
boundary in the $(a,y)$-plane along the curve given by $\kappa(a) = \lambda(y)$.  
This phase transition between the ballistic and adsorbed phases is 
first order \cite{Guttmann2014}.

A \textit{loop} is a positive walk with both vertices of degree 1 in the
adsorbing surface $x_3=0$.  If the loop is pulled at its mid-point but 
does not interact with the adsorbing surface (so that $a=1$), then the 
free energy is $\lambda(y^{1/2})$ \cite{Rensburg2017}.  This free energy 
is unchanged if we require that only the vertices of degree 1 are in 
the surface or if we require that the loop is \textit{unfolded} in the $x_1$-direction 
\cite{HTW,HammersleyWelsh}. Various definitions of unfolded have 
been used in the literature but here we mean the following:  If the vertices
along the loop are labelled $j$ with $j=0,1,2,\ldots n$ then the 
$x_1$-coordinate of the $0$-th vertex is strictly less than that of any 
other vertex and the $x_1$-coordinate of the $n$-th vertex is at least 
as large as that of any over vertex.   That is, if $x_1(j)$ is the $x_1$-coordinate 
of the $j$-th vertex, then $x_1(0) < x_1(j) \leq x_1(n)$ for $0 < j \leq n$.

In the case that the loop (or unfolded loop)  is not subject to a force ($y=1$) 
(but the vertices interact with the adsorbing surface) then the free energy is 
the same as that of positive walks \cite{HTW}, that is, if the partition function 
of loops interacting with the surface is $L_n(a)$ then \cite{HTW} 
$L_n(a)=e^{\kappa(a)\,n + o(n)}$.  If we consider unfolded loops with partition
function $L_n^{\dagger}(a)$ then \cite{HTW}
\begin{equation}
\label{loop-ads}
L_n^{\dagger}(a) \le L_n(a) \le  L_n^{\dagger}(a) e^{O(\sqrt{n})}.
\end{equation}
If the loop is pulled at its mid-point and interacts with the surface, then 
using the arguments of reference \cite[theorem 2]{Rensburg2019} 
the free energy can be shown to be $\psi_e(a,y^{1/2})$.  That is, this is the 
same free energy as a walk pulled from its endpoint but with a weaker force.  

We shall also make use of the properties of \textit{bridges}.  These are 
positive walks with the extra condition that the $x_3$-coordinate of the
last (the $n$-th) vertex is strictly larger than that of any other vertex:  
If the vertices are labelled $j$ with $0\leq j \leq n$, then 
$x_3(0) \leq x_3(j) < x_3(n)$ for $0\leq j <n)$.  Bridges can also be 
unfolded in the $x_1$-direction, with the above definition suitably adapted. 
Denote the number of bridges of length $n$ by $b_n$, and the number of 
unfolded bridges of length $n$ by $b_n^\dagger$.  It is known that 
$\lim_{n\to\infty} \Sfrac{1}{n} \log b_n 
= \lim_{n\to\infty} \Sfrac{1}{n} \log b_n^\dagger 
= \log \mu_3$ \cite{HammersleyWelsh}.

For a general walk pulled at its midpoint and interacting with the surface,  
the free energy is (see \cite{Rensburg2017}  and \cite[Section 3.3]{Rensburg2019})
\begin{equation}
\psi_m(a,y)=\max[\kappa(a),\sfrac{1}{2}(\lambda(y)+\log\mu_3)]
\end{equation}
(the subscript ``$m$" refers to pulling at the midpoint) and the phase boundary is 
determined by  $2\,\kappa(a)=\lambda(y)+\log\mu_3$  
\cite[figure 6]{Rensburg2017}.  It is known that 
$\psi_m(a,y)=\sfrac{1}{2}(\lambda(y)+\log\mu_3)$ for $y\geq 1$ and $a\leq a_c$  \cite{Bradly2019,Rensburg2017}. 

There has not been much work on the adsorption of copolymers where the 
underlying model is a self-avoiding walk, even without an applied force.  
For the case of a block copolymer we do know that the blocks behave 
quasi-independently in that the free energy is the sum of the free energies of 
the separate blocks \cite{Whittington1998}.  There is however rigorous work 
on the force-induced desorption for directed walk models of copolymers \cite{Iliev}.  
The model that was considered is Dyck paths pulled at their mid-point, so 
these are similar to unfolded loops with both vertices of degree 1 constrained 
to be in the surface.  The directed model allows for the determination of closed form 
expressions for the free energy and corresponding phase boundaries.  In the case of 
a diblock directed walk copolymer model, where both blocks are interacting with 
the adsorbing line, and where the walk is pulled from its midpoint, it is known 
that there exists a mixed adsorbed-ballistic phase where on average 
one subwalk  has order $n$ vertices adsorbed in the surface and the other 
subwalk is largely ballistic \cite{Iliev}.  This is in contrast to the homopolymer 
walk case where no such mixed phase exists.   These directed walk studies 
have partly motivated the work presented here by motivating the question: 
under what conditions do such adsorbed-ballistic mixed phases exist?

\section{The free energy $\boldsymbol{\lambda(y)}$ of pulled positive walks}
\label{sec:lambda}

In this section we prove some results about the free energy, $\lambda(y)$, of
endpoint-pulled positive self-avoiding walks.  These are new results which 
establish sufficient conditions for a \textit{weak} strict log-convexity of 
$\lambda(y)$  and confirm some  expected configurational properties of 
pulled walks in the ballistic phase ($y>1$).  The results will be useful in 
section \ref{sec:loops} for establishing the existence of a mixed adsorbed-ballistic 
phase for the loop version of the model in figure \ref{figure1}(b).   We define 
the required notation and establish the relevant results here in a series of lemmas.  

The free energy of pulled positive walks was defined in equation \Ref{FE:lambda}.  
Putting $a=1$ in equation \Ref{Cnay} and summing over $v$ gives
\begin{equation}
C_n(y)\equiv C_n(1,y) = \sum_{h=0}^n c_n(h)\, y^h
\label{8}   
\end{equation}
where $c_n(h)$ is the number of positive self-avoiding walks of length $n$ 
ending in a vertex at height $h$.   It follows from equation \Ref{FE:lambda} 
that $\lambda(y) = \lim_{n\to\infty} \sfrac{1}{n} \log C_n(y)$.

Notice that $c_n(h) \geq \binom{n}{h}$ (this bound is the number of 
directed paths stepping east and north with exactly $h$ north steps).  This 
shows that $\lambda(y) \geq \log (1{+}y)$.
 
More generally it is known that if $y>1$ then $\lambda(y) > \log \mu_3$
\cite{Beaton2015,IoffeVelenik,IoffeVelenik2010}.   
Together these show that $\lambda(y) > \max\{\log \mu_3,\log y \}$ 
for all $y>1$, while it is known that $\lambda(y) = \log \mu_3$ for all $y \leq 1$.  
On the other hand, overcounting all self-avoiding walks of length $n$ ending in height 
$h$ gives $c_n(h) \leq \binom{n}{h}(2d)^{n-h}$ so that 
$\lambda(y) \leq \log(2d{+}y)$ for $y>1$, and dimension $d$.  These bounds 
show that for $d=3$ and $y>1$
\begin{equation}
\log(6+y) \geq \lambda(y) \ge  \max\{\log \mu_3,\log (1+y)\} .
\label{eqn4A}  
\end{equation}
In particular, it follows that $\lambda(y) = \log y + O(\sfrac{1}{y})$ as $y\to\infty$.
 
Since $\lambda(y)$ is a strictly increasing function for all $y>1$,
\begin{equation}
\lambda(y) > \lambda(y^{1/2}) > \lambda(1) = \log \mu_3, \qquad\hbox{for all $y>1$}.
\end{equation}
Moreover, by log-convexity, for all $y>1$: 
\begin{equation}
2\,\lambda(y^{1/2}) \leq \lambda(y) + \lambda(1),
\;\hbox{and}\; \lambda(y^{\alpha}) \leq \alpha\,\lambda(y) 
+ (1-\alpha)\,\lambda(1), \qquad\hbox{for all $\alpha\in[0,1]$};
\end{equation}
and, for any $0\leq \alpha_1 <  \alpha_2 < 1/2$,
\begin{equation}
\label{log-convex}
2\,\lambda(y^{1/2}) \leq \lambda(y^{\alpha_2}) + \lambda(y^{1-\alpha_2})  \leq  \lambda(y^{\alpha_1}) + \lambda(y^{1-\alpha_1}).
\end{equation}
It is not known that $\lambda(y)$ is strictly log-convex, however, if it were, 
all the inequalities above would be strict. Using equation \Ref{eqn4A}, 
the following lemma proves a property weaker than strict log-convexity, 
but stronger than log-convexity, for sufficiently large $y$.

\begin{lemm}
Suppose that $\delta\not=\sfrac{1}{2}$ and $\delta \in[0,1]$.  
Then there is a $y_{\delta}\geq 1$ (a function of $\delta$) such that 
$$2\,\lambda(y^{1/2}) < \lambda(y^\delta) + \lambda(y^{1-\delta}),\,\hbox{for all $y>y_{\delta}$} .$$
\label{lemma4B}   
\end{lemm}

\noindent\textit{Proof:}
For $y>1$ and for any $\alpha\in[0,1]$, by equation \Ref{eqn4A}, 
$\log(6{+}y^{\alpha}) \geq \lambda(y^{\alpha}) \geq \log(1{+}y^{\alpha})$.
Thus, given $\delta\not=\sfrac{1}{2}$ and $\delta \in[0,1]$, and respectively 
considering $\alpha=\delta$ and $\alpha = \shalf$ gives:
$$ \lambda(y^\delta)+\lambda(y^{1-\delta}) \geq
\log(1+y^\delta)+\log(1+y^{1-\delta})
\quad\hbox{and}\quad
 2\, \log(6+y^{1/2}) \geq 2\,\lambda(y^{1/2}) . $$
Thus if $2\log(6{+}y^{1/2})< \log(1{+}y^\delta)+\log(1{+}y^{1-\delta})$, then  
$2\,\lambda(y^{1/2})  <  \lambda(y^\delta)+\lambda(y^{1-\delta})$.
Exponentiating and simplifying shows that 
$$ 2\,\lambda(y^{1/2})  < \lambda(y^\delta)+\lambda(y^{1-\delta})
\;\hbox{provided that}\;
35+12\,y^{1/2} <y^{1-\delta}+y^\delta.$$
If $\delta \not= \sfrac{1}{2}$ then $\max\{\delta, 1{-}\delta\} >\shalf$ and 
hence there is a $y_{\delta}$ (a function of $\delta$) such that for all 
$y>y_{\delta}$ it is the case that $35+12\,y^{1/2} < y^{1-\delta}+y^{\delta}$.
\hqed
 
Notice that it is sufficient to choose $\log y_\delta = (2\log 24)/(2\delta{-}1)$ 
in lemma \ref{lemma4B}.

Since for $\delta$ and $y> y_\delta$ as in lemma \ref{lemma4B} we have 
$2\,\lambda(y^{1/2})  <  \lambda(y^\delta)+\lambda(y^{1-\delta}) $, then it follows 
from the continuity and log-convexity of $\lambda(y)$ that for any 
$\alpha_1, \alpha_2 \in [0,1]$ such that $\max\{\alpha_1, 1{-}\alpha_1\} 
> \max\{\alpha_2, 1{-}\alpha_2\}>  \max\{\delta, 1{-}\delta\}$,
\begin{equation}
\label{log-convex-strict}
 2\,\lambda(y^{1/2})  < \lambda(y^\delta) + \lambda(y^{1-\delta}) 
 <  \lambda(y^{\alpha_2}) + \lambda(y^{1-\alpha_2})  <  \lambda(y^{\alpha_1}) + \lambda(y^{1-\alpha_1}) , 
 \end{equation}
for all $y> y_{\delta}$.  This gives a strict version of equation \Ref{log-convex} 
for sufficiently large $y$ and a reduced range of $\alpha_1,\alpha_2$.  Note also 
that the smallest bound from the proof  is $y_1=y_0\leq  6{+}\sqrt{37}$ and 
hence for all $y> 6{+}\sqrt{37}$,
\begin{equation}
2\,\lambda(y^{1/2})  < \lambda(1)+\lambda(y) = \log \mu_3 +\lambda(y).
\end{equation}
 
While we haven't proved strict convexity of $\lambda(y)$ for $y>1$, numerical
data in the square lattice is consistent with this \cite{Guttmann2014}.  The 
general properties of $\lambda(y)$ in the cubic lattice is similar to the 
square lattice case, and so we expect that $\lambda(y)$ should 
be strictly convex in the ballistic phase in the cubic lattice.
We make the following conjecture in the square and cubic lattices:
\begin{conj}
Suppose that $\delta\not=\sfrac{1}{2}$ and $\delta \in[0,1]$.  
Then $$2\,\lambda(y^{1/2}) 
< \lambda(y^\delta) + \lambda(y^{1-\delta}),\,\hbox{for all $y>1$} . \eqno\hqed$$
\label{conjecture}   
\end{conj}

Next we establish relationships between some configurational properties of 
pulled walks in the ballistic phase.  In particular we prove that, for a given $y$,  
the limiting values of the  ``average height per walk edge" and the 
``most popular height per walk edge" are the same.  We use some known results 
about the (microcanonical) density function of pulled walks to connect the two.   
We establish properties about each of the relevant quantities separately and 
then use them to obtain the final result in lemma \ref{lemma6B}. 

\subsection{The average height of pulled walks}

Since $\lambda(y)$ is log-convex it is continuous and differentiable 
almost everywhere.  Further, the function 
$\beta^*(y) =\sfrac{d}{d{\log y}} \lambda(y)=y\sfrac{d}{dy} \lambda(y)$, 
where it exists,  is monotonic increasing and thus (by Lebesgue's theorem) 
is differentiable almost everywhere.  Since $\lambda(y)$ is constant for 
$y\leq 1$ and strictly increasing for $y>1$, it also follows that $\beta^*(y) > 0$ 
for $y>1$. Note that whenever $\lambda(y)$ is differentiable, then
\begin{equation}
\beta^*(y) = y\sfrac{d}{dy} \lambda(y)
= \lim_{n\to\infty} \Sfrac{1}{n} \, \langle h(y)\rangle_n
= \lim_{n\to\infty} \Sfrac{1}{n}\, 
\frac{\sum_h h\, c_n(h)\,y^h}{\sum_h c_n(h)\,y^h} \qquad \hbox{for $\alev$ $y>0$},
\label{eqnbeta*}
\end{equation}
where $\langle h(y) \rangle_n$ is the average height of the endpoint of 
the walk of length $n$.  (When $y$ is fixed, the notation is simplified 
to $\langle h \rangle_n=\langle h(y) \rangle_n$.)  Thus $0 \leq \beta^*(y) \leq 1$, 
and $\beta^*(y)$ is asymptotic (as $y\to\infty$) to $1$ (this follows, for example, 
from equation \Ref{eqn4A}).

\begin{lemm}
The function $\beta^*(y) < 1$ for almost all $y>0$.
\label{lemmabeta}  
\end{lemm}

\noindent{\it Proof:}
If $y\leq 1$ then $\lambda(y) = \log \mu_3$ and $\beta^*(y) = 0 $.  

In addition, $\lambda(y) > \log y$ by equation \Ref{eqn4A} for all $y>1$, 
and $\lambda(y) \simeq \log y$ in the sense that 
$\lim_{y\to\infty} \lambda(y) / \log y = 1$.

Thus, suppose that $y>1$ and suppose that there exists a smallest $y_1$ 
such that $\beta^*(y_1) = 1$.  Since $\beta^*(y)$ is monotonic increasing, 
this shows that $\beta^*(y) \geq 1$ for all $y\geq y_1$ (except for a set of 
measure zero).  Integration of both sides of the inequality 
$\sfrac{d}{d\log y} \lambda(y) \geq 1$ for $y\geq y_1$ gives:
$$ \lambda(y)  - \lambda(y_1) \geq \int_{y_1}^y d\log y = \log y - \log y_1 . $$
This gives
$$ \lambda(y) \geq \log y + (\lambda(y_1) - \log y_1). $$
Since $\lambda(y_1)-\log y_1 = C_1 > 0$ by equation \Ref{eqn4A}, 
this shows that $\log y + C_1 \leq \lambda(y) \leq \log y+\log(1+\sfrac{6}{y})$
which is a contradiction if $y$ is large enough, and so the assumption that 
there exists a $y_1$ such that  $\beta^*(y_1) = 1$ is false. \hqed

\subsection{The density function of pulled walks}

The (microcanonical) density function of pulled walks, $P_\lambda(\epsilon)$, 
is defined by the Legendre transform (see, for example, section 3.3 in 
reference \cite{Rensburg2015}):
\begin{equation}
\log P_\lambda(\eps) = \inf_{y>0} \left\{ \lambda(y) - \eps\log y \right\}.
\end{equation}
Note also that $P_\lambda(\eps)$ can be related to  the sequence $c_n(h)$ by
(see the methods of \cite{Madras88} and see \cite{Rensburg2015}, theorem 3.9)
\begin{equation*}
\log P_\lambda(\eps) 
= \lim_{n\to\infty} \Sfrac{1}{n} \log c_n(\lfloor \eps n \rfloor).
\end{equation*}

$\log P_\lambda(\eps)$ is a concave function of $\eps$ on $[0,1]$ and so 
differentiable almost everywhere on $[0,1]$.  It is also finite for
$\eps\in [0,1)$, since, for example, it can be shown that 
$\log P_\lambda(0) = \log \mu_3$ and $\log P_\lambda(\eps) \leq \log \mu_3$ 
for $\eps>0$.  The free energy is given by
\begin{equation}
\lambda(y) = \sup_{\eps\in[0,1]} \left\{ \log P_\lambda(\eps) + \eps\log y \right\} .
\label{eqnAA}
\end{equation}
Given $y$, this supremum is realised at a value of $\eps=\eps_*(y)\in[0,1]$ for 
almost every $y$ in the domain of $\lambda(y)$, that is, for almost every $y>0$.  
Hence by equation
\Ref{eqnAA}, for almost every $y>0$,
\begin{equation}
\lambda(y) = \log P_\lambda(\eps_*(y)) + \eps_*(y)\,\log y ,
\label{eqnD}  
\end{equation} 
and by the concavity of $\log P_\lambda(\eps)$, $\eps_*(y)$ is a non-decreasing 
function of $y$ on its domain.   

The next lemma establishes that $\beta^*(y)=\eps_*(y)$ for almost all $y> 0$. 
This result was previously established in \cite[section 3.2]{Rensburg2013} but we are
presenting more details of the proof here towards obtaining the results of lemma \ref{lemma6B}.

\begin{lemm}[Janse van Rensburg and Whittington  \cite{Rensburg2013} ]
 $\beta^*(y) = \eps_*(y)$ for almost all $y>0$.
\label{lemma-density}  
\end{lemm}

\noindent{\it Proof:}
Since $\log P_\lambda(\eps)$ is differentiable almost everywhere and concave in $\epsilon$,
it follows that if the equation
\begin{equation}
\Sfrac{d}{d\eps} \left( \log P_\lambda(\eps) + \eps \log y \right) = 
\Sfrac{d}{d\eps}\log P_\lambda(\eps) + \log y = 0
\label{eqnC}  
\end{equation} 
has a solution for $\epsilon\in[0,1]$, then it must be equal to $\eps_*(y)$, the location of the supremum.  That is, if there's a solution
\begin{equation}
\left[ \sfrac{d}{d\eps}\, \log P_\lambda(\eps) \right] \vert_{\eps=\eps_*(y)} = -\log y .
\label{eqnF} 
\end{equation}

On the other hand, equation \Ref{eqnC} may not have a solution for particular values
of $y$.  This occurs, in particular, if $\log P_\lambda(\epsilon)$ is linear for some $\eps$
in an interval, say $ \log P_\lambda(\eps) = \alpha + \beta \eps$ for $\eps\in[\eps_1,\eps_2]$
(and by concavity, $\log P_\lambda(\eps) < \alpha+\beta \eps$ 
if $\eps \in [0,1]\setminus[\eps_1,\eps_2]$ so that $\eps_1$ and $\eps_2$ are singular
points of $\log P_\lambda(\epsilon)$).  Since $\log P_\lambda(\eps)$ is a concave function 
of $\eps$, it is differentiable almost everywhere, except at isolated singular
points, and so the number of such singular points is countable.  In this event there 
is a jump discontinuity in $\eps_*(y)$ given by
\begin{equation}
\eps_*(y) 
\begin{cases}
\leq \eps_1, & \hbox{if $\log y < - \Sfrac{\log P_\lambda(\eps_2)-\log P_\lambda(\eps_1)}{\eps_2-\eps_1}$}\cr
\geq \eps_2, & \hbox{if $\log y > - \Sfrac{\log P_\lambda(\eps_2)-\log P_\lambda(\eps_1)}{\eps_2-\eps_1}$,}
\end{cases}
\label{eqnFF} 
\end{equation}
since the supremum in equation \Ref{eqnD} has to occur before or at $\eps_1$ 
for small values of $y$, and at or after $\eps_2$ for large values of $y$.   
Substituting $\log P_\lambda(\eps) = \alpha + \beta\eps$ shows that 
$\eps_*(y) \leq \eps_1$ if $\log y < - \beta$ and $\eps_*(y) \geq \eps_2$ 
if $\log y > - \beta$ so that there is a critical value of $y$ at $y_c=-\beta$.
Since the number of singular points in $\log P_\lambda(\eps)$ is countable, there
can only be a countable number of critical points $y_c$, that is, the set of
all these critical points has zero measure.

First, assume $y$ is chosen so that $\eps_*(y)$ is a solution of equation \Ref{eqnC}.  
Taking the derivative of equation \Ref{eqnD} with respect to $\log y$ gives
\begin{equation}
\beta^*(y)= y \sfrac{d}{dy}\, \lambda(y)
= y\left[  \sfrac{d}{dy}\, \eps_*(y)\right]\,
 \left[ \sfrac{d}{d\eps}\, \log P_\lambda(\eps) \right] \vert_{\eps=\eps_*(y)}
+ \eps_*(y) + y \log y \, \left[ \sfrac{d}{dy} \eps_*(y) \right]  .
\end{equation}
By equation \Ref{eqnF}, the above simplifies to $\eps_*(y) = y\sfrac{d}{dy} \lambda(y)=\beta^*(y)$
(for every $y$ where equation \Ref{eqnC} has a solution).

On the other hand, if $y$ is chosen such that $\eps_*(y)$ is not a solution of equation 
\Ref{eqnC}, then it is either a critical point as in equation \Ref{eqnFF}, or there is a
jump-discontinuity in $\sfrac{d}{d\eps} \log P_\lambda(\eps)$ in which case
$\eps_*(y)$ is a constant function for $y$ in an interval.  Suppose that
$\eps_*(y) = C$ for (say) $y\in [y_a,y_b]$.  It follows that for $y$ in this interval,
\begin{equation}
\beta^*(y)= y \sfrac{d}{dy}\, \lambda(y)
= y\sfrac{d}{dy} \L \log P_\lambda (C) + C\log y \R = C = \eps_*(y).
\end{equation}

In other words, $\beta^*(y)=y \sfrac{d}{dy}\, \lambda(y) = \eps_*(y)$ except for 
$y$ equal to a critical point -- that is $y \sfrac{d}{dy}\, \lambda(y) = \eps_*(y)$ 
for almost all $y\in [y_a,y_b]$.  By equation \Ref{eqnF}, the above simplifies to 
$\eps_*(y) = y\sfrac{d}{dy} \lambda(y)=\beta^*(y)$ for almost all $y\in [0,\infty)$ 
so that
\begin{equation}
\eps_*(y) 
= \lim_{n\to\infty} \Sfrac{1}{n} \, \langle h\rangle_n
= \lim_{n\to\infty} \Sfrac{1}{n}\, 
\frac{\sum_h h\, c_n(h)\,y^h}{\sum_h c_n(h)\,y^h} \qquad \hbox{for $\alev$ $y>0$},
\label{eqn16A}
\end{equation}
where $\langle h \rangle_n$ is as introduced in equation \Ref{eqnbeta*}.  By 
the arguments preceding lemma \ref{lemma4B} and by lemmas
\ref{lemma4B} and \ref{lemmabeta}, $0 \leq \eps_*(y) < 1$ for all $y>0$.  
Since $\lambda(y)$ is a convex function of $\log y$ and $\lambda(y)>\lambda(1)$ 
for $y>1$, $\eps_*(y)$ is a strictly increasing function if $y>1$.  \hqed

\subsection{The most popular height of pulled walks} 
Given a $y>0$, let $h^*_n(y)$ (when $y$ is fixed we denote this by $h^*_n$) be a most
popular height of the endpoint of a pulled $n$-step walk 
(so that $h_n^*$ maximizes $c_n(h)\,y^h$):
\begin{equation}
c_n(h^*_n)\, y^{h^*_n} \leq \sum_h c_n(h)\, y^h \leq (n+1)\, c_n(h^*_n)\, y^{h^*_n} .
\label{eqn13C} 
\end{equation}
The free energy is therefore given by
\begin{equation}
\lambda(y) = \lim_{n\to\infty} \Sfrac{1}{n} \log \sum_h c_n(h)\,y^h
= \lim_{n\to\infty} \Sfrac{1}{n} \log c_n(h^*_n)\, y^{h^*_n}.
\label{eqn14C}   
\end{equation}
We show next that there exists an $\eps=\eps^*(y)$ (which is a function of $y$) 
such that the limits
\begin{equation}
\lim_{n\to\infty} \Sfrac{1}{n} h^*_n = \eps^*(y)\, \quad\hbox{and}\quad
\log P_\lambda(\eps^*(y)) = \lim_{n\to\infty}\Sfrac{1}{n} \log c_n(h^*_n)
\label{25A}
\end{equation}
exist, and so that 
\begin{equation}
\lambda(y) = \log P_\lambda(\eps^*(y)) + \eps^*(y) \log y .
\label{eqn16}   
\end{equation}
Comparison to equation \Ref{eqnD} then shows that $\eps^*(y)=\eps_*(y)$, 
and it follows that for almost every $y>0$, 
$\lim_{n\to\infty} \Sfrac{1}{n} h^*_n 
= \lim_{n\to\infty} \Sfrac{1}{n} \, \langle h\rangle_n$.

\begin{lemm}
For almost every $y>0$,
$$
\lim_{n\to\infty} \Sfrac{1}{n} h^*_n 
= \lim_{n\to\infty} \Sfrac{1}{n} \langle h\rangle_n = \eps_*(y).
$$
In addition, if $0<y< 1$, then $\eps_*(y) = 0$, and if $y>1$, then $\eps_*(y)>0$. 
\label{lemma6B}   
\end{lemm}

\noindent{\it Proof:}
Let $y>0$ and let $h_n^*$ be the smallest value of $h$ maximising $c_n(h)\,y^h$
(that is, $h_n^*$ is the smallest most popular value of $h$ and is a function of $y$).  
Clearly, $0 \leq h_n^* \leq n$.

Multiply equation \Ref{eqn13C} by $y^{-\lfloor \eps n\rfloor}$ 
(for $\eps\in [0,1]$), take logarithms, divide by $n$ and take $n\to\infty$.  
Since the limit in equation \Ref{eqn14C} exists, 
this shows that
\begin{equation}
\inf_{y>0} \LC
\lim_{n\to\infty} \Sfrac{1}{n} \log \left(
c_n(h_n^*)\,y^{h_n^* - \lfloor \eps n \rfloor} \right) \RC
= \inf_{y>0} \LC \lambda(y) - \eps \log y \RC 
= \log P_\lambda(\eps)
\label{27}
\end{equation}
and this is finite (for $\eps\in[0,1)$).  In particular, we note that 
$\lambda(y) - \eps \log y$ is a continuous, non-decreasing and 
log-convex function of $y>0$ for $\eps\in[0,1)$ and $\lambda(y)$ 
is asymptotic to $\log y$.  This shows that the infimum of 
$\lambda(y)-\eps\log y$ is realised at a finite value of $y$ so that 
for $\eps\in [0,1)$, it follows that
\begin{equation}
- \infty < \log P_\lambda(\eps)
= \min_{y\geq0} \left\{   \lim_{n\to\infty} \Sfrac{1}{n} \log \left(
c_n(h_n^*)\,y^{h_n^* - \lfloor \eps n \rfloor} \right)\right\} 
= \min_{y\geq 0} \LC \lambda(y) - \eps\log y \RC < \infty .
\label{eqnE}
\end{equation}

We now present two proofs of the lemma.

\smallskip

\textit{First proof:}
$P_\lambda(\eps)$ is given by 
(see the methods of \cite{Madras88} and see \cite{Rensburg2015}, theorem 3.9)
\begin{equation*}
\log P_\lambda(\eps) 
= \lim_{n\to\infty} \Sfrac{1}{n} \log c_n(\lfloor \eps n \rfloor) .
\end{equation*}

Given $y>0$, define $\zeta_y = \limsup_{n\to\infty} \sfrac{1}{n}\, h_n^*$ 
and suppose that $\{ n_k\}$ is a subsequence realising the limsup:
\begin{equation}
\zeta_y = \lim_{k\to\infty} \Sfrac{1}{n_k} \log h_{n_k}^* .
\label{29A}
\end{equation}
Since this limit exists, it follows that
\begin{equation}
\lambda(y) = \lim_{k\to\infty} \Sfrac{1}{n_k} 
\log \L c_{n_k}(h^*_{n_k})\, y^{h^*_{n_k}} \R
= \lim_{k\to\infty} \Sfrac{1}{n_k} \log c_{n_k}(h^*_{n_k}) + \zeta_y \, \log (y) .
\label{35A}
\end{equation}
This will be simplified and then compared to equation \Ref{eqnD}.

Since $h_n^*$ is a most popular height, it is the case that
\begin{equation*}
 \lim_{k\to\infty} \Sfrac{1}{n_k} 
 \log \L c_{n_k}(\lfloor \zeta_y n_k \rfloor)\, y^{\lfloor \zeta_y n_k \rfloor} \R
\leq \lim_{k\to\infty} \Sfrac{1}{n_k} 
\log \L c_{n_k}(h^*_{n_k})\, y^{h_n^*} \R .
\end{equation*}
Since $\lim_{k\to\infty} \sfrac{1}{n_k} \lfloor \zeta_y n_k \rfloor
= \lim_{k\to\infty} \sfrac{1}{n_k} h_{n_k}^* = \zeta_y$ it follows that
\begin{equation}
\log P_\lambda(\zeta_y) 
= \lim_{k\to\infty} \Sfrac{1}{n_k} \log c_{n_k}(\lfloor \zeta_y n_k \rfloor)
\leq \lim_{k\to\infty} \Sfrac{1}{n_k} \log c_{n_k}(h^*_{n_k}) .
\label{37A}
\end{equation}

Let  $\nu>0$ be a small number.  Observe that for large enough (but finite) 
$k$ (say $k\geq K$) it is the case that
$\lfloor (\zeta_y - \nu)n_{k} \rfloor
< h^*_{n_k} 
< \lfloor (\zeta_y + \nu)n_{k} \rfloor$. Thus 
\begin{equation}
\lim_{k\to\infty} \Sfrac{1}{n_k} \log c_{n_k}(h^*_{n_k})
\leq \begin{cases} 
\displaystyle
\lim_{k\to\infty} \sfrac{1}{n_k} 
\log \sum_{\ell=0}^{\lfloor (\zeta_y + \nu)n_{k} \rfloor} c_{n_k}(\ell) 
= \log P_{\lambda}({\leq}(\zeta_y+\nu)) \\
\displaystyle
\lim_{k\to\infty} \sfrac{1}{n_k} 
\log \sum_{\ell=\lfloor (\zeta_y - \nu)n_{k} \rfloor}^n c_{n_k}(\ell)
= \log P_{\lambda}({\geq}(\zeta_y-\nu))
\end{cases}
\label{38A}
\end{equation}
where $P_\lambda({\leq}\epsilon)$ and $P_\lambda({\geq}\epsilon)$ are
\textit{integrated density functions} (\cite{Rensburg2015}, section 3.4) 
defined by
\begin{equation*}
P_{\lambda}({\leq}\nu) = \lim_{n\to\infty} 
\Sfrac{1}{n} \log \L \sum_{m=0}^{\lfloor \nu n \rfloor} c_n(m)\,y^m \R
\hspace{3mm}\hbox{and}\hspace{3mm}
P_{\lambda}({\geq}\nu) = \lim_{n\to\infty} 
\Sfrac{1}{n} \log \L \sum_{m=\lfloor \nu n \rfloor}^{n} c_n(m)\,y^m \R .
\end{equation*}
It is the case that \cite{Rensburg2015} (theorem 3.16)
\begin{equation*}
\log P_\lambda(\epsilon) = \min\{ \log P_\lambda({\leq}\epsilon) , \log P_\lambda({\geq}\epsilon)\}  .
\end{equation*}
Since $P_\lambda({\leq}\epsilon)$ and $P_\lambda({\geq}\epsilon)$ 
are continuous functions, take $\nu\to 0^+$ in equation \Ref{38A} to see that 
\begin{equation*}
\lim_{k\to\infty} \Sfrac{1}{n_k} \log c_{n_k}(h^*_{n_k})
\leq \min\{  \log P_\lambda({\leq}\zeta_y) , \log P_\lambda({\geq}\zeta_y)\}
= \log P_\lambda(\zeta_y).
\end{equation*}
Together with equation \Ref{37A} this shows that
\begin{equation*}
\lim_{k\to\infty} \Sfrac{1}{n_k} \log c_{n_k}(h^*_{n_k}) = \log P_\lambda(\zeta_y) .
\end{equation*}
Substitute this result in equation \Ref{35A} to obtain
\begin{equation*}
\lambda(y) = \log P_\lambda(\zeta_y) + \zeta_y \log y .
\end{equation*}
Put $\eps^*(y)=\zeta_y$ (see equation \Ref{25A}).
Comparison to equation \Ref{eqn16} gives the result that 
$\eps_*(y) = \zeta_y = \eps^*(y)$.  This shows, in particular, that
\begin{equation}
\eps_*(y) = \lim_{n\to\infty} \Sfrac{1}{n} \langle h\rangle_n 
= \limsup_{n\to\infty} \sfrac{1}{n}\, h_n^*, \qquad \hbox{for $\alev$ $y>0$} .
\label{44}
\end{equation}

The above arguments remain unchanged if we defined 
$\zeta_y = \liminf_{n\to\infty} \sfrac{1}{n}\, h_n^*$ instead.  
This shows that the limsup in equation \Ref{44} is a limit, with the result
that
\begin{equation*}
\eps_*(y) = \lim_{n\to\infty} \Sfrac{1}{n} \langle h\rangle_n 
= \lim_{n\to\infty} \sfrac{1}{n}\, h_n^*,  \qquad \hbox{for $\alev$ $y>0$} .
\end{equation*}

\smallskip

\textit{Second proof:}
Define $\zeta_y$ as in equation \Ref{29A}.  Then $\lambda(y)$ is
again given in equation \Ref{35A}.  Moreover, for each fixed value of $y$,
\begin{equation*}
\lim_{k\to\infty} \Sfrac{1}{n_k}
 \log \L c_n(h_n^*)\,y^{h_n^* - \lfloor \eps n \rfloor} \R
= \lim_{k\to\infty} \Sfrac{1}{n_k} \log c_{n_k}(h_{n_k}^*)
  + (\zeta_y - \eps) \log y .
\end{equation*}
Define $f(y) = \lim_{k\to\infty} \sfrac{1}{n_k} \log c_{n_k}(h_{n_k}^*)$.
The function $f(y)$ is bounded for $y\geq 0$ and by equations \Ref{eqnE}
and \Ref{35A},
\begin{equation}
\log P_\lambda(\eps) 
= \min_{y\geq 0} \LC f(y) + (\zeta_y -\epsilon) \log y \RC .
\label{31}
\end{equation}

Suppose that $\eta>0$ is small and fixed, and that 
$\zeta_y > \epsilon + \eta$ for all $y\geq 0$.  In this case
the minimum in equation \Ref{31} is unbounded by either taking
$y\to 0^+$, or by taking $y\to\infty$. Similarly, if $\zeta_y < \epsilon -\eta$
for all $y\geq 0$, then the minimum is again unbounded. This is a
contradiction since $\log P_\lambda(\eps)$ is finite for
fixed $\eps \in [0,1)$.  Thus either (1) $\lim_{y\to\infty} \zeta_y = \epsilon$, or 
(2) there exists a \textit{finite} $y_1$ 
minimising the right hand side of equation \Ref{31},
such that $\zeta_{y_1} = \epsilon$.

We now rule out (1) for $\epsilon \in [0,1)$.  Our claim is that
$\lim_{y\to\infty} \zeta_y = 1 \not= \epsilon$.  To see this, suppose
that $\lim_{y\to\infty} \zeta_y = \nu < 1$.  Let $\tau>0$ be small 
enough (say $\tau<1-\nu$). Then there exists a large but finite $Y$ such that 
$\zeta_y \leq \shalf (\nu+1-\tau)$ for all $y\geq Y$.
This shows that there is a $K$ such that
$\sfrac{1}{n_k} h_{n_k}^* < \shalf (\nu+1-\tau) + \shalf \tau
= \sfrac{1}{2}(1+\nu)$ for all $k > N$ (and $y\geq Y$).  
This shows that for all $k>N$, 
\begin{equation*}
c_n(h_{n_k}^*) y^{h_{n_k}^*} 
< c_n(h_{n_k}^*) y^{(\nu+1){n_k}/2} .
\end{equation*}
By taking logarithms, dividing by $n_k$ and taking $k\to\infty$, 
$\lambda(y) \leq \log \mu_3 + \sfrac{1}{2}(\nu+1)\log y
< \log y$ for large $y$.  This is a contradiction, thus
$\lim_{y\to\infty} \zeta_y = 1 \not= \epsilon$.  In other words,
only (2) remains, so that there is a finite $y_1$ (a function of $\eps$) so that
$\zeta_{y_1}=\epsilon$.  Let this value of $\eps$ corresponding
to $y_1$ be denoted by $\epsilon^*(y_1) = \zeta_{y_1}$.  By equation
\Ref{eqnE} it also follows that $\log P_\lambda(\eps^*(y_1)) = 
\lim_{k\to\infty} \sfrac{1}{n_k} \log c_{n_k}(h_{n_k}^*)$.

By equation \Ref{eqn14C},
\begin{equation*}
\lambda(y_1) 
= \lim_{n\to\infty} \Sfrac{1}{n} \log c_n(h^*_n)\, y_1^{h^*_n}
= \log P_\lambda(\eps^*(y_1)) + \eps^*(y_1) \log y_1 .
\end{equation*}
Comparison to equation \Ref{eqnD} shows that $\eps^*(y) = \eps_*(y)$
for almost every $y>0$ (that is, whenever $\eps_*(y)$ exists).  
Since $\eps_*(y)$ is given by equation \Ref{eqn16A}, this shows that
$\limsup_{n\to\infty} \sfrac{1}{n} h^*_n 
= \lim_{n\to\infty} \sfrac{1}{n} \langle h\rangle_n$ 
for almost every $y>0$. 

Similarly, instead defining $\zeta_y = \liminf_{n\to\infty} \sfrac{1}{n} h_n^*$, 
it follows that $\liminf_{n\to\infty} \sfrac{1}{n} h^*_n 
= \lim_{n\to\infty} \sfrac{1}{n} \langle h\rangle_n$
for almost every $y>0$.  In other words, $\lim_{n\to\infty} \sfrac{1}{n} h^*_n 
= \lim_{n\to\infty} \sfrac{1}{n} \langle h\rangle_n$
for almost every $y>0$.  This completes the second proof.

Finally, since $\lambda(y) = \log \mu_3$ for $0\leq y \leq 1$ and since it 
is a continuous function, it follows that $\eps_*(y) = 0$ if $0< y < 1$.  
Also, since $\lambda(y) > \log \mu_3$ for $y>1$ 
\cite{Beaton2015,IoffeVelenik,IoffeVelenik2010},
$\eps_*(y)$ is strictly increasing for $y>1$.  This completes the proof. \hqed

\section{Pulled adsorbing $\boldsymbol{AB}$-diblock loops}
\label{sec:loops}

\begin{figure}[t!]
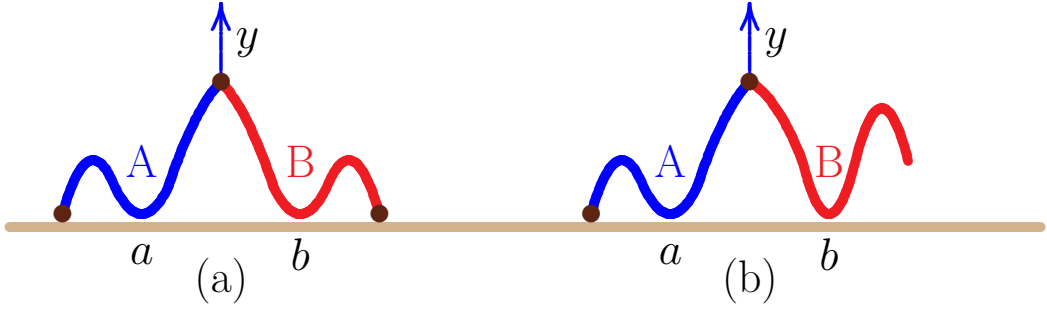

\beginpicture
\setcoordinatesystem units <1pt,1pt>
\setplotarea x from -55 to 300, y from -10 to 90
\setplotarea x from 0 to 300, y from 0 to 90

\setplotsymbol ({\large.})
\arrow <10pt>  [.2,.67] from 60 50 to 60 80
\arrow <10pt>  [.2,.67] from 260 50 to 260 80

\setplotsymbol ({\footnotesize$\bullet$})

\color{Tan}
\plot -20 -5 370 -5 /

\setcoordinatesystem units <1pt,1pt> point at 200 0 

\setquadratic
\color{Blue}
\plot 200 0 210 20 220 10 230 0 240 10 250 35 260 50   /  \put {\LARGE A} at 230 20
\color{Red}
\plot 260 50 270 35 280 10 290 0 300 10 310 20 320 0 /  \put {\LARGE B} at 290 20

\color{Sepia}
\multiput {\LARGE$\bullet$} at 200 0 260 50 320 0 /

\color{black}
\normalcolor

\put {\LARGE$a$} at 230 -15 
\put {\LARGE$b$} at 290 -15 
\put {\LARGE$y$} at 270 65 
\put {\LARGE(a)} at 260 -25

\setcoordinatesystem units <1pt,1pt> point at -200 0 

\setquadratic
\color{Blue}
\plot 0 0 10 20 20 10 30 0 40 10 50 35 60 50   /  \put {\LARGE A} at 30 20
\color{Red}
\plot 60 50 70 40 80 20 90 0 100 20 110 40 120 20 /  \put {\LARGE B} at 90 20

\color{Sepia}
\multiput {\LARGE$\bullet$} at 0 0 60 50 /

\color{black}
\normalcolor

\put {\LARGE$a$} at 30 -15 
\put {\LARGE$b$} at 90 -15 
\put {\LARGE$y$} at 70 65 
\put {\LARGE(b)} at 60 -25

\endpicture
\caption{Two cases of an adsorbing diblock copolymer pulled at its 
midpoint. In both cases comonomers of type $A$ adsorb with activity 
$a$ in the adsorbing line, and comonomers of type $B$ with activity $b$. 
(a) The diblock loop case with both endpoints in the adsorbing plane
(it is fixed in the adsorbing plane at its first vertex, and its last vertex 
is in the adsorbing plane, but can freely move in this plane); 
(b) The general diblock copolymer case with its first vertex fixed in 
the adsorbing plane.}
\label{figure2bp}  
\end{figure}

In this section our aim is to examine the phase diagram of adsorbing diblock 
\textit{loops} pulled in the middle (see figure \ref{figure2bp}(a)).  We will 
determine the phase diagram of this model, and later compare it to our results in
section \ref{sec:diblock}, in particular the phase diagram of a self-avoiding walk 
model of adsorbing diblock copolymers pulled in the middle, with one 
endpoint fixed at the origin and the other free (see figure \ref{figure1}(b)).   
The models in figure \ref{figure1} (namely (a), (c) and (d) of diblock 
and triblock copolymers) are simpler to analyse, and so we first focus 
on the pulled adsorbing diblock loop.

The models in figure \ref{figure2bp} are of linear copolymers with two blocks, 
labelled $A$ and $B$ (and with vertices or monomers of types $A$ or $B$
respectively).  Each block in the copolymer has length $n$ (see figures 
\ref{figure1}(a) and \ref{figure1}(b)).  The walks are positive walks with
$2n{+}1$ vertices  labelled $j=0,1, \ldots 2n$.  The vertex $0$ is 
fixed at the origin and is not weighted.   Vertices $1 \le j \le n$ are $A$-vertices 
while vertices $n{+}1 \le j \le 2n$ are $B$-vertices.  

As before, we use a positive self-avoiding walk model in the cubic lattice 
$\IntZ^3$ and coordinate system $(x_1,x_2,x_3)$.   Figure \ref{figure2bp}(a) 
shows a loop in the positive half-lattice with $x_3\geq 0$.  The adsorption 
of the copolymer model is defined by counting the numbers $v_A$ and 
$v_B$ of $A$- and $B$-vertices in the (adsorbing) plane $x_3=0$.  These
are \textit{$A$- or $B$-visits}, and they have associated weights $a$ and 
$b$ respectively (where $a = e^{-\epsilon_A/k_BT}$ and 
$b= e^{-\epsilon_B/k_BT}$ with $k_B$ the Boltzmann's constant, $T$ the 
absolute temperature, and $\epsilon_A$ and $\epsilon_B$ the 
energies associated with $A$- and $B$-visits).  

The walk adsorbs if either $a$ or $b$ is sufficiently large, and an applied 
force $F$ pulling the walk from the adsorbing plane will pull the walk
into a ballistic phase if $F$ is large enough.  $F$ is related to an activity 
$y$ by $y=e^{F/k_BT}$ (and the weight of a loop with middle vertex
of height $h$ is $y^h$ so that $F$ is conjugate to $h$ in this model).

The loop in figure \ref{figure2bp}(a) has partition function given by
\begin{equation}
\W{U}_{2n}(a,b,y) = \sum_{h=0}^n \sum_{v_A,v_B} \ell_{2n}^{AB}(v_A,v_B,h)\,a^{v_A}b^{v_B}\,y^h,
\label{eqn25C}
\end{equation}
where $\ell_{2n}^{AB}(v_A,v_B,h)$ is the number of loops of length $2n$ 
with $n$ $A$-vertices, $n$ $B$-vertices, $v_A$ and $v_B$ vertices that are
$A$- or $B$-visits, and with midpoint (the location of the last $A$ vertex) 
at height $h$ above the adsorbing plane $x_3=0$.   
General bounds on $\W{U}_{2n}(a,b,y)$ 
in terms of the partition functions of bridges and loops are obtained next 
using arguments similar to those leading up to theorem 2 in reference 
\cite{Rensburg2019}. 

First, the partition function $\W{U}_{2n}(a,b,y)$ is  bounded in the limit by
$\psi_e(a,y^{1/2})$ and $\psi_e(b,y^{1/2})$ (see equation \Ref{eqn:psicondition}).  
The bounds are obtained by comparing $\W{U}_{2n}(a,b,y)$ to homopolymer 
loops, that is, if $a\geq b$, then 
$\W{U}_{2n}(b,b,y)\leq \W{U}_{2n}(a,b,y)\leq \W{U}_{2n}(a,a,y)$.
This gives the following lemma, since, as discussed after equation \Ref{loop-ads},
a homopolymer loop pulled at its mid-point and interacting with a surface 
(with activities $a$ and $y$) has free energy given by $\psi_e(a,{y}^{1/2})$.

\begin{lemm}
Suppose that $a\geq b$. Then, by monotonicity, since the $AB$-block copolymer 
has length $2n$, 
$$
\psi_e(b,y^{1/2})
\leq \liminf_{n\to\infty} \Sfrac{1}{2n} \log \W{U}_{2n}(a,b,y)
\leq \limsup_{n\to\infty} \Sfrac{1}{2n} \log \W{U}_{2n}(a,b,y)
\leq \psi_e(a,y^{1/2})
$$
where $\psi_e(b,y^{1/2}) = \max\{\kappa(b),\lambda(y^{1/2})\}$ and
$\psi_e(a,y^{1/2}) =  \max\{\kappa(a),\lambda(y^{1/2})\}$. \hqed
\label{lemma1}   
\end{lemm}

Next, a lower bound is obtained in terms of unfolded bridges.  
Let $b_n^{\dagger}(v,h)$ denote the number of unfolded bridges with 
$v$ surface visits and endpoint at height $h$, where necessarily $v,h\geq 1$.
Then by reference
\cite[theorem 1]{Rensburg2019}, for $a,y>0$, 
\begin{equation}
\lim_{n\to\infty} \Sfrac{1}{n} \log \sum_{v,h} b_n ^{\dagger}(v,h)\, a^vy^h
= \psi_e(a,y).
\label{eqn2}  
\end{equation}
Note that 
for $0\leq y\leq 1$,
$\psi_e(a,y)=\kappa(a)$ since trivially  
$\psi_e(a,0)=\psi_e(a,1)=\kappa(a)$ and $\psi_e$ is a non-decreasing function 
of $a$ and of $y$.  In addition, for any fixed positive integer $h$ and 
for $0 < y$,
\begin{equation}
\lim_{n\to\infty} \Sfrac{1}{n} \log \sum_{v} b_n ^{\dagger}(v,h)\, a^vy^h
= \kappa(a).
\label{eqn2A}  
\end{equation}
Also, for any $y> 0$,
\begin{equation}
\lim_{n\to\infty} \Sfrac{1}{n} \log \sum_{h} b_n ^{\dagger}(1,h)\, y^h
= \lambda(y).
\label{eqn2Asecond}  
\end{equation}
Finally, a result that will be useful in section \ref{4.3}: Given $h_n^*(y)$,  
a most popular height (as defined in section \ref{sec:lambda}) for an 
$n$-step positive walk at activity $y$, standard unfolding arguments 
(which do not change the height of any vertex in any bridge) 
\cite{HammersleyWelsh} establish that for any $y> 0$, 
\begin{equation}
\lim_{n\to\infty} \Sfrac{1}{n} \log b_n ^{\dagger}(h_n^*(y))\, y^{h_n^*(y)} =
\lim_{n\to\infty} \Sfrac{1}{n} \log \sum_{h=1}^{n-1} b_n ^{\dagger}(h)\, y^{h} = 
\lambda(y),
\label{eqn-pulledbridge}  
\end{equation}
where $b_n ^{\dagger}(h) \equiv  \sum_{v} b_n ^{\dagger}(v,h)$.
In fact, and more generally, this is also true for unfolded bridges with 
exactly 1 visit, or for unfolded bridges with visits weighted by $a$ where $0 < a\leq a_c$:
\begin{equation}
\lim_{n\to\infty} \Sfrac{1}{n} \log b_n ^{\dagger}(1,h_n^*(y))\, y^{h_n^*(y)}  
= \lim_{n\to\infty} \Sfrac{1}{n} 
   \log \sum_{v,h} b_n ^{\dagger}(v,h)\, a^v y^{h}  
= \lambda(y) .
\label{eqn-pulledbridge-2}  
\end{equation}

The above results lead to the standard lower bounds given by lemma \ref{lemma2}.

\begin{lemm}[Standard (lower) bounds 1]
In the cubic lattice, for any $\delta\in[0,1]$
and $a,b,y>0$,
\begin{align*} 
\W{U}_{2n}(a,b,y) & \geq 
b\sum_h y^h \left(\sum_v b_{n-1}^{\dagger}(v,h)\, a^v \right) 
\left(\sum_w b_{n-1}^{\dagger}(w,h)\, b^w \right) \\
& =  b\sum_h \left(\sum_v b_{n-1}^{\dagger}(v,h)\, a^vy^{\delta h} \right) 
\left(\sum_w b_{n-1}^{\dagger}(w,h)\, b^wy^{(1-\delta)h} \right) . &
\label{lemma2-eqn1}
\end{align*} 

Indeed, by considering separately the two cases, 
$h=h_n^*(y)$ with $v=w=1$ and $\delta=\sfrac{1}{2}$, or just $h=1$,
in the  equation above  it follows that for all $a,b,y>0$,
\begin{equation}
\liminf_{n\to\infty} \Sfrac{1}{2n} \log \W{U}_{2n} (a,b,y) \geq
\max \{ \lambda(y^{1/2}),\sfrac{1}{2}(\kappa(a)+\kappa(b))\} .
\label{eqn10}   
\end{equation}

\hfill
\label{lemma2}  
\end{lemm}

\begin{figure}[t!]
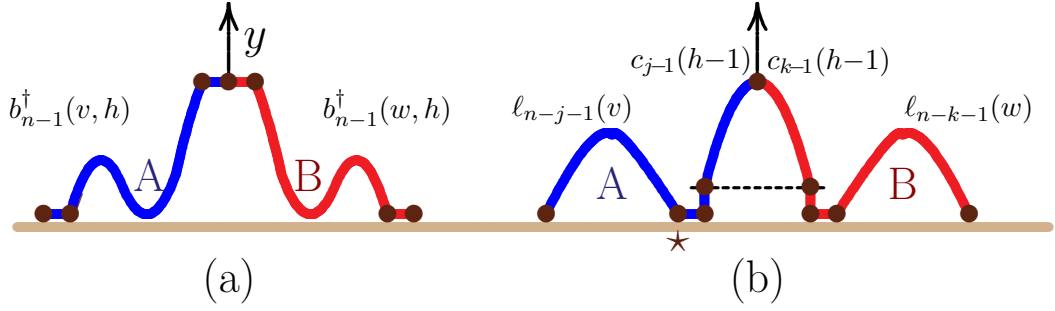

\beginpicture
\setcoordinatesystem units <1pt,1pt>
\setplotarea x from -50 to 300, y from -10 to 90
\setplotarea x from 0 to 300, y from 0 to 90

\color{black}
\normalcolor

\setplotsymbol ({\large.})
\arrow <10pt>  [.2,.67] from 60 50 to 60 80
\arrow <10pt>  [.2,.67] from 260 50 to 260 80

\setplotsymbol ({\footnotesize$\bullet$})

\color{Tan}
\plot -20 -5 370 -5 /

\setquadratic
\color{Blue}
\plot 0 0 10 20 20 10 30 0 40 15 45 35 50 50   / 
\color{Red}
\plot 70 50 75 35 80 15 90 0 100 10 110 20 120 0 /  
\setlinear
\color{Blue}
\plot -10 0 0 0 /
\plot 50 50 60 50 /
\color{Red}
\plot 60 50 70 50 /
\plot 120 0 130 0 /

\color{Sepia}
\multiput {\LARGE$\bullet$} at -10 0 0 0 50 50 60 50 70 50 120 0  130 0 /

\color{black}
\normalcolor

\put {\LARGE\textcolor{DarkBlue} A} at 30 15 
\put {$b_{n-1}^{\dagger}(v,h)$} at 0 40 
\put {\LARGE\textcolor{DarkRed} B} at 90 15 
\put {$b_{n-1}^{\dagger}(w,h)$} at 120 40 
\put {\LARGE$y$} at 70 65 
\put {\LARGE(a)} at 60 -25
\put {\LARGE(b)} at 260 -25

\put {$\ell_{n{-}j{-}1}(v)$} at 190 40 
\put {$c_{j\!{-}\!1}(h{-}1)$} at 235 58 
\put {$c_{k\!{-}\!1}(h{-}1)$} at 287 58 
\put {$\ell_{n{-}k{-}1}(w)$} at 340 40 
\put {\LARGE\textcolor{DarkBlue} A} at 205 12
\put {\LARGE\textcolor{DarkRed} B} at 315 12

\setlinear
\setplotsymbol ({$\cdot$})
\setdashes <2pt> \plot 235 10 285 10 / \setsolid
\setplotsymbol ({\footnotesize$\bullet$})
\color{Blue}
\plot 230 0 240 0 240 10 /
\color{Red}
\plot 280 10 280 0 290 0 /

\setquadratic
\color{Blue}
\plot 180 0 195 25 205 30 212.5 25 230 0 /
\plot 240 10 250 40 260 50 /
\color{Red}
\plot 260 50 270 40 280 10 /
\plot 290 0 307.5 25 315 30 325.5 25 340 0  /

\color{Sepia}
\multiput {\LARGE$\bullet$} at 180 0 230 0 240 0 240 10  260 50  
280 10 280 0 290 0 340 0  /
\put{\LARGE$\star$} at 230 -10 /

\color{black}
\normalcolor

\endpicture
\caption{(a) By reversing an unfolded bridge in the horizontal direction, 
two bridges with final vertex at height $h$ can be concatenated as shown 
to form an $AB$-diblock loop with endpoints in the adsorbing plane, and 
pulled in the middle.  This proves lemma \ref{lemma2}. (b)  The $A$-block 
of an $AB$-diblock loop with first vertex at the origin is a self-avoiding walk 
ending at height $h$ in the middle of the loop.  This $A$-block has a last 
return to the adsorbing plane (marked by $\star$) whereafter it steps 
horizontal then vertical, then to continue as a positive self-avoiding walk 
with no visits to the adsorbing plane and to end in the midpoint of the loop 
at height $h$.  The part of the $A$-block from the origin to the last return 
is an adsorbing loop.  The $B$-block can be similarly analysed.  Putting 
these parts together gives an upper bound on the number of $AB$-diblock 
loops in terms of loop and positive walk partition functions.}
\label{figurelemma23}
\end{figure}

\noindent{\it Proof:}
A loop can be built by connecting two unfolded bridges with the 
same endpoint height; a proof-by-picture is shown in 
figure \ref{figurelemma23}(a). \hqed

In figure \ref{figurelemma23}(b) a proof-by-picture is given of some 
standard upper bounds on $\W{U}_{2n}(a,b,y)$.
These bounds are constructed by using loops and positive walks to
create a pulled loop as shown.  Since these component walks
do not avoid each other, an upper bound is obtained.  To construct the $A$-block,
a loop of length $n{-}j{-}1$ is concatenated with a positive walk of length
$j{-}3$, using two steps to join the loop from the vertex marked with a star
to the first vertex of the positive walk (which is placed at height $1$ so that
the walk is completely above the adsorbing surface and makes no visits to it).
The $B$-block is similarly constructed.  The two blocks are then concatenated
into a loop by choosing the heights of the positive walks to be equal and then
placing the two endpoints together.

Taken together, these arguments show the following.
In the first place, one may consider each of the blocks to be composed of
a loop and a positive walk (each positive walk ending at the same height).
These are joined together by a few edges.  This construction gives, 
in terms of the loop partition function $L_n(a) = \sum_v \ell_n(v)\, a^v$, 
\begin{equation}
\W{U}_{2n}(a,b,y) 
\leq 4(d{-}1)^2b\sum_h y^h \left( \sum_j c_{j-1}(h{-}1) \, L_{n-j-1}(a) \right)
\left( \sum_k c_{k-1}(h{-}1) \, L_{n-k-1}(b) \right). 
\label{upperbound2}
\end{equation}
On the other hand, one may instead consider the blocks to be positive 
adsorbing walks ending at the same height. 
Then, given any $\epsilon\in[0,1]$, using $y=y^{\epsilon}\, y^{1-\epsilon}$, 
and then summing over $h$ independently in each block, 
gives the bound
\begin{eqnarray} 
& \hspace{-1.0cm}
\W{U}_{2n}(a,b,y) \leq 
b \left(\sum_h y^{\epsilon h} \sum_v c_{n} (v,h)\, a^v \right) 
\left(\sum_h y^{(1-\epsilon) h}\sum_w c_{n} (w,h)\, b^w \right) 
=  b~C_{n}(a,y^{\epsilon})\, C_{n}(b,y^{1-\epsilon})  . 
\label{upperbound1}
\end{eqnarray} 

Taking logarithms of equation \Ref{upperbound1}, dividing by $n$ and
taking $n\to\infty$, the following lemma is proven:

\begin{lemm}[Standard (upper) bound 2]
In the cubic lattice, for any $\epsilon\in[0,1]$ and $a,b,y>0$,
\begin{eqnarray}
\limsup_{n\to\infty} \Sfrac{1}{n} \log \hat{U}_{2n}(a,b,y) 
&\leq& \Sfrac{1}{2} \min_{\epsilon\in[0,1]}  \left\{ \psi_e(a,y^\epsilon) 
                                          +  \psi_e(b,y^{1-\epsilon})\right\}  \nonumber\\
&=& \Sfrac{1}{2} \min_{\epsilon\in[0,1]}  
\left\{ \max\{\kappa(a), \lambda(y^\epsilon)\}  +  
                                       \max\{\kappa(b),\lambda(y^{1-\epsilon})\}\right\} .
\label{eqn37A}  
\end{eqnarray}
\hfill\hqed
\label{lemma3}  
\end{lemm}

In the remainder of this section we show  that in fact the free energy upper 
bound in equation \Ref{eqn37A} of lemma \ref{lemma3} in all cases gives 
the free energy, that is we show that:
\begin{equation}
\W{\rho}^{AB}(a,b,y) = \lim_{n\to\infty} \Sfrac{1}{2n} 
\log \W{U}_{2n}(a,b,y) =  \Sfrac{1}{2} \min_{\epsilon\in[0,1]}  \left\{ \max\{\kappa(a), \lambda(y^\epsilon)\}  +  \max\{\kappa(b),\lambda(y^{1-\epsilon})\}\right\}.
\label{generalform}
\end{equation}

The proof of equation \Ref{generalform} starts next with the case $y\leq 1$.

\subsection{The case $\boldsymbol{y\leq 1}$}
\label{section41}
Suppose that $0<y\leq 1$. Since $y\leq1$, for all $\epsilon\in[0,1]$, $\lambda(y^{\epsilon})=\log\mu_3$. Hence there is no ballistic phase and 
the phase diagram will be determined by considering to what extent the 
$A$ and/or $B$ vertices are adsorbed in the surface.  In terms of the 
free energy, since $\kappa(b),\kappa(a)\geq \log\mu_3$, 
equations \Ref{eqn10} and \Ref{eqn37A} together give that for all $y\leq 1$:
$$\W{\rho}^{AB}(a,b,y) = \lim_{n\to\infty} \Sfrac{1}{2n} 
\log \W{U}_{2n}(a,b,y) = \sfrac{1}{2} (\kappa(a) + \kappa(b)),$$
and  this satisfies equation \Ref{generalform}.

In terms of the phase diagram, 
if $b\leq a\leq a_c$ (with $a_c$ as defined after equation \Ref{walk-ads}), 
then we expect the walk to be desorbed.  Indeed, since 
$\kappa(b)=\kappa(a)=\log\mu_3$
\begin{equation}
\W{\rho}^{AB}(a,b,y) =  \log \mu_3, 
\qquad\hbox{if $y\leq 1$ and $a,b\leq a_c$.}
\end{equation}
This establishes the existence of the $AB$-free phase. 

Next, consider the case $b \leq a_c < a$.   In this case, we expect the 
$A$-block to be adsorbed and the $B$-block to be desorbed. Since now 
$\kappa(a)>\log\mu_3$, this establishes the existence of the $A$-adsorbed 
($B$-free) phase, with free energy
\begin{equation}
\W{\rho}^{AB}(a,b,y) = \lim_{n\to\infty} \Sfrac{1}{2n} \log \W{U}_{2n}(a,b,y) = 
\Sfrac{1}{2} ( \kappa(a) +  \log \mu_3), \qquad\hbox{if $y\leq 1$ and $b\leq a_c < a$.}
\end{equation}
This also shows, by symmetry, that $\W{\rho}^{AB}(a,b,y) 
= \sfrac{1}{2} ( \kappa(b) +  \log \mu_3)$ if $y\leq 1$ and $a\leq a_c < b$.

The final case (for $a>b$)  is $a_c < b<  a$.  It follows that $\kappa(a)
> \kappa(b) > \log \mu_3$ and this establishes the existence of the 
$AB$-adsorbed phase with free energy
\begin{equation}
\W{\rho}^{AB}(a,b,y) = 
\Sfrac{1}{2} (\kappa(a) + \kappa(b)).
\label{eqn7}  
\end{equation}

Collecting the results above gives the following theorem.

\begin{theo}
Suppose that $y\leq 1$.  Then the free energy of an adsorbing 
$AB$-diblock loop pulled at its middle vertex is 
$\W{\rho}^{AB}(a,b,y) = \sfrac{1}{2} (\kappa(a) + \kappa(b))$, 
consistent with equation \Ref{generalform}.  To indicate the locations of 
phase boundaries, this can also be expressed as:
\begin{equation*}
\W{\rho}^{AB}(a,b,y) =
\begin{cases}
\log \mu_3, & \hbox{if $a\leq a_c$  $\&$ $b\leq a_c$}; \\
\sfrac{1}{2} (\kappa(a) + \log \mu_3), & \hbox{if $a>a_c$ $\&$ $b \leq a_c$}; \\
\sfrac{1}{2} (\kappa(b) + \log \mu_3), & \hbox{if $a\leq a_c$ $\&$ $b > a_c$}; \\
\sfrac{1}{2} (\kappa(a) + \kappa(b)), & \hbox{if $a > a_c$ $\&$ $b > a_c$}. 
\end{cases}
\end{equation*} 
\hfill\hqed
\label{thmyl1}
\end{theo}

\begin{figure}[t]
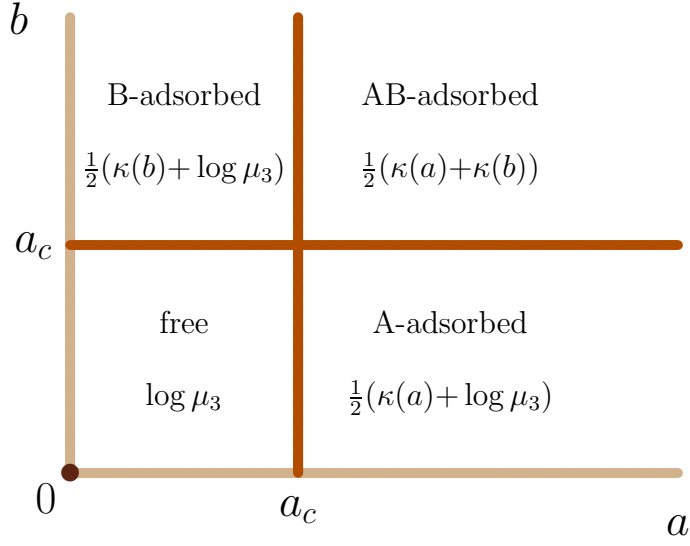

\beginpicture
\setcoordinatesystem units <1.15pt,1.15pt>
\setplotarea x from -100 to 300, y from -10 to 150
\setplotarea x from 0 to 300, y from 0 to 150

\setplotsymbol ({\footnotesize$\bullet$})

\color{Tan}
\plot 0 150 0 0 200 0 /
\color{Brown}
\plot 0 75 200 75 / \plot 75 0 75 150 /

\color{Sepia}
\put {\LARGE$\bullet$} at 0 0 

\color{black}
\normalcolor

\put {\LARGE$a_c$} at 75 -12  \put {\LARGE$a$} at 200 -17
\put {\LARGE$a_c$} at -12 75  \put {\LARGE$b$} at -17 150
\put {\LARGE$0$} at -8 -8 

\put {\large free} at 37.5 50  \put {\large$\log \mu_3$} at 37.5 25
\put {\large B-adsorbed} at 37.5 125  \put {\large$\sfrac{1}{2}(\kappa(b) {+} \log \mu_3)$} at 37.5 100
\put {\large A-adsorbed} at 125 50  \put {\large$\sfrac{1}{2}(\kappa(a) {+} \log \mu_3)$} at 125 25
\put {\large AB-adsorbed} at 125 125  \put {\large$\sfrac{1}{2}(\kappa(a) {+} \kappa(b))$} at 125 100

\endpicture
\caption{The three adsorbed phases (and one free phase) of the pulled 
adsorbing diblock loop when $y\leq 1$.}
\label{figure5}   
\end{figure}

By comparing the free energies, there are phase boundaries for 
$a=a_c$ and $b\geq 0$,  and $b=a_c$ and $a\geq 0$.  This is illustrated 
in figure \ref{figure5}.  These phases are characterised by whether none, 
one or both arms of the loop are adsorbed.

\subsection{The case $\boldsymbol{y>1}$}

For $y>1$, in addition to  phases that can occur for $y\leq 1$ 
(with free energy $\sfrac{1}{2} (\kappa(a)+\kappa(b))$),  we expect there to 
be a ballistic phase with the same free energy as a mid-point pulled loop 
($\lambda(y^{1/2})$).  In order to delineate between the different phases 
and the corresponding solutions to the right hand side of 
equation \Ref{generalform}, we introduce a useful function first.  Given fixed $y>1$ 
and $x\geq 0$, because $\lambda(y)$ is a strictly increasing continuous function 
and because $\kappa(x)\geq \log\mu_3=\lambda(1)$, there exists a unique 
$\delta_x(y)\geq 0$ such that $\kappa(x)=\lambda(y^{\delta_x(y)})$.  That is, 
the function 
\begin{equation}
\delta_x(y) =\log_y (\lambda^{-1}(\kappa(x))),
\label{deltadef}
\end{equation}
with $\lambda^{-1}(\log\mu_3)\equiv 1$, is well defined. When $y$ is fixed, 
this is simplified to $\delta_x=\delta_x(y)$.  Thus given a fixed $y>1$ and 
any $a,b$, we can define $\delta_a$ and $\delta_b$ such that 
$\kappa(a)=\lambda(y^{\delta_a})$ and $\kappa(b)=\lambda(y^{\delta_b})$.  
We note also that for fixed $y>1$ and a given $\delta >0$, the equation $\kappa(x)=\lambda(y^{\delta})>\log\mu_3$ has a unique solution for 
$x>a_c$, since by convexity $\kappa(x)$ is continuous and strictly increasing 
in $x>a_c$;  thus $\delta=\delta_x$. For $\delta =0$, the 
equation $\kappa(x)=\lambda(y^{\delta})=\log\mu_3$ holds for all $x\leq a_c$.
We now break down the determination of the free energy into subcases 
dependent on the values of $\delta_a$ and $\delta_b$.

First, for $\shalf \geq \max\{\delta_a,\delta_b\}$  (equivalently 
$\lambda(y^{1/2}) \geq \max\{\kappa(a),\kappa(b)\}$), taking 
$\epsilon = \shalf$ in equation \Ref{eqn37A} together with equation \Ref{eqn10} 
gives 
\begin{equation}
\W{\rho}^{AB}(a,b,y) = 
\lim_{n\to\infty} \Sfrac{1}{2n} \log \W{U}_{2n}(a,b,y) = \lambda(y^{1/2}),
\quad\hbox{for $\shalf \geq \max\{\delta_a,\delta_b\}$}.
\label{eqn11}  
\end{equation}

Next,  for $\sfrac{1}{2} < \max\{\delta_a,\delta_b\}$  (equivalently $\lambda(y^{1/2})
< \max\{\kappa(a),\kappa(b)\}$), we consider $\delta_a+\delta_b\geq 1$ 
(equivalently $\lambda^{-1}(\kappa(b))\, \lambda^{-1}(\kappa(a)) \geq y$).
Without loss of generality, consider $\delta_a=\max\{\delta_a,\delta_b\}>1/2$.  
If $\delta_a\geq 1$ then for any $\delta_b\geq 0$, $\delta_a+\delta_b\geq 1$. 
Further $\kappa(a)\geq \lambda(y)$ and since $\kappa(b)\geq \log\mu_3=\lambda(1)$, 
by log-convexity $\sfrac{1}{2} (\kappa(a)+\kappa(b))
\geq \sfrac{1}{2} (\lambda(y)+\lambda(1))\geq \lambda(y^{1/2})$. 
Thus taking $\epsilon =1$ in equation \Ref{eqn37A} (from lemma \ref{lemma3}) 
together with \Ref{eqn10}  shows  that 
\begin{equation}
\W{\rho}^{AB}(a,b,y) = 
\sfrac{1}{2} (\kappa(a)+\kappa(b)),
\quad\hbox{for $1\leq  \max\{\delta_a,\delta_b\}$}.
\label{eqn12new}  
\end{equation}
On the other hand, if $\sfrac{1}{2} <\delta_a<1$, then $\delta_a+\delta_b\geq 1$ 
implies $\delta_a\geq \delta_b \geq 1-\delta_a> 0$. This ensures by 
log-convexity  that $\sfrac{1}{2} (\lambda(y^{\delta_a})
+\lambda(y^{\delta_b}))=\sfrac{1}{2} (\kappa(a)+\kappa(b))\geq \sfrac{1}{2} 
(\lambda(y^{\delta_a})+\lambda(y^{1-\delta_a})) \geq \lambda(y^{1/2})$.
Thus, taking $\epsilon=\delta_a$ in equation \Ref{eqn37A} together with 
equation \Ref{eqn10}, gives again that $\W{\rho}^{AB}(a,b,y) = 
\sfrac{1}{2} (\kappa(a)+\kappa(b))$. In summary
\begin{equation}
\W{\rho}^{AB}(a,b,y) = 
 \sfrac{1}{2} (\kappa(a)+\kappa(b)),
\quad\hbox{for $\sfrac{1}{2} < \max\{\delta_a,\delta_b\}$ and $1\leq \delta_b +\delta_a$}.
\label{eqn12}  
\end{equation}
These results leave the free energy unexplored for $a, b$ such that
$\sfrac{1}{2} < \max\{\delta_a,\delta_b\}$ (equivalently $\lambda(y^{1/2}) 
< \max\{\kappa(a),\kappa(b)\}$) and $\delta_b +\delta_a< 1$ 
(equivalently $\lambda^{-1}(\kappa(b))\,\lambda^{-1}(\kappa(a)) < y$).
This latter region will be explored in the next subsection.  Before doing that, 
however, we explore the regions defined by equations \Ref{eqn11}-\Ref{eqn12} 
further first. 

Note that  for any $a,b\leq a_c$, we have 
$\lambda(y^{1/2}) > \max\{\kappa(a),\kappa(b)\}
=\sfrac{1}{2} (\kappa(a)+\kappa(b))=\log \mu_3$, 
so the free energy region of equation \Ref{eqn11} is nonempty.  This 
establishes the existence of the ballistic phase.  We focus next on the region of
equation \Ref{eqn12}. If  $\lambda(y^{1/2}) \leq \min\{\kappa(a), \kappa(b)\}$, 
then $\min\{\delta_a,\delta_b\}\geq \sfrac{1}{2} $ and hence $\delta_b\geq 1-\delta_a$.
Also $\min\{\kappa(a), \kappa(b)\}>\log\mu_3$ and hence $a,b\geq a_c$.  
Since above $a_c$ the function $\kappa$ is strictly increasing in its 
argument (by log-convexity), there exists large enough $a,b$ such that 
$\lambda(y^{1/2}) \leq \min\{\kappa(a), \kappa(b)\}$.  This establishes 
the existence of the $AB$-adsorbed phase.  Lastly, consider the region 
of equation \Ref{eqn12new}.  Consider the subcase $b\leq a_c < a$. 
Hence $\kappa(b) = \log \mu_3=\lambda(1)<\lambda(y^{1/2})$ and 
$\kappa(a) \geq \lambda(y) > \log \mu_3$.  Then by equation \Ref{eqn12new}, 
\begin{equation}
\W{\rho}^{AB}(a,b,y) = \sfrac{1}{2} ( \kappa(a)+\log \mu_3),
\quad\hbox{for $b\leq a_c <a $ and $\lambda(y) \leq  \kappa(a)$} .
\label{eqn13A}  
\end{equation}
This region is non-empty for $b<a_c$ and $a$ sufficiently large, 
hence this establishes the existence of the $A$-adsorbed ($B$-free) phase. 
Similarly, the following free energy region, corresponding to a 
$B$-adsorbed ($A$-free) phase, is non-empty:
\begin{equation}
\W{\rho}^{AB}(a,b,y) = \sfrac{1}{2} ( \kappa(b)+\log \mu_3),
\quad\hbox{if $a\leq a_c <b $ and $\lambda(y) \leq \kappa(b)$} .
\label{eqn13}  
\end{equation}
In summary, we have shown that all the phases from $y\leq 1$ exist 
except for the phase with free energy $\log\mu_3$, that part of phase space 
is now a ballistic phase.  

We collect the above results in the following lemma.

\begin{lemm}
If $y>1$, then for the following non-empty subregions of the $(a,b)-plane$, 
\begin{equation*}
\W{\rho}^{AB}(a,b,y) =
\begin{cases}
\lambda(y^{1/2}), &  \hbox{for $1/2\geq \max\{\delta_a,\delta_b\}$}; \\
\sfrac{1}{2} (\kappa(a) + \kappa(b)), 
        & \hbox{for $1/2< \max\{\delta_a,\delta_b\}$ with $1\leq \delta_b +\delta_a$},
\end{cases}
\end{equation*}
consistent with equation \Ref{generalform}.  Here 
$\delta_x =\log_y (\lambda^{-1}(\kappa(x))$ 
with $\lambda^{-1}(\log\mu_3)\equiv 1$.

To indicate the possible locations of phase boundaries, this can be 
further expressed as follows, where each listed subregion of the 
$(a,b)$-plane is non-empty:
\begin{equation*}
\W{\rho}^{AB}(a,b,y) =
\begin{cases}
\lambda(y^{1/2}), 
  & \hbox{for $\lambda(y^{1/2}) \geq \max\{\kappa(a),\kappa(b)\}$}; \\
\sfrac{1}{2} (\kappa(a) + \log \mu_3), 
  & \hbox{for  $b \leq a_c<a$ $\&$ $\lambda(y) \leq \kappa(a)$}; \\
\sfrac{1}{2} (\kappa(b) + \log \mu_3), 
  & \hbox{for $a \leq a_c<b$ $\&$ $\lambda(y) \leq \kappa(b)$}; \\
\sfrac{1}{2} (\kappa(a) + \kappa(b)), 
  & \hbox{for  $1/2< \max\{\delta_a,\delta_b\}$ $\&$ $1\leq \delta_b +\delta_a$}. 
\end{cases}
\end{equation*}
\hfill\hqed
\label{lemma4}   
\end{lemm}

As mentioned above, however, this theorem leaves the free energy unexplored 
for $a, b$ such that $\shalf < \max\{\delta_a,\delta_b\}$ (equivalently 
$\lambda(y^{1/2}) < \max\{\kappa(a),\kappa(b)\}$) and 
$\delta_b +\delta_a< 1$ (equivalently 
$\lambda^{-1}(\kappa(b))\, \lambda^{-1}(\kappa(a)) < y$).  We consider this region 
in the  next subsection.

\subsection{Mixed phases in the $AB$-block copolymer phase diagram}
\label{4.3}

In this section the exception to lemma \ref{lemma4} is considered.
This is the region in three dimensional phase space (with dimensions $(a,b,y)$) 
where $\shalf < \max\{\delta_a,\delta_b\}$ (equivalently $\lambda(y^{1/2}) 
< \max\{\kappa(a),\kappa(b)\}$) and $\delta_b +\delta_a< 1$ (equivalently 
$\lambda^{-1}(\kappa(b)) \, \lambda^{-1}(\kappa(a)) < y$).

In the directed version of this model \cite{Iliev}, Dyck paths (these
are loops) are pulled from the  middle vertex and both the $A$- and $B$-blocks 
have an endpoint in the surface.  This model exhibits a mixed adsorbed-ballistic 
phase where one block is considered ``adsorbed" and the other ``ballistic".
However when this occurs, the adsorbed side is only partially 
adsorbed (only a portion of the block interacts with the surface) and the ballistic 
side is not fully ballistic (the strength of the pulling force is not fully felt).  
Further, the pulling and adsorbing forces balance on the adsorbed side. 

We show now that a
similar situation occurs here, namely that
there is a mixed adsorbed-ballistic phase in our model which is similar 
to the directed model mixed phase.

\begin{figure}[t]
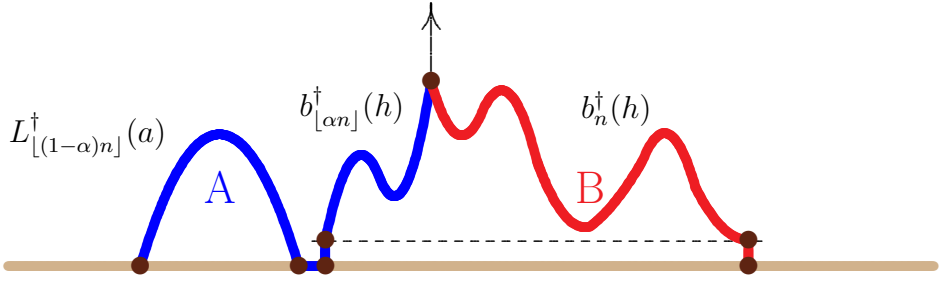

\beginpicture
\setcoordinatesystem units <1pt,1pt>
\setplotarea x from -100 to 300, y from -10 to 100
\setplotarea x from 0 to 300, y from 0 to 100

\color{black}
\arrow <10pt> [.1,.67] from 110 70 to 110 100

\setdashes <3pt>
\plot 65 10 235 10 /
\setsolid

\setplotsymbol ({\footnotesize$\bullet$})

\color{Tan}
\plot -50 0 300 0 /

\setquadratic
\color{Blue}
\plot 0 0 30 50 60 0 /
\setlinear
\plot 60 0 70 0 70 10 /
\setquadratic
\plot 70 10 80 40 90 35 100 30 110 70 /

\color{Red}
\plot 110 70 120 50 130 60 140 65 150 40 160 20 170 15 180 25 190 40 
200 50 210 30  220 15 230 10   /
\setlinear
\plot 230 10 230 0 /

\color{Sepia}
\multiput {\LARGE$\bullet$} at 0 0 60 0 70 0 70 10 110 70 230 10 230 0 /

\color{Blue} \put {\LARGE A} at 30 30 
\color{Red} \put {\LARGE B} at 170 30

\color{black}
\put {\large$L^\dagger_{\lfloor (1-\alpha)n\rfloor}(a)$} at -20 50 
\put {\large$b^\dagger_{\lfloor \alpha n \rfloor}(h)$} at 80 60
\put {\large$b^\dagger_n(h)$} at 180 60

\color{black}
\normalcolor

\endpicture
\caption{By concatenating unfolded walks, loops and bridges in this way, the
lower bound in equation \Ref{eqn26} is obtained.}
\label{figure6}  
\end{figure}

In figure \ref{figure6} an idealized walk conformation in a mixed
adsorbed-ballistic phase is illustrated (with the $A$-block partially adsorbed, and
the $B$-block partially ballistic).  
For $a>b$, we
shall show that these types of conformations
occur when
the first $(1{-}\alpha)n$ 
vertices of the $A$-block behave like an adsorbing loop and then the remaining 
portion of the conformation, including the $B$-block, is ballistic.
In this situation the ballistic portion of the $A$-block can only be pulled as high 
as $\alpha n$ so that the height of the middle vertex is constrained (and the first vertex
of the $B$-block cannot be pulled any higher).  To accomodate the competing forces, 
the activity $y$ is partitioned between the two blocks so that a higher weight 
$y^{\delta}$ (with $\delta \geq \sfrac{1}{2}$) is applied on the $A$-block to pull 
the shorter (length $\alpha n$) segment as high as possible and the lower weight, 
$y^{1-\delta}$, pulls the longer $B$-block to the same height.  In other words, the pulling 
force is partitioned between the two blocks so that the $A$-block feels a stronger 
pull than the $B$-block.  A similar situation is encountered when the $A$-block 
is ballistic, and the $B$-block is adsorbed.

In addition, for a mixed adsorbed-ballistic phase to exist, it must be that 
neither the adsorbing nor the pulling forces ``win" on the $A$-block.  This 
suggests that the associated free energies are equal, namely 
$\kappa(a) = \lambda(y^{\delta})$, ie $\delta=\delta_a$. Indeed, we will show 
that taking $\delta=\delta_a$ enables us to determine the free energy as  
$\sfrac{1}{2} (\kappa(a)+\lambda(y^{1-\delta_a}))$ in a region where the
$A$-block is both adsorbed and ballistic, and the $B$-block is ballistic.
Using lemma  \ref{lemma4B}, this will establish the existence of a 
mixed adsorbed-ballistic phase.  Since our model is symmetric in $a$ and $b$, 
there is a similar result where the $B$-block is both adsorbed and ballistic,
and the $A$-block is ballistic.

By symmetry in $a$ and $b$, we assume, without loss of generality
that $a>b$.   By lemma \ref{lemma4} for any $y>1$ we note that the region 
of interest in the $(a,b)$-plane  is given by
$\delta_b <1-\delta_a < \shalf < \delta_a <1$ which is equivalent to
\begin{equation}
\kappa(b) < \lambda(y^{1-\delta_a}) < \lambda(y^{1/2}) < \kappa(a) < \lambda(y).
\label{55}  
\end{equation}
Given $a$ such that $\lambda(y^{1/2}) < \kappa(a) < \lambda(y)$, 
equation \Ref{55} holds for any $b$ such  that $\kappa(a)\,\kappa(b) 
< \lambda(y^{\delta_a}) \, \lambda(y^{1-\delta_a})$ or equivalently 
any $b$ such that  $\lambda^{-1}(\kappa(a))\,\lambda^{-1}(\kappa(b)) < y$.

Alternatively, we can re-parameterize this region in terms of $b$ and a 
parameter $\delta$ (instead of $a$) as follows.
Introduce a \underbar{fixed parameter} $\delta\in (\shalf,1)$.
For any $y>1$, there is an $a$ so that $\kappa(a) = \lambda(y^\delta)$. 
(Note that because of uniqueness this means that $a$ is such that $\delta_a=\delta$.) 
Since $\lambda(y^{1/2}) < \lambda(y^\delta) = \kappa(a) < \lambda(y)$ the point
$(a,b,y)$ is in the region in equation \Ref{55} provided that $b$ is fixed so
that $\kappa(a)\,\kappa(b) < \lambda(y^\delta) \, \lambda(y^{1-\delta})$, 
or equivalently $\delta_b <1-\delta$.

\begin{figure}[t]
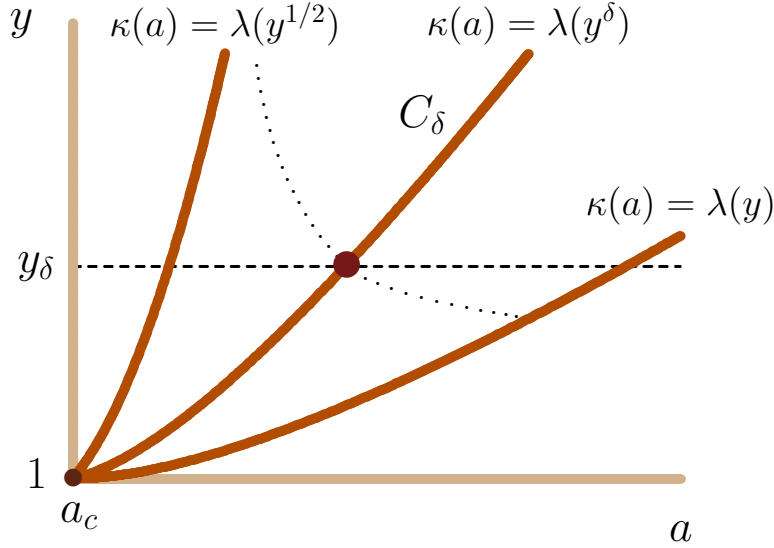

\beginpicture
\setcoordinatesystem units <1.15pt,1.15pt>
\setplotarea x from -100 to 300, y from -10 to 150
\setplotarea x from 0 to 300, y from 0 to 150

\setlinear
\setdashes <3pt>
\setplotsymbol ({.})
\color{black}
\plot 0 70 200 70 /

\setdots
\setquadratic
\plot 90 70 70 100 60 140 /
\plot 90 70 110 61 150 53 /
\setlinear
\setsolid

\setplotsymbol ({\footnotesize$\bullet$})

\color{Tan}
\plot 0 150 0 0 200 0 /
\color{Brown}

\setquadratic
\plot 0 0 25 50 50 140 /
\plot 0 0 80 20 200 80 /
\plot 0 0 60 40 150 140 /

\color{Sepia}
\put {\LARGE$\bullet$} at 0 0 

\color{black}
\normalcolor

\put {\LARGE$a$} at 200 -17
\put {\LARGE$y$} at -17 150
\put {\LARGE$a_c$} at 2 -12
\put {\LARGE$1$} at -12 2 
\put {\LARGE$y_\delta$} at -12 70
\put {\LARGE$C_\delta$} at 115 120

\put {\Large$\kappa(a)=\lambda(y^{1/2})$} at 50 150
\put {\Large$\kappa(a)=\lambda(y)$} at 200 90
\put {\Large$\kappa(a)=\lambda(y^\delta)$} at 150 150

\color{Maroon}
\put {\scalebox{1.5}{\LARGE$\bullet$}} at 90 70 

\color{black}
\normalcolor

\endpicture
\caption{A schematic drawing of the curve $C_\delta(b)$ projected onto the 
$ay$-plane.  Increasing $y$ also increases $a$ along $C_\delta(b)$, and
we show, using lemma \ref{lemma4B}, that for $y>y_\delta$, the
point $(a,b,y)$ is located in a mixed adsorbed-ballistic phase.
Notice that $y_\delta \to\infty$ as $\delta$ approaches $\shalf$ from above.
The function $y_\delta$ as a function of $\delta$ is indicated by the dotted curve for
$\delta\in(\sfrac{1}{2},1)$.}
\label{figureCdelta}   
\end{figure}

Note next that every fixed value of $\delta \in (\shalf,1)$ and fixed $b>0$ 
defines a curve in three dimensional
phase space given by 
\begin{equation}
C_\delta(b) = \{(a,b,y)\, \vert\, \hbox{$\kappa(a) = \lambda(y^\delta)$}\}.  
\label{56}  
\end{equation}
See figure \ref{figureCdelta}. Parametrizing $C_\delta(b)$ by $y$ shows that it is within 
the region defined in equation \Ref{55} for all $y>1$ provided that $\delta_b< 1- \delta$.  
Moreover, for any $b>0$, $C_\delta(b)$ is either in the region defined in equation
\Ref{55}, or there is a $y_b$, such that $C_\delta(b)$ is in this region for all $y\geq y_b$
(that is, $y_b=0$ if $\delta_b<1{-}\delta$).  Thus, every point in  the region defined in 
equation \Ref{55} is located on a given $C_\delta(b)$ for some value of 
$\delta \in (\shalf,1)$ and a value of $b$ such that 
$\lambda^{-1}(\kappa(a))\,\lambda^{-1}(\kappa(b)) < y$.
Below we prove that for a fixed $\delta \in (\shalf,1)$, with $b$ satisfying
$\lambda^{-1}(\kappa(a))\,\lambda^{-1}(\kappa(b)) < y$,
there exists a $y_\delta$ so that the point $(a,b,y)$ is located in
a mixed adsorbed-ballistic phase provided that $y>y_\delta$ 
(with $y_\delta$ defined in lemma \ref{lemma4B}).  Note that $a$ and
$b$ depend on $y$ and $\delta$ but that $y_\delta$ is a function only of $\delta$.

 Proceed then by fixing $\delta \in (\shalf,1)$ and given a $y>1$, fix $b$ so that 
the conditions in equation \Ref{55} are satisfied, ie $\delta_b<1-\delta$.

Construct a lower bound on the partition function as shown in figure \ref{figure6}.
Use $x_1$-unfolded loops and bridges.  This gives for any $0 \leq \alpha \leq 1$ and any integer $1\leq h\leq \alpha n$
\begin{equation}
\W{U}_{2n} (a,b,y) \geq
b~ L^\dagger_{\lfloor (1-\alpha)n\rfloor}(a)\, b^\dagger_{\lfloor \alpha n \rfloor}(h)\,
y^h\, b^\dagger_n(h) .
\label{eqn26}   
\end{equation}

Our goal now is to choose a sequence of  $h$'s ($h_n$)  and $\alpha$'s ($\alpha_n$) in equation \Ref{eqn26} in such a way that the free energy from the lower bound will be $\sfrac{1}{2} (\kappa(a)+\lambda(y^{1-\delta}))$.
To find the appropriate sequences, first introduce $\delta$ in this bound by writing $y = y^\delta\, y^{1-\delta}$ 
and consider the factors $b^\dagger_{\lfloor \alpha n \rfloor}(h)\, y^{\delta h}$ 
and $b^\dagger_n(h)\,y^{(1-\delta)h}$.  
We next use $\delta$ to determine the choice of $h$'s. 
Specifically, choose $h=h^*_n(y^{1-\delta})$ to 
be a most popular height (see section \ref{sec:lambda} and equation \Ref{eqn-pulledbridge}) of an endpoint pulled $n$ step positive walk  
with activitiy $y^{1-\delta}$; this will correspond to a most popular height of the first vertex of the $B$-block (independent of the $A$-block).  From equation \Ref{eqn-pulledbridge} we have
\begin{equation}
\lim_{n\to\infty} \Sfrac{1}{n} \log \left(b^\dagger_n(h^*_n(y^{1-\delta}))
\,y^{(1-\delta)h^*_n(y^{1-\delta})}   \right)
= \lambda(y^{1-\delta}) .
\end{equation}

We now choose the $\alpha$'s so that the last vertex of the $A$-block has the 
same  height as the first vertex of the $B$-block, namely $h_n^*(y^{1-\delta})$, while at the same time  
 $\alpha_n$ satisfies $h_{\lfloor \alpha_n n \rfloor}^*(y^\delta) = h_n^*(y^{1-\delta})+o(n)$ 
(and the last part of the $A$-block after its last $A$-visit has length 
$\lfloor \alpha_n n \rfloor$).  The limit of the $\alpha_n$ as $n\to\infty$
will be shown to exist.

Specifically, by lemma \ref{lemma6B}, $h_n^*(y^{1-\delta}) = \eps_*(y^{1-\delta})\, n + o(n)$.
Similarly, $h_{\lfloor \alpha n \rfloor}^*(y^\delta) = \eps_*(y^\delta)\, \lfloor \alpha n \rfloor + o(n)$.
For values of $y>1$ the log-convexity of $\lambda(y)$ implies that $ \eps_*(y)$ is a non-decreasing function of $y$ and hence
$\eps_*(y^{\delta})\geq \eps_*(y^{1-\delta})$.   Thus there exists $\alpha \in [0,1]$ such that $\eps_*(y^\delta)\, \lfloor \alpha n \rfloor = \eps_*(y^{1-\delta})\, n$.
Set $\alpha_n$ to be such a value of $\alpha$.

Thus
\begin{equation}
\eps_*(y^\delta)\, \lfloor \alpha_n n \rfloor = \eps_*(y^{1-\delta})\, n 
\label{epsilon-relate}
\end{equation}
and
\begin{equation}
h_{\lfloor \alpha_n n \rfloor}^*(y^\delta) =  h_n^*(y^{1-\delta}) + o(n). 
\label{alphan-relate}  
\end{equation}
Our aim is to replace $h$ in equation \Ref{eqn26} by most popular
heights, but this shows that the most popular 
height $h_{\lfloor \alpha_n n \rfloor}^*(y^\delta)$ of the
final vertex of the $A$-block in figure \ref{figure6}, and the most popular
height $h_n^*(y^{1-\delta})$ of the $B$-block do not coincide, but
differ by $o(n)$. We show below how to compensate for this.

It follows from equations \Ref{epsilon-relate} and  \Ref{alphan-relate} that
\begin{equation}
\alpha_{\delta} \equiv \lim_{n\to\infty} \alpha_n = \lim_{n\to\infty}  \frac{\lfloor \alpha_n n\rfloor}{n} = \lim_{n\to\infty} \left( \frac{\lfloor \alpha_n n\rfloor}{n} 
\times
\frac{h_n^*(y^{1-\delta})}{h_{\lfloor \alpha_n n \rfloor}^*(y^\delta)}  \right)
 =  \frac{\eps_*(y^{1-\delta})}{\eps_*(y^\delta)}
\end{equation}
where this limit exists by lemma \ref{lemma6B}.
From this and 
equation \Ref{alphan-relate}, it follows that 
\begin{eqnarray}
\lim_{n\to\infty} \Sfrac{1}{n} \log \left(
b_{\lfloor \alpha_n n \rfloor}^\dagger (h_n^*(y^{1-\delta}))\, 
                                                       y^{h_n^*(y^{1-\delta})}\right) 
& = & \lim_{n\to\infty} \Sfrac{1}{n} \log \left(
b_{\lfloor \alpha_n n \rfloor}^\dagger (h_{\lfloor \alpha_n n \rfloor}^*(y^\delta))
\, y^{h_{\lfloor \alpha_n n \rfloor}^*(y^\delta)} \right) \nonumber \\
& = & \alpha_{\delta}\, \lambda(y^\delta) = \alpha_{\delta}\, \kappa(a) . 
\end{eqnarray}
This shows that we can replace $h$ in equation \Ref{eqn26} by
$h_n^*(y^{1-\delta})+o(n)$, take logarithms, divide by $n$ and
let $n\to\infty$ to obtain a lower bound on the free energy. This result is given below in theorem \ref{thm2}.

In terms of determining the phase diagram, we will need to determine when $\alpha_{\delta}<1$. For values of $y>1$ the log-convexity of $\lambda(y)$ implies that
$\eps_*(y^{1/2})\geq \eps_*(y^{1-\delta})$ 
(which is the same as $\eps_*(y^{1/2})/\eps_*(y^{1-\delta})\geq 1$), since $\delta>\shalf$.
For large values of $y$ (that is, for $y>y_\delta$) lemma \ref{lemma4B} shows 
that $\eps_*(y^\delta) > 2\,\eps^*(y^{1/2}) - \eps_*(y^{1-\delta})$.
Dividing by $\eps_*(y^{1-\delta})$ gives
\begin{equation}
\frac{\eps_*(y^\delta)}{\eps_*(y^{1-\delta})} 
> 2\frac{\eps_*(y^{1/2})}{\eps_*(y^{1-\delta})} - 1
\geq \frac{\eps_*(y^{1/2})}{\eps_*(y^{1-\delta})} \geq 1
\end{equation}
since $\eps_*(y)$ is a non-decreasing function of $y$.  In particular,
this shows that 
\begin{equation}
\alpha_\delta 
=\frac{\eps_*(y^{1-\delta})}{\eps_*(y^\delta)} < 1, 
\quad\hbox{if $y>y_{\delta}$}. 
\end{equation}
By lemma \ref{lemma4B} it is sufficient to choose $y_\delta = 24^{2/(2\delta-1)}$, 
and we note that this is finite, but unbounded, for $\delta\in(\shalf,1]$.

If our conjecture \ref{lemma4B} that $\lambda(y)$ is strictly log-convex 
for $y>1$ is true,  then this strict bound on $\alpha$ would be true for 
all values of $y>1$ (that is, we could choose $y_\delta = 1$).

The above results give the following theorem.

\begin{theo}
Given $\delta\in(\shalf,1)$ and $y>1$, for $a,b>0$ such that $\kappa(a) = \lambda(y^\delta)$ and
$\kappa(b) <  \lambda(y^{1-\delta}) < \lambda(y^{1/2}) < \kappa(a) < \lambda(y)$
(equivalently $\lambda^{-1}(\kappa(b))\, \lambda^{-1}(\kappa(a)) < y$
or equivalently, $\kappa(a)\,\kappa(b) < \lambda(y^\delta) \, \lambda(y^{1-\delta})$):  
\begin{eqnarray*}
\liminf_{n\to\infty} \Sfrac{1}{2n} \log \W{U}_{2n}(a,b,y) 
&\geq & \sfrac{1}{2} 
  \left( (1-\alpha) \kappa(a) 
  + \alpha \lambda (y^{\delta}) + \lambda(y^{1-\delta}) \right)\\
&=& \sfrac{1}{2} ( \kappa(a) + \lambda(y^{1-\delta}) )
 = \sfrac{1}{2} (\lambda(y^\delta) + \lambda(y^{1-\delta}) )
\end{eqnarray*}
where $\alpha = \Sfrac{\eps_*(y^{1-\delta})}{\eps_*(y^\delta)}$.  If
$y>y_\delta$, then $\alpha< 1$. 
\hfill\hqed
\label{thm2}  
\end{theo}

Notice that for $\kappa(b) \geq \lambda(y^{1-\delta})$ this lower bound 
is beaten by $\sfrac{1}{2} (\kappa(a) + \kappa(b))$ as seen in 
lemma \ref{lemma4}.  Moreover, given the conditions of theorem \ref{thm2},
$2\,\lambda(y^{1/2}) < \lambda(y^\delta) + \lambda(y^{1-\delta})$
(when $y > y_\delta$), and $\lambda(y^\delta) + \lambda(y^{1-\delta}) 
> \kappa(a) + \log \mu_3$,  since $\lambda(y^\delta) = \kappa(a)$ and $\lambda(y^{1-\delta})>\log \mu_3$.  This shows that the lower bound in 
theorem \ref{thm2} exceeds both the free energy in lemma \ref{lemma4} 
in the ballistic regime ($\W{\rho} (a,b,y) = \lambda(y^{1/2})$) and the 
free energy in the $A$-adsorbed phase ($\W{\rho}(a,b,y) 
= \shalf(\kappa(a) + \log \mu_3)$).  In other words, the model is not in a 
fully ballistic, nor a fully adsorbed, phase.

For a fixed $\delta\in(\shalf,1)$, a point $(a,b,y)$ of theorem \ref{thm2} moves along the curve $C_\delta(b)$ with increasing
$y$ (see equation \Ref{56}).  When $y>y_\delta$, $\alpha < 1$, and
$C_\delta$ is in the mixed phase described in the last paragraph.  This is
also shown schematically in figure \ref{figureCdelta}.

We proceed by determining an upper bound on $\W{U}_{2n}(a,b,y)$.  
Recall that $\kappa(b)=\lambda(y^{\delta_b})<  \lambda(y^{1-\delta})< \lambda(y^{1/2})<\kappa(a)=\lambda(y^{\delta})$ 
where $\delta_b\in[0, 1-\delta)$.

Consider the upper bound of equation \Ref{eqn37A} again
\begin{eqnarray}
\limsup_{n\to\infty} \Sfrac{1}{n} \log \hat{U}_{2n}(a,b,y) 
&\leq& \Sfrac{1}{2} \min_{\epsilon\in[0,1]}  \left\{ \psi_e(a,y^\epsilon) 
                                          +  \psi_e(b,y^{1-\epsilon})\right\}  \nonumber\\
&=& \Sfrac{1}{2} \min_{\epsilon\in[0,1]}  
\left\{ \max\{\kappa(a), \lambda(y^\epsilon)\}  
      +  \max\{\kappa(b),\lambda(y^{1-\epsilon})\}\right\} .
\label{eqn37Aagain}  
\end{eqnarray}

Firstly, due to the log-convexity and continuity of $\lambda(y)$, 
for any $0\leq\epsilon_1\leq \epsilon_2\leq 1$,
\begin{equation}
\min_{\epsilon\in[\epsilon_1,\epsilon_2]}  
\left\{ \lambda(y^\epsilon)+ \lambda(y^{1-\epsilon}) \right\} 
= \lambda(y^{\epsilon_1})+ \lambda(y^{1-\epsilon_1}).
\label{eqn47A}  
\end{equation}
The minimum in equation \Ref{eqn37A} can now be determined.
Define $\delta_b$ as before by $\kappa(b) =\lambda(y^{\delta_b})$.
If $\epsilon <\delta$, then  
$\kappa(a)=\lambda(y^{\delta}) > \lambda(y^\epsilon)$ so that
$\kappa(b) =\lambda(y^{\delta_b}) < \lambda(y^{1-\delta})
   < \lambda(y^{1-\epsilon})$.  
Thus, $\max\{\kappa(a), \lambda(y^\epsilon)\} = \kappa(a)=\lambda(y^{\delta})$ 
and $\max\{\kappa(b),\lambda(y^{1-\epsilon})\} = \lambda(y^{1-\epsilon})$.  
It follows that
\begin{eqnarray}
\min_{\epsilon\in[0,\delta]}  
\left\{ \max\{\kappa(a), \lambda(y^\epsilon)\}  
   + \max\{\kappa(b),\lambda(y^{1-\epsilon})\}\right\}
&=& \lambda(y^{\delta}) + \min_{\epsilon\in[0,\delta]} 
                 \lambda(y^{1-\epsilon}) \nonumber \\
&=& \lambda(y^{\delta}) + \lambda(y^{1-\delta}).
\end{eqnarray}

Secondly, if $1-\delta_b\geq \epsilon \geq \delta > \sfrac{1}{2}$, then
$\kappa(a)=\lambda(y^{\delta})\leq \lambda(y^\epsilon)$ and $\kappa(b)=\lambda(y^{\delta_b})\leq \lambda(y^{1-\epsilon})$.  
Thus $\max\{\kappa(a), \lambda(y^\epsilon)\} = \lambda(y^\epsilon)$ 
and $\max\{\kappa(b),\lambda(y^{1-\epsilon})\}=\lambda(y^{1-\epsilon})$ 
and so it follows by equation \Ref{eqn47A} that
\begin{eqnarray}
\min_{\epsilon\in[\delta,1-\delta_b]}  
\left\{ \max\{\kappa(a), \lambda(y^\epsilon)\}  + \max\{\kappa(b),\lambda(y^{1-\epsilon})\}\right\}
&=& \min_{\epsilon\in[\delta,1-\delta_b]}  \L \lambda(y^{\epsilon}) 
        + \lambda(y^{1-\epsilon}) \R \nonumber \\
&=& \lambda(y^{\delta}) + \lambda(y^{1-\delta}).
\end{eqnarray}

Finally, for $1\geq \epsilon > 1- \delta_b > \delta >\sfrac{1}{2}$, then
$\kappa(a)=\lambda(y^{\delta})< \lambda(y^\epsilon)$ and 
$\kappa(b)=\lambda(y^{\delta_b}) >\lambda(y^{1-\epsilon})$.  
Thus $\max\{\kappa(a), \lambda(y^\epsilon)\} = \lambda(y^\epsilon)$ and 
$\max\{\kappa(b),\lambda(y^{1-\epsilon})\} 
  = \lambda(y^{\delta_b})=\kappa(b)$ and so it follows by 
equation \Ref{eqn47A} that
\begin{eqnarray}
\min_{\epsilon\in[1-\delta_b,1]}  
\left\{ \max\{\kappa(a), \lambda(y^\epsilon)\}  + \max\{\kappa(b),\lambda(y^{1-\epsilon})\}\right\}
&=& \min_{\epsilon\in[1-\delta_b,1]}  \L \lambda(y^{\epsilon}) 
         + \lambda(y^{1-\delta_b}) \R  \nonumber \\
&=& \lambda(y^{\delta_b}) + \lambda(y^{1-\delta_b}).
\end{eqnarray}

Taking the minimum over all the intervals gives (since $1-\delta_b>\delta$)
the upper bound
\begin{equation}
\limsup_{n\to\infty} \frac{1}{n} \log \W{U}_{2n}(a,b,y) 
  \leq  \Sfrac{1}{2} \left( \lambda(y^{\delta}) + \lambda(y^{1-\delta}) \right).
\end{equation}
Combining this with the result in theorem \ref{thm2}
(and then using the symmetry in $a$ and $b$), and using the
continuity of $\kappa(a)$ and $\lambda(y)$, give the following result.

\begin{theo}
Suppose that $y>1$ and $a,b$ are such that both 
$\max\{\kappa(a),\kappa(b)\}>\lambda(y^{1/2})$\\
 and
$\lambda^{-1}(\kappa(b))\, \lambda^{-1}(\kappa(a)) < y$.\\

For $\kappa(b) \leq \lambda(y^{1/2}) \leq \kappa(a) \leq \lambda(y)$, consistent with equation \Ref{generalform},
\begin{eqnarray*}
\W{\rho}^{AB}(a,b,y)=\lim_{n\to\infty} \Sfrac{1}{2n} \log \W{U}_{2n}(a,b,y) 
&=& \Sfrac{1}{2} 
  \left( (1-\alpha_a)\,\kappa(a) + \alpha_a\,\lambda (y^{\sigma(a)}) 
  + \lambda(y^{1-\sigma(a)}) \right) \\
&=& \Sfrac{1}{2} (\kappa(a) + \lambda(y^{1-\sigma(a)}) )
 = \Sfrac{1}{2} (\lambda(y^{\sigma(a)}) + \lambda(y^{1-\sigma(a)}) )
\end{eqnarray*}
where $\sigma(x)$ is the solution of $\kappa(x) = \lambda(y^{\sigma(x)})$,
$\shalf \leq \sigma(x) \leq 1$, and
$\alpha_x = \Sfrac{\eps_*(y^{1-\sigma(x)})}{\eps_*(y^{\sigma(x)})}$.\\
If $\shalf < \sigma(a) \leq 1$, then it follows from lemma \ref{lemma4B} there exists a finite 
$y_a \equiv y_{\sigma(a)} \geq 1$,  such that 
$\alpha_a < 1$ if $y\geq y_a$. 

Similarly, 
for $\kappa(a) \leq \lambda(y^{1/2}) \leq \kappa(b) \leq \lambda(y)$, consistent with equation \Ref{generalform},
\begin{eqnarray*}
\W{\rho}^{AB}(a,b,y)=\lim_{n\to\infty} \Sfrac{1}{2n} \log \W{U}_{2n}(a,b,y) 
&=& \Sfrac{1}{2} 
  \left( (1-\alpha_b)\,\kappa(b) + \alpha_b\,\lambda (y^{\sigma(b)}) 
  + \lambda(y^{1-\sigma(b)}) \right) \\
&=& \Sfrac{1}{2} (\kappa(b) + \lambda(y^{1-\sigma(b)}) )
 = \Sfrac{1}{2} (\lambda(y^{\sigma(b)}) + \lambda(y^{1-\sigma(b)}) ) .
\end{eqnarray*}
If $\shalf < \sigma(b) \leq 1$, then it follows from lemma \ref{lemma4B} there exists a finite 
$y_b \equiv y_{\sigma(b)} \geq 1$,  such that 
$\alpha_b < 1$ if $y\geq y_b$. 
\hfill\hqed
\label{thm3}  
\end{theo}

%

\begin{figure}[t]
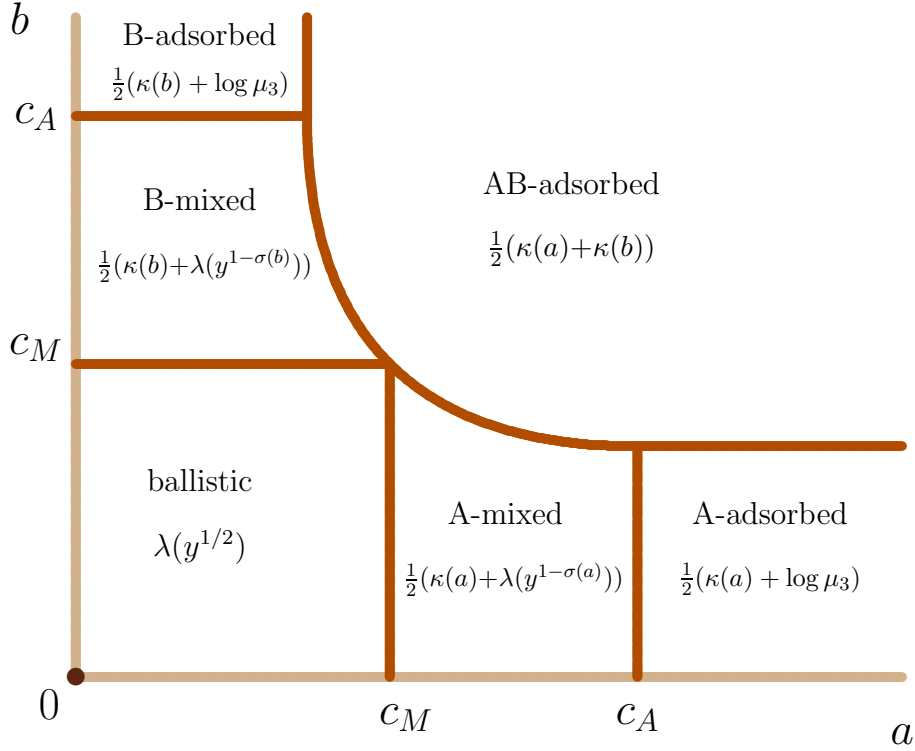

\beginpicture
\setcoordinatesystem units <1.25pt,1.25pt>
\setplotarea x from -75 to 300, y from -10 to 200
\setplotarea x from 0 to 300, y from 0 to 200

\setplotsymbol ({\footnotesize$\bullet$})

\color{Tan}
\plot 0  200 0 0 250 0 /
\color{Brown}
\setquadratic
\plot  70 170 95 95 170 70 /
\setlinear
\plot 0 95 95 95 / \plot 95 0 95 95 /
\plot 0 170 70 170 / \plot 170 0 170 70 /
\plot 70 170 70 200 / \plot 170 70 250 70 /

\color{Sepia}
\put {\LARGE$\bullet$} at 0 0 

\color{black}

\put {\LARGE$c_M$} at 100 -12  \put {\LARGE$c_A$} at 170 -12  \put {\LARGE$a$} at 250 -17
\put {\LARGE$c_M$} at -12 100  \put {\LARGE$c_A$} at -12 170  \put {\LARGE$b$} at -17 200
\put {\LARGE$0$} at -8 -8 

\put {\large ballistic} at 37.5 60  \put {\large$\lambda(y^{1/2})$} at 37.5 40
\put {\large B-mixed} at 37.5 145  
    \put {\scalebox{0.9}{$\sfrac{1}{2}(\kappa(b) {+} \lambda(y^{1-\sigma(b)}))$}} at 39 125
\put {\large A-mixed} at 130 50  
    \put {\scalebox{0.9}{$\sfrac{1}{2}(\kappa(a){+}\lambda(y^{1-\sigma(a)}))$}} at 132.5 30
\put {\large B-adsorbed} at 37.5 195  
    \put {\scalebox{0.9}{$\sfrac{1}{2}(\kappa(b) + \log \mu_3)$}} at 37.5 180
\put {\large A-adsorbed} at 210 50  
    \put {\scalebox{0.9}{$\sfrac{1}{2}(\kappa(a) + \log \mu_3)$}} at 210 30 
\put {\large AB-adsorbed} at 150 150  \put {$\sfrac{1}{2}(\kappa(a) {+} \kappa(b))$} at 150 130

\normalcolor
\color{black}

\endpicture
\caption{The phase diagram of the pulled $AB$-diblock loop for $y >1$ provided
that $\lambda(y)$ is strictly log-convex (see conjecture 1).  Otherwise, given any 
$\delta< \shalf$, we know there exists $y_{\delta}$ such that the mixed phases 
exist for all $y>y_{\delta}$ (ie $c_M < c_A$ but these values may depend on $\delta$).}
\label{figure7}   
\end{figure}

This theorem shows that, for points on the curve $C_\delta(b)$ in figure 
\ref{figureCdelta}, the function $\alpha_a < 1$ provided that $y$ is 
large enough (that is with $\delta=\sigma(a)$, $y$ exceeds $y_{\sigma(a)} \equiv y_a$). 
By symmetry in $a$ and $b$, the same is true when $a$ is replaced by $b$.
The complete free energy of pulled adsorbing diblock loops when $y>1$ can be obtained
from lemma \ref{lemma4} and by theorem \ref{thm3}.  We take this together
in the next section, and discuss the phase diagram of this model.

\subsection{The pulled and adsorbing $\boldsymbol{AB}$-block 
loop phase diagram for $y>1$}

The phase diagram of the pulled adsorbing loop is readily found if we assume that
$\lambda(y)$ is \textit{strictly} log-convex (see conjecture \ref{conjecture}).
In this section we will first assume strict log-convexity of $\lambda(y)$, and
then consider the phase diagram if we have the weaker form of log-convexity
shown in lemma \ref{lemma4B}.

Thus, assume that $\lambda(y)$ is strictly log-convex.  The phase diagram can 
then be determined from lemma \ref{lemma4} and theorem \ref{thm3}
and it is shown for $y>1$ in figure \ref{figure7}. This phase diagram is 
similar to the phase diagram calculated for a Dyck path model 
of an adsorbing block copolymer pulled in the middle 
(see figure 6 in reference \cite{Iliev}).  There are $6$ distinct phases, 
namely a ballistic phase, two ballistic-adsorbed mixed phases
($A$-mixed and $B$-mixed), 
and three adsorbed phases (one $A$-adsorbed, another $B$-adsorbed, 
and the third $AB$-adsorbed).  The free energies in each of the
phases are indicated in figure \ref{figure7} and are given in the 
following theorem.

\begin{theo}
If for all $y>1$ it is the case that $\lambda(y)$ is strictly log-convex, then, for $y>1$,
\begin{equation*}
\W{\rho}^{AB}(a,b,y) =
\begin{cases}
\lambda(y^{1/2}), & \hbox{for $\lambda(y^{1/2}) > \max\{\kappa(a),\kappa(b)\}$}; \\
\sfrac{1}{2} (\kappa(a) + \log \mu_3), & \hbox{for $\lambda(y) < \kappa(a)$ $\&$ $b \leq a_c$}; \\
\sfrac{1}{2} (\kappa(b) + \log \mu_3), & \hbox{for $\lambda(y) < \kappa(b)$ $\&$ $a \leq a_c$}; \\
\sfrac{1}{2} (\kappa(a)+\lambda(y^{1-\sigma(a)})),& \hbox{for $\kappa(b)<\lambda(y^{1-\sigma(a)})<\lambda(y^{1/2})<\kappa(a)<\lambda(y)$} \\
\sfrac{1}{2} (\kappa(b)+\lambda(y^{1-\sigma(b)})),& \hbox{for $\kappa(a)<\lambda(y^{1-\sigma(b)})<\lambda(y^{1/2})<\kappa(b)<\lambda(y)$} \\
\sfrac{1}{2} (\kappa(a) + \kappa(b)), &  \hbox{for $\lambda(y^{1/2}) <  \max\{\kappa(a),\kappa(b)\}$ $\&$ $\sigma(b) +\sigma(a)\geq 1$}. 
\end{cases}
\end{equation*}
where $\sigma(x)$ is the solution of $\kappa(x) = \lambda(y^{\sigma(x)})$. 
\hfill\hqed
\label{thmphases}   
\end{theo}

Notice that this theorem verifies equation \Ref{generalform}.

The phase boundary separating the ballistic and $A$-mixed phase is determined 
by putting $2\,\lambda(y^{1/2}) = \kappa(a)+\lambda(y^{1-\sigma(a)}) = 
\lambda(y^{\sigma(a)})+\lambda(y^{1-\sigma(a)})$, 
since $\kappa(a) = \lambda(y^{\sigma(a)})$.  Under the strict log-convexity 
assumption, this occurs only when $\sigma(a) = \sfrac{1}{2}$,  so that 
$a=c_M = \kappa^{-1}(\lambda(y^{1/2}))$.
A similar argument shows that the phase boundary separating the ballistic 
and $B$-mixed phase is determined by $b=c_M = \kappa^{-1}(\lambda(y^{1/2}))$.

The phase boundary separating the $A$-mixed and $A$-adsorbed phases 
occurs when $\kappa(a)+\log \mu_3 = \kappa(a)+\lambda(y^{1-\sigma(a)})$.  
Since $\lambda(y^{1-\sigma(a)}) > \log \mu_3$ when $\sigma(a)<1$, 
the phase boundary is evidently determined by $\sigma(a)=1$, in which 
case it is along the line $\kappa(a) =\lambda(y)$ or 
$a=c_A = \kappa^{-1}(\lambda(y))$.  Similarly, for the corresponding $B$ phases, 
the boundary is $b=c_A=\kappa^{-1}(\lambda(y))$.

Since $\sigma(x)$ is the solution of $\kappa(x) = \lambda(y^{\sigma(x)})$
for $x=a$ or $x=b$, the free energy in the two mixed phases is also given by
\begin{equation}
\W{\rho}^{AB}(a,b,y) 
= \Sfrac{1}{2} \left( \lambda(y^{\sigma(x)})+\lambda(y^{1-{\sigma(x)}}) \right) 
\label{rhox}
\end{equation}
where $x=a$ in the $A$-mixed phase, and $x=b$ in the $B$-mixed phase.
Notice that $\sfrac{1}{2}\leq {\sigma(x)} \leq 1$ (as discussed in section
\ref{4.3}) and that $\sigma(x)$ increases from $\sigma(x)=\shalf$ if
$x=c_M$, to $\sigma(x)=1$ when $x=c_A$.  Due to the strict log-convexity, the strict version of 
equation \Ref{log-convex} holds and thus we also have that $\W{\rho}^{AB}(a,b,y) $ of equation \Ref{rhox} is strictly increasing in $x$.

Existence of the $A$-mixed phase is proven if it is shown that $c_M < c_A$.  
This is so if
\begin{equation}
2\,\lambda(y^{1/2}) 
< \lambda(y^{\sigma(a)})+\lambda(y^{1-\sigma(a)}),
\label{74}
\end{equation}
for $\sigma(a) \in (\shalf,1)$.  This follows from our conjecture
of strict log-convexity of $\lambda(y)$.  (Notice that
$\lambda(y^{1-\sigma(a)}) \leq \lambda(y)$ so that the inequality 
\Ref{74} is sufficient for showing that $c_M < c_A$).

Since the phase diagram is symmetric in $a$ and $b$, this argument
also proves the existence of the $B$-mixed phase.

The phase boundary separating the $A$-adsorbed from the $AB$-adsorbed phases
is determined by $\kappa(b) = \log\mu_3$, so it is at $b=a_c$,  where $a_c$ is the 
adsorption critical point in $\kappa(a)$.  Similarly, the phase boundary separating 
the $B$-adsorbed phase from the $AB$-adsorbed phase is along the line $a=a_c$.

This leaves the curved phase boundary separating the mixed phases from the
$AB$-adsorbed phase.  Along the phase boundary separating the $A$-mixed and
the $AB$-adsorbed phases, 
\begin{equation}
\lambda(y^{\sigma(a)})+\lambda(y^{1-{\sigma(a)}}) = \kappa(a) + \kappa(b),
\quad\hbox{where $\sfrac{1}{2} \leq {\sigma(a)} \leq 1$}.
\end{equation}
Since $\kappa(a)  = \lambda(y^{\sigma(a)})$, it follows that
$\kappa(b)=\lambda(y^{1-{\sigma(a)}})$.  In other words,
\begin{equation}
\lambda^{-1}(\kappa(b)) \, y^{{\sigma(a)}} = y,
\quad\hbox{and}\; 
\lambda^{-1}(\kappa(a)) = y^{\sigma(a)}.
\end{equation}
Eliminating $y^{\sigma(a)}$ gives the expression
\begin{equation}
\lambda^{-1}(\kappa(a)) \; \lambda^{-1}(\kappa(b)) = y,
\label{eqn52}  
\end{equation}
which is the curve separating these phases for a given $y>1$.  Notice that if
$a=b$ then $\lambda^{-1}(\kappa(a)) = \lambda^{-1}(\kappa(b))=y^{1/2}$ so that
both $\kappa(a) =\lambda(y^{1/2})$ and $\kappa(b)=\lambda(y^{1/2})$.  This
gives $a=a_M$ and $b=b_M$ so that the point $(a_M,b_M)$ is on this phase
boundary.

Similarly, the phase boundary separating the $B$-mixed phase from the $AB$-adsorbed
phase is determined by $\kappa(a) = \lambda(y^{1-{\sigma(b)}})$ and this simplifies
again to equation \Ref{eqn52}. This completes the description of the phase diagram
for the case that $\lambda(y)$ is strictly log-convex.

Next, relax the assumption of strict log-convexity to the (proven)
weaker log-convexity in lemma \ref{lemma4B}.  Suppose that
$\xi \in (\shalf,1)$ is fixed.  Then there exists a finite $y_\xi>1$
such that
\begin{equation}
2\,\lambda(y^{1/2}) 
< \lambda(y^{\xi})+\lambda(y^{1-\xi}),
\qquad\hbox{for all $y>y_\xi$}.
\label{75}
\end{equation}
In other words, for any $x$ such that $\sigma(x)$ is the solution
of $\kappa(x) = \lambda(y^{\sigma(x)})$ and 
$\xi \leq \sigma(x) < 1$ there is a $y_\xi$ such that
for all $y>y_\xi$ equation \Ref{75} holds and therefore
\begin{equation}
2\,\lambda(y^{1/2}) 
< \lambda(y^{\xi})+\lambda(y^{1-\xi})
< \lambda(y^{\sigma(x)})+\lambda(y^{1-\sigma(x)})
\qquad\hbox{for all $y>y_\xi$ and $\sigma(x)\in(\xi,1)$},
\label{76}
\end{equation}
since for fixed $y$, $ \lambda(y^{\delta})+\lambda(y^{1-\delta})$
is a non-decreasing function of $\delta\in(\shalf,1)$ and by the consequences of lemma \ref{lemma4B}, a strictly increasing function for $\delta\in(\xi,1)$.

Fix $y>y_\xi$, and let $c_X$ be given by the solution of $\sigma(x) = \xi$.
Since $\xi<1$ this proves that $c_X < c_A$ and for any $x\in(c_X,c_A)$,
\begin{equation} 
2\,\lambda(y^{1/2}) 
< \lambda(y^{\sigma(x)})+\lambda(y^{1-\sigma(x)}).
\end{equation}
Since $x<c_A$ and $\kappa(x) = \lambda(y^{\sigma(x)})$ it also 
follows that $\sigma(x) < 1$ for $x\in(c_X,c_A)$, so that
\begin{equation}
 \lambda(y^{\sigma(x)})+\lambda(y^{1-\sigma(x)})
 > \kappa(x) + \log \mu_3,
\qquad\hbox{for $x\in(c_X,c_A)$.}
\end{equation}
In other words, $\lambda(y^{\sigma(x)})+\lambda(y^{1-\sigma(x)})$
exceeds both $2\,\lambda(y^{1/2})$ and 
$\kappa(x) + \log \mu_3$ if $c_X < x < c_A$.  This is the mixed
ballistic-adsorbed phase, and there exists a $C_M \leq C_X < C_A$ such that
the $A$-mixed phase has free energy 
$\lambda(y^{\sigma(a)})+\lambda(y^{1-\sigma(a)})$
for $C_M < a < C_A$ and $\kappa(b) < \lambda(y^{1-\sigma(a)})$, 
and the $B$-mixed phase has free energy
$\lambda(y^{\sigma(a)})+\lambda(y^{1-\sigma(a)})$
for $C_M < a < C_A$ and $\kappa(a) < \lambda(y^{1-\sigma(b)})$.
By lemma \ref{lemma4B}, as $\xi$ approaches $\shalf$ from above, 
$y_\xi$ increases to $\infty$, and $c_X \to c_M$ and the phase diagram in 
figure \ref{thmphases} is recovered in this limit.

\section{Pulled adsorbing diblock copolymers}
\label{sec:diblock}

In this section we continue by examining the behaviour of $AB$-diblock copolymers
(see figures \ref{figure1}(a) and \ref{figure1}(b)). Similar to the case examined
in section \ref{sec:loops} these models are positive walks with
$2n{+}1$ vertices labelled $j=0,1, \ldots 2n$.  The vertex $0$ is fixed at 
the origin and is not weighted.   Vertices $1 \le j \le n$ are $A$-vertices and weighted
by $a$, while vertices $n{+}1 \le j \le 2n$ are $B$-vertices and weighted by $b$.  
We consider two cases:  the vertical force is applied at vertex $2n$, 
namely at the end of the walk, or at vertex $n$ (the middle vertex of the walk).  
There are interesting differences between the two cases and the 
second is more difficult to treat (we shall rely, in that case, on the results for the
pulled loops in section \ref{sec:loops}).

\subsection{Diblock copolymers pulled at an end-point}
\label{sec:diblock:diblockendpoint}

The model is defined similarly to loops in section \ref{sec:loops}.  A positive
walk from the origin of length $2n$ with vertices $j=0,1,\ldots, 2n$ has vertex
$0$  fixed at the origin.  Vertices $1,2,\ldots, n$ are $A$-vertices and these
have $v_A$ visits in the adsorbing plane and each visit is weighted by $a$.  Vertices 
$n{+}1,n{+}2,\ldots, 2n$ are $B$-vertices and there are $v_B$ vertices in 
the adsorbing plane weighted by $b$.  The partition function of this model 
is given by
\begin{equation}
D_{2n}^{(e)}(a,b,y)= \sum_{v_A,v_B,h} 
d_{2n}^{(e)}(v_A,v_B,h)\, a^{v_A}b^{v_B}\,y^h 
\end{equation}
where $d_{2n}^{(e)}(v_A,v_B,h)$ counts walks with the above labelling and length $2n$, with $v_A$ $A$-visits, $v_B$ $B$-visits and 
with the $x_3$-coordinate of the last vertex equal to $h$.
We shall write 
\begin{equation}
\Delta_e(a,b,y)= \lim_{n\to\infty} \sfrac{1}{2n} \log D_{2n}^{(e)}(a,b,y)
\end{equation}
for the free energy of this model when we can prove that the limit exists.
Note that this free energy does not exist when $b=y=0$ -- this exception is assumed for
this model in what follows below.

\begin{lemm}
For $y \le 1$ and any $a\geq 0$ and $b\geq 0$, the free energy $\Delta_e(a,b,y)$ 
is given by 
$$\Delta_e(a,b,y) =  \Sfrac{1}{2} \L\kappa(a) + \kappa(b)\R.$$
Thus, for any $y\geq 0$ and $a<a_c$ and $b<a_c$,  $\Delta_e(a,b,y) = \lambda(y)$.
Hence for $y \leq 1$ and  $a<a_c$ and $b<a_c$, $\Delta_e(a,b,y) = \log \mu_3$.
\label{lemma:abysmall}
\end{lemm}

\noindent\textit{Proof:}
When $y \le 1$ we can use monotonicity to establish that
$$D_{2n}^{(e)}(a,b,0) \le D_{2n}^{(e)}(a,b,y) \le D_{2n}^{(e)}(a,b,1).$$
Treating the first $n$ edges and the second $n$ edges as independent, the partition 
function for the first $n$ edges is bounded above by the partition function of a positive 
walk with all vertices labelled $A$, that is by $e^{n \kappa(a) + o(n)}$.  
The final $n$ edges might not visit the surface; their contribution to the partition function is $\mu_3^{n+o(n)}$.  If the final $n$ edges do visit the surface 
then their partition function is bounded above
by a product of partition functions that together are bounded above by 
$e^{n \kappa(b) + o(n)}$.  This gives the upper bound
$$D_{2n}^{(e)}(a,b,0) \le D_{2n}^{(e)}(a,b,y) \le D_{2n}^{(e)}(a,b,1) 
=e^{(\kappa(a)+\kappa(b))n + o(n)}.$$

To complete the lower bound, consider concatenating two unfolded adsorbing loops, 
each with $n$ edges, one loop labelled with all $A$'s and the other all $B$'s.  Then, 
assuming that $a,b>0$,
$$D_{2n}^{(e)}(a,b,0) \ge L_n^{\dagger}(a)\, L_n^{\dagger}(b) 
= e^{(\kappa(a)+\kappa(b))n + o(n)},$$
where the construction can also be modified to show that the final lower bound 
here holds for the cases that $a \geq 0$ and $b\geq 0$ (using an unfolded walk 
without visits instead, and noting that $\kappa(0)=\log \mu_3$).
Taking logarithms, dividing by $2n$ and letting $n \to\infty$ gives 
$\Delta_e(a,b,y) \geq  \Sfrac{1}{2} \L\kappa(a) + \kappa(b)\R$, as required.
If $b=0$ then a similar lower bound can be obtained, but using an unfolded loop, 
and an $x_1$-unfolded pulled walk.  This gives, $D_{2n}^{(e)}(a,0,y)
\geq L_n^{\dagger}(a)\, C_n^{\dagger} (y)$,  where $C_n^{\dagger} (y)$ 
is the $x_1$-unfolded version of the partition function given in equation \Ref{8},
obtained by using the Hammersley-Welsh construction \cite{HammersleyWelsh}.  
Taking logarithms, dividing by $2n$ and letting $n \to\infty$ gives 
$\Delta_e(a,0,y) \geq  \sfrac{1}{2} (\kappa(a) + \log \mu_3)
= \sfrac{1}{2} (\kappa(a) + \kappa(0))$.

For any $y\geq 0$, when $a,b \leq a_c$ we have by monotonicity that 
$$C_{2n}(0,y) = D_{2n}^{(e)}(0,0,y) \leq D_{2n}^{(e)}(a,b,y) 
\leq D_{2n}^{(e)}(\max\{a,b\},\max\{a,b\},y) = C_{2n}(\max\{a,b\},y).$$
Thus, if $y>0$, then by equation \Ref{eqn:psicondition}, 
since $\max\{\kappa(a),\lambda(y)\}=\lambda(y)$ 
for $a<a_c$, both $C_{2n}(0,y)$ and  $C_{2n}(\max\{a,b\},y)$  
$= e^{2n\,\lambda(y) + o(n)}$.  Taking logarithms, dividing by $2n$ 
and letting $n\to\infty$ shows that $\Delta_e(a,b,y) = \lambda(y)$ if
$y>0$ and $a,b \leq a_c$.  In the event that $y=0$, then we know that 
the free energy is equal to $\lambda(0)$ since 
$\Delta_e(a,b,0)=\log \mu_3 = \lambda(0)$ if $a,b\leq a_c$. \hqed

The free energy for the cases $y\leq 1$ or  $a,b< a_c$ are completed in 
lemma \ref{lemma:abysmall}.  This leaves the case $y>1$ and $a,b \geq a_c$.  
Consider an endpoint pulled walk and observe that it either has some 
$B$-vertices in the adsorbing surface, or it has no $B$-vertices in the 
adsorbing surface.  If there are $B$-vertices in the surface then the $A$-block 
is not pulled at all and  contributes $\sfrac{1}{2}\kappa(a)$  to the free energy 
while the $B$-block contributes $\sfrac{1}{2}\max\{\kappa(b),\lambda(y)\}$.  
If there are no $B$-vertices in the surface the $A$-block contributes 
$\sfrac{1}{2}\max\{\kappa(a), \lambda(y)\}$ and the $B$-block contributes 
$\sfrac{1}{2}\lambda(y)$. This gives the upper bound 
\begin{eqnarray}
\!
\limsup_{n\to\infty} \Sfrac{1}{2n} \log D_{2n}^{(e)}(a,b,y) & \leq &
\max\LC \Sfrac{1}{2} \L \kappa(a) + \max\LC\kappa(b),\lambda(y) \RC \R, 
   \Sfrac{1}{2} \L \max\LC \kappa(a),\lambda(y)\RC +\lambda(y) \R   \RC
\nonumber \\
& = & \max\LC \Sfrac{1}{2} \L\kappa(a)+\kappa(b)\R, 
                             \Sfrac{1}{2} \L\kappa(a)+\lambda(y)\R, \lambda(y)  \RC.
\label{eqn57}
\end{eqnarray}
In the first line of the upper bound, the second term dominates when 
$\kappa(a) > \lambda(y)$ and $\lambda(y) > \kappa(b)$ which implies 
that $\kappa(a) > \kappa(b)$.  In this case we have three terms in the 
second line of the  upper bound.  If $\kappa(a) \le \kappa(b)$ we have only 
two terms, namely the first and third.

Next, we construct corresponding lower bounds.  By lemma 
\ref{lemma:abysmall} and monotonicity, 
\begin{equation}
\liminf_{n\to\infty} \Sfrac{1}{2n} \log D_{2n}^{(e)}(a,b,y) \geq 
\max \LC \Sfrac{1}{2} \L \kappa(a)+\kappa(b) \R, \lambda(y)\RC .
\label{eqn58}
\end{equation}
If $\kappa(a) \le \kappa(b)$ these two lower bounds match the two upper 
bounds derived above.  If $\kappa(a) > \kappa(b)$ we need a third lower 
bound which we construct as follows.  Concatenate an $n$-edge loop 
(for the $A$-block) which has been unfolded in the $x_1$-direction 
in such a way that its first and last edges are in the surface, to an 
$n$-edge positive walk which has been unfolded in the $x_1$-direction 
in such a way that only its $0$-th vertex is in the surface.  Since the 
two subwalks only have the vertex in common where they meet, this 
gives the lower bound
\begin{equation}
\liminf_{n\to\infty} \Sfrac{1}{2n} \log D_{2n}^{(e)}(a,b,y) \geq 
\Sfrac{1}{2} \L \kappa(a)+\lambda(y) \R,
\label{eqn59}
\end{equation}
which can be shown to hold for $b\geq 0$.

The upper and lower bounds in equations \Ref{eqn57}, \Ref{eqn58}
and \Ref{eqn59} prove the following theorem giving the
phases of an adsorbing diblock copolymer pulled at its endpoint.
\begin{theo}
Suppose that $a,b \geq a_c$ and $y > 1$.  When $\kappa(a) > \kappa(b)$
$$\Delta_e(a,b,y) = \max \LC \Sfrac{1}{2} \L \kappa(a)+\kappa(b) \R,
\Sfrac{1}{2} \L \kappa(a)+\lambda(y) \R, \lambda(y)  \RC$$
and when $\kappa(a) \le \kappa(b)$
$$\Delta_e(a,b,y)=\max\LC \Sfrac{1}{2} \L \kappa(a)+\kappa(b) \R,  
\lambda(y)  \RC.$$
\label{thm5} \hfill\hqed
\end{theo}

\begin{figure}[t]
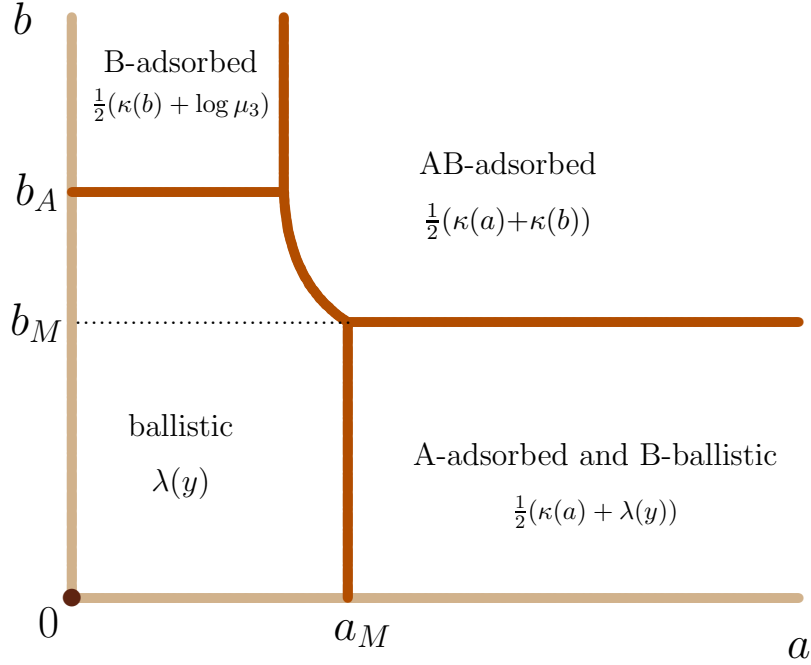

\beginpicture
\setcoordinatesystem units <1.1pt,1.1pt>
\setplotarea x from -90 to 300, y from -10 to 200
\setplotarea x from 0 to 300, y from 0 to 200

\setplotsymbol ({\footnotesize$\bullet$})

\color{Tan}
\plot 0  200 0 0 250 0 /
\color{Brown}
\setquadratic
\plot  73 140 79 113 95 95  /
\setlinear
\plot 95 0 95 95 /
\plot 0 140 73 140 / 
\plot 73 140 73 200 /
\plot 95 95 250 95 /

\color{Sepia}
\put {\LARGE$\bullet$} at 0 0 

\color{black}
\normalcolor

\setplotsymbol ({\footnotesize.})
\setdots <3pt>
\plot 95 95 0 95 /

\put {\LARGE$a_M$} at 100 -12  \put {\LARGE$a$} at 250 -17
\put {\LARGE$b_A$} at -12 140  \put {\LARGE$b$} at -17 200
\put {\LARGE$b_M$} at -12 95  
\put {\LARGE$0$} at -8 -8 

\put {\large ballistic} at 37.5 60  \put {\large$\lambda(y)$} at 37.5 40
\put {\large B-adsorbed} at 37.5 185  
    \put {\scalebox{0.9}{$\sfrac{1}{2}(\kappa(b) + \log \mu_3)$}} at 37.5 170
\put {\large A-adsorbed and B-ballistic} at 180 50  
    \put {\scalebox{0.9}{$\sfrac{1}{2}(\kappa(a) + \lambda(y))$}} at 180 30 
\put {\large AB-adsorbed} at 150 150  
    \put {$\sfrac{1}{2}(\kappa(a) {+} \kappa(b))$} at 150 130

\endpicture
\caption{The phase diagram of the pulled $AB$-diblock walk pulled at its end-point $y >1$.}
\label{figure7C}   
\end{figure}

The phase diagram of this model can be determined from the results
above.  The case $y\leq 1$ is given by lemma \ref{lemma:abysmall}
and it can be checked that the phase diagram is identical to the diagram
shown in figure \ref{figure5}.  The more complex situation is encountered
for $y>1$ and the phases are given in theorem \ref{thm5}.  The phase
diagram for this case is shown in figure \ref{figure7C}, and we identify
four phases by noting, in addition to the results in theorem \ref{thm5},
that $\kappa(a)$ is singular at the adsorption critical point $a=a_c$.
For large $a$ and $b$ the copolymer is adsorbed, and if both
$a$ and $b$ are small, then it is ballistic (since $y>1$).  For small
$a<a_c$ and large $b$ the adsorption of the $B$-block overcomes the
ballistic phase, and the copolymer is in a phase with the $B$-block
adsorbed, and the $A$-block free.  On the other hand, if $b$ is small
so that $\kappa(b) < \lambda(y)$, and $a>a_c$ is large, then the
adsorption of the $A$-block overcomes the ballistic phase (only
in the $A$-block), while the $B$-block remains ballistic, it being
pulled at its end-point.

The phase boundary separating the ballistic and B-adsorbed phases is given
by the solution $b_A$ of $\kappa(b) = 2\,\lambda(y) {-} \log \mu_3$ so that
$b_A>a_c$ (the adsorption critical point).  The $B$-adsorbed phase is separated
by the phase boundary $a=a_c$ from the $AB$-adsorbed phase.  The
ballistic and $AB$-adsorbed phases are separated by the curve
$\kappa(a){+}\kappa(b) = 2\,\lambda(y)$, while the ballistic and 
mixed $A$-adsorbed and $B$-ballistic phases are separated by the
solution $a=a_M$ of $\kappa(a) = \lambda(y)$.  Similarly, this mixed
$A$-adsorbed and $B$-ballistic phase is also separated from the
$AB$-adsorbed phase by the solution $b=b_M$ of 
$\kappa(b)=\lambda(y)$.

\subsection{Diblock copolymers pulled at a mid-point}

\begin{figure}[t!]
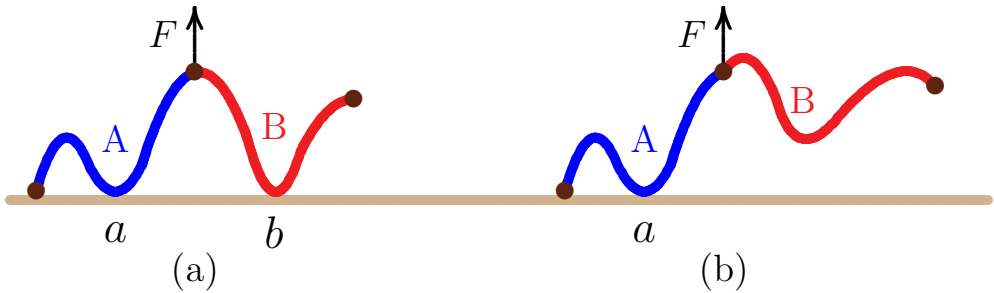

\beginpicture
\setcoordinatesystem units <1pt,1pt>
\setplotarea x from -50 to 250, y from -5 to 80
\setplotarea x from 0 to 250, y from 0 to 80

\setcoordinatesystem units <1pt,1pt> point at 0 0 

\setplotsymbol ({\footnotesize$\bullet$})
\color{Tan}
\plot -10 -3 360 -3  /

\setplotsymbol ({\large.})
\color{black}
\arrow <8pt>  [.2,.67] from 60 45 to 60 70
\put {\Large$F$} at 48 60

\setplotsymbol ({\footnotesize$\bullet$})

\setquadratic
\color{Blue}
\plot 0 0 10 20 20 10 30 0 40 10 50 35 60 45   /  \put {\Large A} at 30 20
\color{Red}
\plot 60 45 70 40 80 20 90 0 100 15 110 30 120 35 /  \put {\Large B} at 90 25

\color{Sepia}
\multiput {\LARGE$\bullet$} at 0 0 60 45 120 35  /

\color{black}
\normalcolor

\put {\LARGE$a$} at 30 -15 
\put {\LARGE$b$} at 90 -15 
\put {\Large (a)} at 60 -30

\setcoordinatesystem units <1pt,1pt> point at -200 0 

\setplotsymbol ({\large.})
\arrow <8pt>  [.2,.67] from 60 45 to 60 70
\put {\Large$F$} at 48 60

\setplotsymbol ({\footnotesize$\bullet$})

\setquadratic
\color{Blue}
\plot 0 0 10 20 20 10 30 0 40 10 50 35 60 45   /  \put {\Large A} at 30 20
\color{Red}
\plot 60 45 70 50 80 35 90 20 105 30 125 45 140 40 /  \put {\Large B} at 90 35

\color{Sepia}
\multiput {\LARGE$\bullet$} at 0 0 60 45 140 40  /

\color{black}
\normalcolor

\put {\LARGE$a$} at 30 -15 
\put {\Large (b)} at 60 -30

\endpicture
\caption{The cases of a $AB$-diblock copolymer pulled in the middle.
(a) Both the $A$-block and the $B$-block have visits in the adsorbing plane.
(b) Only the $A$-block has visits in the adsorbing plane.}
\label{figure8} 
\end{figure}

Adsorbing $AB$-diblock copolymers pulled in the middle (see figure \ref{figure1}(b)) 
can be analysed identifying two subcases, and using the results for pulled
adsorbing $AB$-diblock loops in section \ref{sec:loops}.  The two subcases are 
shown in figure \ref{figure8}, and are (a) the case where the right-most 
half of the walk (the $B$-block) interacts with the adsorbing surface and 
(b) the case where the $B$-block does not touch the adsorbing surface at all. 
Case (b) is easier to treat, so we proceed by analysing it first. 

Before doing so, however, we introduce the general notation needed.
Let $d_{2n}^{(m)}(v_A,v_B,h)$ be the number of positive walks starting from the 
origin, with the first vertex inert, followed by $n$ $A$-vertices 
and then $n$ $B$-vertices, with  $v_A$  $A$-visits
and $v_B$ $B$-visits and with
the midpoint (the $n$th vertex)  of the walk at height $h$. 
Define the partition function
\begin{equation}
D_{2n}^{(m)} (a,b,y)
= \sum_{h=0}^n \sum_{v_A,v_B} d_{2n}^{(m)}(v_A,v_B,h)\, 
a^{v_A}b^{v_B}\, y^h 
\end{equation}
of adsorbing $AB$-diblock copolymers pulled in the middle vertex. The free 
energy of this model is defined by
\begin{equation}
\Delta_m (a,b,y) = \lim_{n\to\infty} \Sfrac{1}{2n} \log D_{2n}^{(m)} (a,b,y)
\end{equation}
and we shall show that this limit exists.  Notice the exceptional case when $a=y=0$
where this free energy does not exist;  this exception is assumed for this model.

As discussed above we analyse case (b) first, the case where the $B$-block does not touch the adsorbing surface.

\subsubsection{Case (b)}
Note first that this case is equivalent to setting $b=0$ in $D_{2n}^{(m)} (a,b,y)$. 
If the $B$-block is constrained to be disjoint with the adsorbing line, then its
contribution to the free energy of the model is $\log \mu_3$, since it is
a non-interacting walk (it does not interact with the surface and it feels no effect 
from the pulling force $F$ at the midpoint).  The $A$-block, on the other hand, is
a pulled adsorbing walk.  By using strategy bounds similar to those introduced
in references \cite{Rensburg2017}, the free energy of case (b) walks is given by 
the following lemma.
\begin{lemm}
The free energy of case (b) walks is given by
\[ \Delta_m (a,0,y)= \Sfrac{1}{2} \L\psi_e(a,y)+\log\mu_3 \R 
= \Sfrac{1}{2}\L \max\LC \kappa(a),\lambda(y) \RC + \log \mu_3\R . \eqno\hqed \]
\label{lemma9}
\end{lemm} 
This leaves case (a) walks to be considered.  For that we need to establish
bounds on the free energy.

\subsubsection{Case (a)}

This case is only relevent when $b>0$, so we assume throughout this subsection.

Observe by lemma \ref{lemma9} and monotonicity
that $\Delta_m(a,b,y) \geq \Delta_m(a,0,y) 
= \sfrac{1}{2}(\max\LC \kappa(a),\lambda(y) \RC + \log \mu_3)$.  
Further, since the $AB$-diblock loops of section \ref{sec:loops} are a 
subset of the walks counted here in $d_{2n}^{(m)}(v_A,v_B,h)$, 
we have that $\Delta_m(a,b,y) \geq \W{\rho}^{AB}(a,b,y)$ where 
$\W{\rho}^{AB}(a,b,y)$ is given in theorem \ref{thmyl1} for $y\leq 1$, 
and in theorem \ref{thmphases} for $y>1$ and more generally for any 
$y$ by equation \Ref{generalform}.

\begin{figure}[t!]
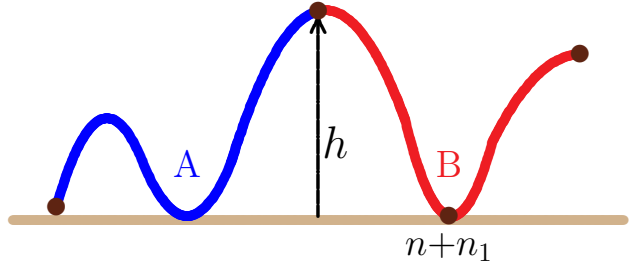

\beginpicture
\setcoordinatesystem units <1.65pt,1.65pt>
\setplotarea x from -70 to 150, y from -5 to 50
\setplotarea x from 0 to 150, y from 0 to 50

\setcoordinatesystem units <1.65pt,1.65pt> point at 0 0 

\setplotsymbol ({\footnotesize$\bullet$})
\color{Tan}
\plot -10 -3 130 -3  /

\setplotsymbol ({\large.})
\color{black}
\arrow <8pt>  [.2,.67] from 60 -2  to 60 45

\setplotsymbol ({\footnotesize$\bullet$})

\setquadratic
\color{Blue}
\plot 0 0 10 20 20 10 30 -2 40 10 50 35 60 45   /  \put {\Large A} at 30 10
\color{Red}
\plot 60 45 70 40 80 20 90 -2 100 15 110 30 120 35 /  \put {\Large B} at 90 10

\color{Sepia}
\multiput {\LARGE$\bullet$} at 0 0 60 45 120 35 90 -2  /

\color{black}
\normalcolor

\put {\LARGE$h$} at 64 15 
\put {\Large$n{+}n_1$} at 90 -9 

\endpicture
\caption{Cutting an adsorbing $AB$-diblock copolymer in its last visit
to the adsorbing plane (with label $(n{+}n_1)$) gives the inequality
in equation \Ref{eqn62A}.}
\label{figure9} 
\end{figure}

To determine an upper bound for case (a), note that 
\begin{equation}
D_{2n}^{(m)}(a,b,y)-D_{2n}^{(m)}(a,0,y) \leq  b\,  \sum_h y^h 
\left(\sum_{v} c_n (v,h)\,a^v \right) \left( \sum_w 
 \sum_{h\leq n_1\leq n}c_{n_1} (w,h)\,b^w c_{n-n_1}\right).
\label{eqn62A}
\end{equation}
This bound is obtained by cutting the walk at its midpoint into two 
independent blocks and then cutting the $B$-block walk again at the 
last vertex (with label $n+n_1$) visiting the adsorbing surface $x_3=0$ 
(in case (a) $n_1\geq \max\{1,h\}$).  See figure \ref{figure9}.  If the walk 
is a loop, then $n_1=n$ and this gives the loop partition function
$\W{U}_{2n}(a,b,y)$ defined in equation \Ref{eqn25C}.  Notice that
not all diblock walks are represented by conformations such as shown 
in figure \ref{figure9}.  Conformations where the $B$-block is disjoint
with the adsorbing plane have no value for $n_1$ on the right hand
side of equation \Ref{eqn62A} and so these are subtracted out on 
the left hand side by substracting $D_{2n}^{(m)}(a,0,y)$.

Choosing the value of $n_1$ that maximizes the upper bound in 
equation \Ref{eqn62A} gives
\begin{eqnarray}
& & D_{2n}^{(m)}(a,b,y) -D_{2n}^{(m)}(a,0,y) \nonumber \\
& & \leq b\, 
(n+1)\left[\sum_h y^h \left(\sum_{v} c_n (v,h)\,a^v \right) 
\left( \sum_w \max_{h\leq n_1\leq n} 
\left( c_{n_1} (w,h)\,b^w c_{n-n_1}\right) \right)\right].
\end{eqnarray}
Summing independently over $h$ in the two factors gives, for any choice 
of $\eps\in[0,1]$, a larger upper bound:
\begin{eqnarray}
& & D_{2n}^{(m)}(a,b,y) - D_{2n}^{(m)}(a,0,y) \nonumber \\ 
& &\leq b\,  (n+1) \left(\sum_h y^{\epsilon h}\sum_{v} c_n (v,h)\,a^v  \right)
\;  \max_{n_1}\left(c_{n-n_1}
\sum_h y^{(1-\epsilon) h}  \sum_w c_{n_1} (w,h)\,  b^w \right).
\end{eqnarray}
We may assume, without loss of generality, that $\limsup_{n\to\infty}
\sfrac{n_1}{n} = \alpha \in [0,1]$.  Taking logarithms, dividing by $2n$ and 
letting $n\to\infty$ gives that for any $\eps\in[0,1]$
\begin{eqnarray}
& &\hspace{-1cm}
\limsup_{n\to\infty} \Sfrac{1}{2n} \log [D_{2n}^{(m)}(a,b,y) - D_{2n}^{(m)}(a,0,y)] \nonumber \\
&\leq& 
 \Sfrac{1}{2} \max_{\alpha\in[0,1]} \left\{(1-\alpha) \log\mu_3 +\psi_e(a,y^\epsilon) +  \alpha\psi_e(b,y^{1-\epsilon})\right\} .
\label{eqn37AA}  
\end{eqnarray}
Because of convexity, the maximum over $\alpha$ will be achieved either for
$\alpha=0$ or $\alpha=1$, so that we have 
\begin{eqnarray}
& &\hspace{-2cm}
\limsup_{n\to\infty} \Sfrac{1}{2n} \log [D_{2n}^{(m)}(a,b,y) - D_{2n}^{(m)}(a,0,y)]  \nonumber \\
&\leq&  \max\left\{ 
 \Sfrac{1}{2}\L \psi_e(a,y^\epsilon)+\log \mu_3 \R, 
  \Sfrac{1}{2}\L \psi_e(a,y^\epsilon) +  \psi_e(b,y^{1-\epsilon}) \R \right\} .
\label{eqn37BB}  
\end{eqnarray}
But since $\psi_e(a,y^{1-\epsilon})\geq \log\mu_3$ for any $\eps\in[0,1]$, this simplifies to
\begin{eqnarray}
& & \hspace{-2cm}
\limsup_{n\to\infty} 
  \Sfrac{1}{2n} \log [D_{2n}^{(m)}(a,b,y) - D_{2n}^{(m)}(a,0,y) ] \nonumber \\
& & \leq \min_{\epsilon\in[0,1]}\left\{ 
 \Sfrac{1}{2}\L \psi_e(a,y^\epsilon) 
                   +  \psi_e(b,y^{1-\epsilon})\R \right\} = \W{\rho}^{AB}(a,b,y).
\label{eqn37CC}
\end{eqnarray}
Thus
\begin{equation}
\lim_{n\to\infty} \Sfrac{1}{2n} \log [D_{2n}^{(m)}(a,b,y) - D_{2n}^{(m)}(a,0,y) ]
= \W{\rho}^{AB}(a,b,y).
\label{eqn37CC2}
\end{equation}

Combining cases (a) and (b) results gives the following.

\begin{lemm}
For $y\geq 0$,
\[
\Delta_m (a,b,y)= \lim_{n\to\infty} \Sfrac{1}{n} \log D_{2n}^{(m)}(a,b,y)
= 
 \max\LC \W{\rho}^{AB}(a,b,y), \Sfrac{1}{2}\L \psi_e(a,y)
+\log \mu_3\R \RC , 
\] 
where for $b=0$, the second term on the right gives the maximum.  \hqed
\label{lemma11} 
\end{lemm}

\subsubsection{The phase diagram of adsorbing $AB$-diblock copolymers pulled in
the middle}

Consider first the case that $y\leq 1$.   In that case for $b> 0$
 $\W{\rho}^{AB}(a,b,y) = \sfrac{1}{2} (\kappa(a) + \kappa(b))
 \geq \Sfrac{1}{2}\L \psi_e(a,y)+\log \mu_3\R$ and for $b=0$ 
 $\Delta_m (a,b,y)= \Sfrac{1}{2}\L \psi_e(a,y)+\log \mu_3\R 
 = \sfrac{1}{2} (\kappa(a) + \kappa(b)) $. Thus  for $y\leq 1$, 
 $\Delta_m (a,b,y)=\sfrac{1}{2} (\kappa(a) + \kappa(b))$ and we can 
use the cases for $\W{\rho}^{AB}(a,b,y)$ in theorem \ref{thmyl1} to 
obtain the following theorem.

\begin{theo}
For $y\leq 1$,   
\begin{equation*}
\Delta_m(a,b,y) =
\begin{cases}
\log \mu_3, & \hbox{for $a\leq a_c$ $\&$ $b\leq a_c$}; \\
\sfrac{1}{2} (\kappa(a) + \log \mu_3), & \hbox{\rm{for} $a>a_c$ $\&$ $b \leq a_c$}; \\
\sfrac{1}{2} (\kappa(b) + \log \mu_3), & \hbox{\rm{for} $a\leq a_c$ $\&$ $b > a_c$}; \\
\sfrac{1}{2} (\kappa(a) + \kappa(b)), & \hbox{\rm{for} $a > a_c$ $\&$ $b > a_c$}. 
\end{cases}
\end{equation*}
\hfill\hqed 
\end{theo}

This gives the phase diagram identical to figure \ref{figure5}.

The case $y>1$ is slightly more complicated. For $a,b$ in either of the 
two loop mixed phases, $\kappa(x)<\lambda(y)$ for both $x\in\{a,b\}$ and 
\begin{eqnarray}
\W{\rho}^{AB}(a,b,y) 
&=& \sfrac{1}{2} ( \kappa(x) + \lambda(y^{1-\sigma(x)}) )
=\sfrac{1}{2} (\lambda(y^{\sigma(x)}) + \lambda(y^{1-\sigma(x)}) ) \nonumber \\
&\leq& \Sfrac{1}{2}\L \lambda(y) +\log \mu_3\R= \Sfrac{1}{2}\L \psi_e(a,y)+\log \mu_3\R
\end{eqnarray}
(by log-convexity), so in this regime $\Delta_m(a,b,y)=\Sfrac{1}{2}\L \lambda(y)
+\log \mu_3\R$.  Similarly for $a,b$ in the loop ballistic phase, 
\begin{equation}
\W{\rho}^{AB}(a,b,y)=\lambda(y^{1/2}) \leq \Sfrac{1}{2}\L \lambda(y)
+\log \mu_3\R= \Sfrac{1}{2}\L \psi_e(a,y)+\log \mu_3\R.
\end{equation}
So the loop ballistic and  mixed ballistic-adsorbed phases are replaced by a mixed ballistic-free phase where $\Delta_m(a,b,y)=\sfrac{1}{2}\L \lambda(y)
+\log \mu_3\R$. Outside these regimes, $\shalf \leq \max\{\sigma(a),\sigma(b)\}$, $\sigma(a)+\sigma(b)\geq 1$, and $\W{\rho}^{AB}(a,b,y)= \sfrac{1}{2} (\kappa(a) + \kappa(b))$.
Comparing this to $\Sfrac{1}{2}\L \psi_e(a,y)+\log \mu_3\R$ leads to the following theorem.

\begin{theo}
For $y > 1$,   
\begin{equation*}
\Delta_m(a,b,y) =\begin{cases}
\sfrac{1}{2} (\lambda(y) + \log \mu_3), & \hbox{for $\kappa(a)\leq \lambda(y)$, $\kappa(b)\leq \lambda(y)$ 
$\&$ $\lambda(y) \ge \kappa(a) + \kappa(b) - \log \mu_3$}; \\
\sfrac{1}{2}(\kappa(b)+\log \mu_3), & \hbox{for $\lambda(y) \le \kappa(b)$ $\&$ $a \le a_c$}; \\
\sfrac{1}{2}(\kappa(a)+\log \mu_3), & \hbox{for $\lambda(y) \le \kappa(a)$ $\&$ $b \le a_c$}; \\
\sfrac{1}{2}(\kappa(a)+\kappa(b)), & \hbox{for $\lambda(y) \le \kappa(a) + \kappa(b) - \log \mu_3$, 
$a \ge a_c$ $\&$ $b \ge a_c$}. 
\end{cases}
\end{equation*}
\hfill\hqed 
\end{theo}

The phase diagram for $y>1$ is shown in figure \ref{figure8}.

\begin{figure}[t]
\beginpicture
\setcoordinatesystem units <1.5pt,1.5pt>
\setplotarea x from -70 to 300, y from -10 to 150
\setplotarea x from 0 to 300, y from 0 to 150

\setdots <2pt>
\plot 0 60 95 60 /
\plot 60 0 60 95 /
\setsolid

\setplotsymbol ({\footnotesize$\bullet$})

\color{Tan}
\plot 0 150 0 0 200 0 /
\color{Brown}
\plot 0 95 60 95 / \plot 95 0 95 60 /
\plot 95 60 200 60 /
\plot 60 95 60 150 /
\setquadratic
\plot 60 95 73 73 95 60 /
\setlinear

\color{Sepia}
\put {\LARGE$\bullet$} at 0 0 

\color{black}
\normalcolor

\put {\LARGE$a_c$} at 60 -9  \put {\LARGE$a$} at 200 -14
\put {\LARGE$a_c$} at -9 60  \put {\LARGE$b$} at -14 150
\put {\LARGE$0$} at -5 -5 
\put {\LARGE$a_M$} at 95 -9
\put {\LARGE$a_M$} at -9 95

\put {\large ballistic} at 37.5 50  
\put {\large$\sfrac{1}{2}(\lambda(y) {+} \log \mu_3)$} at 37.5 30
\put {\large B-adsorbed} at 30 125  
\put {\large$\sfrac{1}{2}(\kappa(b) {+} \log \mu_3)$} at 30 110
\put {\large A-adsorbed} at 145 40  
\put {\large$\sfrac{1}{2}(\kappa(a) {+} \log \mu_3)$} at 145 25
\put {\large AB-adsorbed} at 125 125  
\put {\large$\sfrac{1}{2}(\kappa(a) {+} \kappa(b))$} at 125 110

\endpicture
\caption{The phase diagram of diblock copolymers pulled in a middle vertex
for $y> 1$.  The phase boundary separating the ballistic and $A$-adsorbed
phases is given by $\lambda(y)=\kappa(a)$, of the ballistic and $B$-adsorbed
phases is $\lambda(y)=\kappa(b)$, of the $A$- or $B$-adsorbed phases and
the $AB$-adsorbed phase by $b=a_c$ and $a=a_c$ respectively, and
of the ballistic and $AB$-adsorbed phases by $\kappa(a)+\kappa(b) = 
\lambda(y)+\log \mu_3$.  For asymptotic values of $y$, this curve is
approximated by $ab= (\mu_3/\mu_2^2)\,y$. The critical value
$a_M$ is the solution of $\kappa(a)=\lambda(y)$.}
\label{figure8X}   
\end{figure}

\section{Triblock copolymers}
\label{sec:triblock}
In this section we consider triblock copolymers $ABA$.  The two outer 
blocks consist of $A$ monomers while the central block consists of $B$ 
monomers (see figures \ref{figure1}(c) and \ref{figure1}(d)).  All three blocks
are the same length.  These copolymers are of particular interest since they 
are often used as steric stabilizers of dispersions where either the $A$-blocks
adsorb on the dispersed particles and the $B$-blocks are desorbed and so 
extend into the dispersing phase, or \textit{vice versa} \cite{Fleer}.  We 
model them as a self-avoiding walk with $3n{+}1$ vertices labelled 
$j=0,1,2, \ldots 3n$.  Vertex 0 is fixed at the origin and is not weighted.  
Vertices $1 \le j \le n$ and vertices $2n{+}1 \le j \le 3n$ are $A$-vertices and 
vertices $n{+}1 \leq j \le 2n$ are $B$-vertices.    Thus, the walk starts at the 
origin and every vertex has non-negative $x_3$-coordinate, so that the 
walk is confined to the half-space $x_3\geq 0$.  The number of $A$-visits
($A$-vertices with coordinate $x_3=0$) is denoted by $v_A$, and the
number of $B$-visits is denoted $v_B$.

\begin{figure}[t]
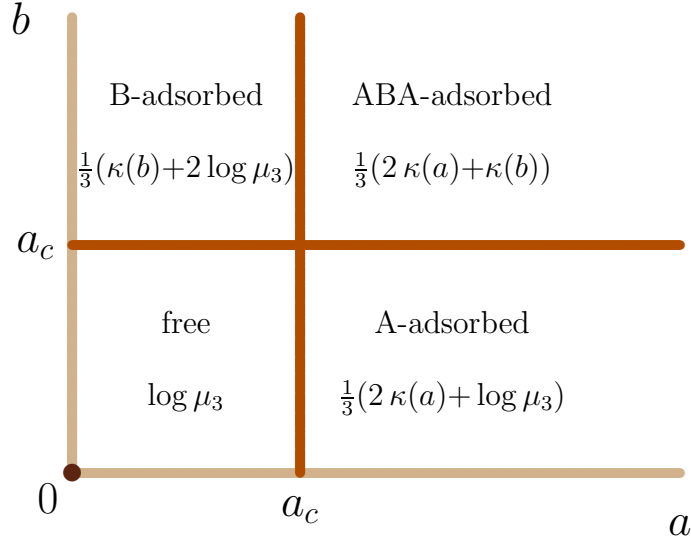

\beginpicture
\setcoordinatesystem units <1.15pt,1.15pt>
\setplotarea x from -100 to 300, y from -10 to 150
\setplotarea x from 0 to 300, y from 0 to 150

\setplotsymbol ({\footnotesize$\bullet$})

\color{Tan}
\plot 0 150 0 0 200 0 /
\color{Brown}
\plot 0 75 200 75 / \plot 75 0 75 150 /

\color{Sepia}
\put {\LARGE$\bullet$} at 0 0 

\color{black}
\normalcolor

\put {\LARGE$a_c$} at 75 -12  \put {\LARGE$a$} at 200 -17
\put {\LARGE$a_c$} at -12 75  \put {\LARGE$b$} at -17 150
\put {\LARGE$0$} at -8 -8 

\put {\large free} at 37.5 50  \put {\large$\log \mu_3$} at 37.5 25
\put {\large B-adsorbed} at 37.5 125  
\put {\large$\sfrac{1}{3}(\kappa(b) {+} 2\log \mu_3)$} at 37.5 100
\put {\large A-adsorbed} at 125 50  
\put {\large$\sfrac{1}{3}(2\,\kappa(a) {+} \log \mu_3)$} at 125 25
\put {\large ABA-adsorbed} at 125 125  
\put {\large$\sfrac{1}{3}(2\,\kappa(a) {+} \kappa(b))$} at 125 100

\endpicture
\caption{The four adsorbed phases of the pulled adsorbing $ABA$-triblock copolymer
when $y\leq 1$.}
\label{figure11}   
\end{figure}

\subsection{Triblock copolymers pulled at an end-point}
\label{sec:TriblockEnd}

We write $t_{3n}^{(e)}(v_A,v_B,h)$ for the number of self-avoiding walks 
with  $3n$ edges, with the above labelling and restrictions, and having the 
$x_3$-coordinate of the last vertex equal to $h$.  The corresponding partition 
function is 
\begin{equation}
T_{3n}^{(e)}(a,b,y) = \sum_{v_A,v_B,h} 
 t_{3n}^{(e)}(v_A,v_B,h)\, a^{v_A}b^{v_B} \, y^h.
\label{eqn:tribloclPFend}
\end{equation}
The free energy is
\begin{equation}
\tau_e(a,b,y) = \lim_{n\to \infty} \sfrac{1}{3n} \log T_{3n}^{(e)}(a,b,y)
\end{equation}
when we can prove the existence of the limit.  There is an exceptional situation 
when $a=y=0$ where the free energy does not exist for this model; 
this exception is assumed for this model.

If $y\leq 1$ then the force is zero or directed towards the adsorbing surface
$x_3=0$, and the free energy is given in the next theorem.

\begin{theo}
When $y \leq 1$ then
$$\tau_e(a,b,y) = \Sfrac{1}{3}(2\,\kappa(a)+\kappa(b)) .$$
\label{theo:triblockendpush}
\end{theo}

\noindent\textit{Proof:}
An upper bound is found by noting that 
$$T_{3n}^{(e)}(a,b,y) \le T_{3n}^{(e)}(a,b,1) 
= e^{(2\,\kappa(a)+\kappa(b))n+o(n)}$$
by a similar argument to that used in lemma \ref{lemma:abysmall}, 
and a lower bound from concatenating three unfolded loops to give
$$T_{3n}^{(e)}(a,b,y) \ge T_{3n}^{(e)}(a,b,0 ) \ge L_n^{\dagger}(a)^2 
L_n^{\dagger}(b) =e^{(2\,\kappa(a)+\kappa(b))n+o(n)}$$
for $a>0$, 
where $L_n^\dagger(a)$ is the partition function of unfolded loops. 
Taking logarithms, dividing by $3n$ and letting $n \to \infty$ completes 
the proof for $a>0$.  When $a=0$ (and $y\leq 1$), then 
$\kappa(0)=\lambda(y)=\log\mu_3$ and arguments similar to those 
leading to equation \Ref{eqn59} can be used to show the above 
lower bound for $T_{3n}^{(e)}(a,b,y)$ still holds.
\hfill \hqed

Since $\kappa(a)$ has critical point $a=a_c$, the free energy of the $ABA$-triblock
copolymer is given by 
\begin{equation}
\tau_e(a,b,y) =
\begin{cases}
\log \mu_3, & \hbox{if $a\leq a_c$ \& $b\leq a_c$}; \\
\sfrac{1}{3} (2\,\kappa(a) + \log \mu_3), & \hbox{if $a>a_c$ \& $b \leq a_c$}; \\
\sfrac{1}{3} (\kappa(b) + 2\log \mu_3), & \hbox{if $a\leq a_c$ \& $b > a_c$}; \\
\sfrac{1}{3} (2\,\kappa(a) + \kappa(b)), & \hbox{if $a > a_c$ \& $b > a_c$}. 
\end{cases}
\label{eqn76}   
\end{equation} 
This shows adsorption transitions when $a=a_c$ and $b=a_c$.  The phase
diagram is shown in figure \ref{figure11}.

Next, consider the case $y>1$. 
We first look at upper bounds on the free energy.  Suppose that the 
$A$-vertices adsorb at least as strongly as the $B$-vertices so that 
$\kappa(a) \ge \kappa(b)$.  
\begin{lemm}
When $y \ge 1$ and $\kappa(a) \ge \kappa(b)$ 
$$\limsup_{n\to \infty} \Sfrac{1}{3n} \log T_{3n}^{(e)}(a,b,y) \leq \max\LC
\Sfrac{1}{3}(2\,\kappa(a)+\kappa(b)), \lambda(y)\RC .$$
\label{lemm:triblockendupper1}
\end{lemm}

\noindent\textit{Proof:}
The proof proceeds by an exhaustive case analysis where we treat the blocks as behaving independently (to obtain upper bounds).   In the following we include the 0-vertex in the first $A$-block, so that the first $A$-block always has at least one vertex (the 0-vertex) in the surface.  
We consider the four cases:
\begin{enumerate}
\item
Only the first block has vertices in the surface.
\item
Only the first two blocks ($A$ and $B$) have vertices in the surface.
\item
Only the two $A$-blocks have vertices in the surface.
\item
All three blocks have vertices in the surface.
\end{enumerate}
In the four cases the free energy is bounded above by:
\begin{enumerate}
\item
$\sfrac{1}{3} (\max\LC\kappa(a), \lambda(y)\RC+2\,\lambda(y))$,
\item
$\sfrac{1}{3} (\kappa(a) + \max\LC\kappa(b),\lambda(y)\RC+\lambda(y))$,
\item
$\sfrac{1}{3} (\kappa(a)+\log \mu_3+\max\LC\kappa(a),\lambda(y)\RC)$, and
\item
$\sfrac{1}{3} (\kappa(a)+\kappa(b)+\max\LC\kappa(a), \lambda(y)\RC)$.
\end{enumerate}
The upper bound is the maximum of these four expressions.  Note that
case (4) always gives a bound at least as large as that of case (3).  
Recall that $\kappa(a) \ge \kappa(b)$ so if $\lambda(y) > \kappa(a)$ the 
maximum of the four expressions is $\lambda(y)$.  If $\kappa(a) > \lambda(y)$ 
the maximum of the four expressions in $\sfrac{1}{3}(2\,\kappa(a)+\kappa(b))$, 
which completes the proof. \hfill \hqed

In the next part we construct appropriate lower bounds to prove the following theorem.

\begin{theo}
When $y \ge 1$ and $\kappa(a) \ge \kappa(b)$ (that is, $a \geq b$)
$$\tau_e(a,b,y) = \lim_{n\to \infty} \Sfrac{1}{3n} \log T_{3n}^{(e)}(a,b,y) 
= \max\LC \Sfrac{1}{3} (2\,\kappa(a)+\kappa(b)), \lambda(y)\RC.$$
\label{theo:triblockend1}
\end{theo}

\noindent\textit{Proof:} 
We use the upper bound in lemma \ref{lemm:triblockendupper1} together 
with strategy lower bounds to prove the theorem.  If we consider the subset 
of walks with only the 0-vertex in the surface then there are no surface energy terms and 
$$\liminf_{n\to\infty} \Sfrac{1}{3n} \log T_{3n}^{(e)}(a,b,h) \ge \lambda(y).$$
To get the other lower bound we concatenate three loops, unfolded in the 
$x_1$-direction, each with $n$ edges, the first and third labelled $A$ and the 
second labelled $B$.  Since adsorbed loops have the same free energy as 
walks and since unfolding in the $x_1$-direction doesn't change the free energy 
\cite{HTW,HammersleyWelsh}, this gives the lower bound 
$$\liminf_{n\to\infty} \Sfrac{1}{3n} \log T_{3n}^{(e)}(a,b,h) 
\ge \Sfrac{1}{3}(2\,\kappa(a)+\kappa(b)).$$
Together with lemma \ref{lemm:triblockendupper1} this completes
the proof. \hfill \hqed

\begin{figure}[t]
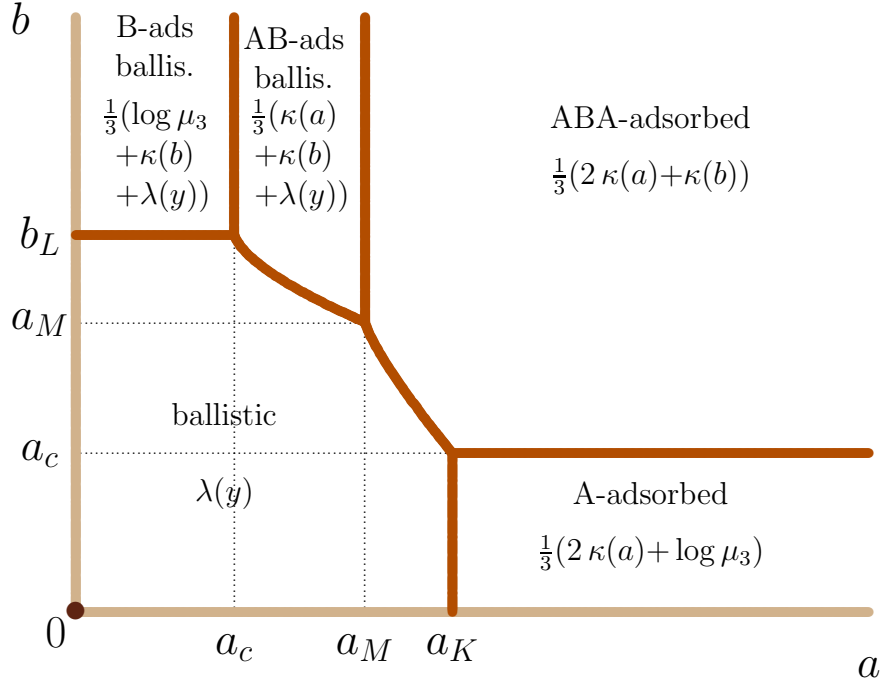

\beginpicture
\setcoordinatesystem units <1.5pt,1.5pt>
\setplotarea x from -70 to 300, y from -10 to 150
\setplotarea x from 0 to 300, y from 0 to 150

\setdots <2pt>
\plot 0 40 95 40 /
\plot 40 0 40 95 /
\plot 73 0 73 73 /
\plot 0 73 73 73 /
\setsolid

\setplotsymbol ({\footnotesize$\bullet$})

\color{Tan}
\plot 0 150 0 0 200 0 /
\color{Brown}
\plot 0 95 40 95 / \plot 95 0 95 40 /
\plot 95 40 200 40 /
\plot 40 95 40 150 /
\plot 73 73 73 150 /
\setquadratic
\plot 40 95 50 85 73 73 /
\plot 73 73 80 60 95 40 /
\setlinear

\color{Sepia}
\put {\LARGE$\bullet$} at 0 0 

\color{black}
\normalcolor

\put {\LARGE$a_c$} at 40 -9  \put {\LARGE$a$} at 200 -14
\put {\LARGE$a_c$} at -9 40  \put {\LARGE$b$} at -14 150
\put {\LARGE$0$} at -5 -5 
\put {\LARGE$a_M$} at 73 -9
\put {\LARGE$a_M$} at -9 73 
\put {\LARGE$a_K$} at 95 -9
\put {\LARGE$b_L$} at -9 95

\put {\large ballistic} at 37.5 50  
\put {\large$\lambda(y)$} at 37.5 30
\put {\large A-adsorbed} at 145 30  
\put {\large$\sfrac{1}{3}(2\, \kappa(a) {+} \log \mu_3)$} at 145 15
\put {\large ABA-adsorbed} at 145 125  
\put {\large$\sfrac{1}{3}(2\, \kappa(a) {+} \kappa(b))$} at 145 110

\put {\large AB-ads} at 55 145
\put {\large ballis.} at 55 135
\put {\large$\sfrac{1}{3}(\kappa(a)$} at 55 125
\put {\large$+\kappa(b)$} at 55 115
\put {\large$+\lambda(y))$} at 57 105
\put {\large B-ads} at 20 148
\put {\large ballis.} at 20 138
\put {\large$\sfrac{1}{3}(\log \mu_3$} at 20 125
\put {\large$+\kappa(b)$} at 20 115
\put {\large$+\lambda(y))$} at 22 105

\endpicture
\caption{The phase diagram of triblock $ABA$-copolymers pulled at the end-vertex.
There are five phases, including phases which are mixed.  The $ABA$-adsorbed
phase is separated from the $A$-adsorbed phase at $b=a_c$, when the middle
part of the adsorbed $ABA$-copolymer releases.  The $ABA$-adsorbed
phase is separated from the $AB$-adsorbed and ballistic phase when the 
last $A$-block is pulled from its adsorbed state at $a=a_M$, where
$a_M$ is the solution of $\kappa(a)=\lambda(y)$.  This mixed phase is
further separated from the $B$-adsorbed and ballistic phase when $a=a_c$
when the first $A$-block desorbs leaving only the middle $B$-block adsorbed
and the last $A$-block ballistic.  For both $a$ and $b$ small the $ABA$-copolymer
is ballistic, and this phase shares phase boundaries with the other four phases.
It is separated at the critical value of $b=b_L$ given by the solution of
$\kappa(b)+\log\mu_3=2\,\lambda(y)$ from the $B$-adsorbed and ballistic
phase and by the solution $a=a_K$ of $2\,\kappa(a)+\log\mu_3 = 3\,\lambda(y)$
from the $A$-adsorbed phase.  The phase boundary separating it from
the $AB$-adsorbed and ballistic phase is given by the solution of 
$\kappa(a)+\kappa(b)=2\,\lambda(y)$ and this is asymptotic to
$a\,b=(y/\mu_2)^2$.  The phase boundary separating it from the
$ABA$-adsorbed phase is given by the solution of 
$2\,\kappa(a)+\kappa(b)=3\,\lambda(y)$ and this is asymptotic to
$a^2\,b=(y/\mu_2)^3$.  The curved phase boundaries can be shown
to pass through the points $(a_c,b_L)$, $(a_M,a_M)$ and $(a_K,a_c)$,
as shown.}
\label{figure12}   
\end{figure}

Next, consider $y>1$ and $\kappa(a)<\kappa(b)$.  In this case the
B-vertices adsorb more strongly than the A-vertices.  The following upper
bounds hold:

\begin{lemm}
When $y \ge 1$ and $\kappa(a) < \kappa(b)$ 
$$\limsup_{n\to \infty} \Sfrac{1}{3n} \log T_{3n}^{(e)}(a,b,y) 
\leq \max\LC \Sfrac{1}{3}(2\,\kappa(a)+\kappa(b)),
\Sfrac{1}{3}(\kappa(a)+\kappa(b)+\lambda(y)), 
\lambda(y)\RC.$$
\label{lemm:triblockendupper2}
\end{lemm}

\noindent\textit{Proof:}
The proof proceeds by the same case analysis as in the proof of lemma
\ref{lemm:triblockendupper1} with the same free energies but now $\kappa(b)
> \kappa(a)$.  Bound (4) is always at least as large as bound (3) so we only need to 
consider cases (1), (2) and (4).  These are equivalent to the following upper bound on the free
energy:
$$\limsup_{n\to\infty} \Sfrac{1}{3n} \log T^{(e)}(a,b,y) 
\leq \max\LC \Sfrac{1}{3} (\kappa(a)+2\,\lambda(y)), 
\Sfrac{1}{3} (\kappa(a)+\kappa(b)+\lambda(y)), 
\Sfrac{1}{3} (2\,\kappa(a)+\kappa(b)), 
\lambda(y)  \RC ,$$
but, since $\kappa(a) < \kappa(b)$,
$$\kappa(a)+ 2\,\lambda(y) 
\leq \max\LC \kappa(a)+\kappa(b)+\lambda(y),3\,\lambda(y)\RC$$
which completes the proof. \hfill\hqed 

The proof of the following theorem uses this lemma together with strategy lower bounds.
\begin{theo}
When $y \geq 1$ and $\kappa(a) < \kappa(b)$  
$$\tau_e(a,b,y)
=\lim_{n\to \infty} \Sfrac{1}{3n} \log T_{3n}^{(e)}(a,b,y) 
= \max\LC \Sfrac{1}{3}(2\,\kappa(a)+\kappa(b)),
\Sfrac{1}{3}(\kappa(a)+\kappa(b)+\lambda(y)), 
\lambda(y)\RC.$$
\label{theo:triblockend2}
\end{theo}

\noindent\textit{Proof:}
The first and third lower bounds come from the  constructions used in the proof of
theorem \ref{theo:triblockend1}.  For the second bound consider the subset of walks 
constructed by concatenating three sub-walks of length $n$, the first being a loop
unfolded in the $x_1$-direction labelled $A$, the second being a loop unfolded
in the $x_1$-direction labelled $B$ and the third being a positive walk unfolded 
in the $x_1$-direction with only the 0-vertex in the surface,  Again, unfolding 
doesn't change the free energy \cite{HTW} so this subset of walks gives the bound
$$\liminf_{n\to \infty} \Sfrac{1}{3n} \log T_{3n}^{(e)}(a,b,y) 
\geq \Sfrac{1}{3} (\kappa(a)+\kappa(b)+\lambda(y)).$$
Together with the two lower bounds from theorem \ref{theo:triblockend1}
and the upper bound from lemma \ref{lemm:triblockendupper2}, this 
completes the proof. \hfill\hqed

Next, we need to determine the phase behaviour in this model for $y>1$.  There
are the cases to consider in theorems \ref{theo:triblockend1} and 
\ref{theo:triblockend2}. In addition to these, there are also phase boundaries
due to adsorption transitions in $\kappa(a)$ and $\kappa(b)$ when either
$a=a_c$ or $b=a_c$.  The free energy is explicitly  given by 
\begin{equation}
\tau_e(a,b,y) =
\begin{cases}
\sfrac{1}{3} (\log \mu_3+\kappa(b)+\lambda(y)), 
& \hbox{if $a<a_c$ \& $\kappa(b)+\log\mu_3>2\,\lambda(y)$}; \\
\sfrac{1}{3} (\kappa(a)+\kappa(b)+\lambda(y)), 
& \hbox{if $a_c<a<a_M$ \& $\kappa(a)+\kappa(b)>2\,\lambda(y)$}; \\
\sfrac{1}{3} (2\,\kappa(a)+\kappa(b)), 
& \hbox{if $a>a_M$, $b>a_c$ \& $2\,\kappa(a)+\kappa(b)>3\,\lambda(y)$}; \\
\sfrac{1}{3} (2\,\kappa(a)+\log \mu_3), 
& \hbox{if $b<a_c$ \& $2\,\kappa(a)+\log \mu_3>3\,\lambda(y)$}; \\
\lambda(y), & \hbox{otherwise},
\end{cases}
\label{eqn77}   
\end{equation} 
where $a_M$ is the solution of $\kappa(a)=\lambda(y)$. The phase diagram 
is shown in figure \ref{figure12}.  

\subsection{Triblock copolymers pulled at a mid-point}
In this section the $ABA$-block copolymer is pulled at its mid-point (which is 
also the mid-point of the B-block).  The setup is very similar to that of the
previous section but now $h$ is the $x_3$-coordinate of the middle vertex
of the walk.  We write $t_{3n}^{(m)}(v_A,v_B,h)$ for the number of 
self-avoiding walks with  $3n$ edges, with the same labelling and restrictions, 
and having the $x_3$-coordinate of the middle vertex equal to $h$.  The 
corresponding partition function is 
\begin{equation}
T_{3n}^{(m)}(a,b,y) = \sum_{v_A,v_B,h} t_{3n}^{(m)}(v_A,v_B,h)\,
 a^{v_A}b^{v_B}\, y^h.
\label{eqn:tribloclPFmid}
\end{equation}
We write 
\begin{equation}
\tau_m(a,b,y) = \lim_{n\to \infty} \sfrac{1}{3n} \log T_{3n}^{(m)}(a,b,y)
\end{equation}
for the free energy, whenever this limit exists.  When $b=y=0$ there is 
an exceptional situation where the free energy does not exist; 
this exception is assumed for this model.

Suppose first that $y\leq 1$.  In this case the force is either zero or
directed towards the adsorbing surface.  The free energy is given in
the next theorem.

\begin{theo}
When $y \leq 1$ 
$$\tau_m(a,b,y) = \Sfrac{1}{3} (2\,\kappa(a)+\kappa(b)) . $$
\label{theo:triblockmidpush}
\end{theo}

\noindent\textit{Proof:}
The proof is similar to the proof of theorem \ref{theo:triblockendpush}.  
For the lower bound (at $y=0$) we consider the case where the middle 
vertex is in the surface and we concatenate four unfolded loops, the 
first and fourth of length $n$ and the second and third of length $\shalf\,n$.
For the special case when $b=0$, even though $y=0$ is not permitted, 
a similar lower bound can be constructed that replaces the loop for the 
$B$-block by an unfolded positive walk that does not intersect the surface 
(see arguments leading to equation \Ref{eqn59} for example).
\qed

The result in theorem \ref{theo:triblockmidpush} is identical to
theorem \ref{theo:triblockendpush}, and the free energy is given by 
equation \Ref{eqn76}.  The phase diagram in this case is shown in 
figure \ref{figure11}.

Consider next the case $y>1$.  If $\kappa(a)\geq \kappa(b)$ (that is, when
$a \geq b$), then upper bounds are given in the next lemma.

\begin{lemm}
When $y \ge 1$ and $\kappa(a) > \kappa(b)$ 
$$\limsup_{n\to \infty} \Sfrac{1}{3n} \log T_{3n}^{(m)}(a,b,y) 
\leq \max\LC \Sfrac{1}{3}(2\,\kappa(a)+\kappa(b)),
\Sfrac{1}{3}(2\,\kappa(a)+\lambda(y^{1/2})), 
\Sfrac{1}{2}(\lambda(y)+\log \mu_3)\RC.$$
\label{lemm:triblockmidupper1}
\end{lemm}

\noindent\textit{Proof:}
The proof uses the same strategy as that used in 
lemma \ref{lemm:triblockendupper1}.  We consider the
same four cases but now the upper bounds on the free energies are:
\begin{enumerate}
\item
$\sfrac{1}{3} 
({\max\LC \kappa(a),\lambda(y)\RC
  +\sfrac{1}{2}\lambda(y)+\sfrac{3}{2}\log \mu_3})$,
\item
$\sfrac{1}{3}
(\kappa(a)+\max\LC \kappa(b),\sfrac{1}{2}\lambda(y)+\sfrac{1}{2}\log \mu_3\RC
+\log \mu_3 )$,
\item
$\max\LC \lambda(y^{1/2}),\sfrac{1}{3}(2\,\kappa(a)+\lambda(y^{1/2})) \RC$, and
\item
$\max\LC\sfrac{1}{3}(2\,\kappa(a)+\kappa(b)), 
\sfrac{1}{3}(2\,\kappa(a)+\lambda(y^{1/2}))\RC$.
\end{enumerate}
Since $\lambda(y)$ is a convex function of $\log y$ \cite{Rensburg2009},
\begin{equation}
\lambda(y^{1/2}) \le \Sfrac{1}{2}(\lambda(y) + \log \mu_3).
\label{eqn:logconvex}
\end{equation}
We note that
\begin{equation}
\Sfrac{1}{3}(\kappa(a) + \sfrac{1}{2} \lambda(y) + \sfrac{3}{2} \log \mu_3)
> \Sfrac{1}{2} (\lambda(y) + \log \mu_3)
\end{equation}
if and only if $\kappa(a) > \lambda(y)$ or $a>a_M$ where $a_M$ is the 
solution of $\kappa(a) = \lambda(y)$.  But if $\kappa(a) > \lambda(y)$ then
\begin{eqnarray}
&\hspace{-8cm}
 (2 \kappa(a) + \kappa(b)) 
  - (\kappa(a) + \sfrac{1}{2} \lambda(y) 
     + \sfrac{3}{2} \log \mu_3) \nonumber \\
= & \sfrac{1}{2}(\kappa(a) - \lambda(y)) 
 +(\sfrac{1}{2} \kappa(a) + \kappa(b) - \sfrac{3}{2} \log \mu_3) > 0.
\end{eqnarray}
Therefore 
\begin{equation}
\Sfrac{1}{3}(\kappa(a) + \sfrac{1}{2}\lambda(y) + \sfrac{3}{2} \log \mu_3)
\leq \max\LC \Sfrac{1}{2}(\lambda(y)+  \log \mu_3), 
   \Sfrac{1}{3}(2\,\kappa(a) + \kappa(b))  \RC .
\label{eqn:sidecond}
\end{equation}
Using (\ref{eqn:logconvex}) and (\ref{eqn:sidecond}) together with 
the four upper bounds above completes the proof. \hfill \hqed

The free energy for $y \ge 1$ and $\kappa(a) \ge \kappa(b)$ is given by the
next theorem. 
\begin{theo}
When $y \ge 1$ and $\kappa(a) \ge \kappa(b)$  
$$\lim_{n\to \infty} \Sfrac{1}{3n} \log T_{3n}^{(m)}(a,b,y) 
= \max\LC \Sfrac{1}{3}(2\,\kappa(a)+\kappa(b)),
\Sfrac{1}{3}(2\,\kappa(a)+\lambda(y^{1/2})), 
\Sfrac{1}{2}(\lambda(y)+\log \mu_3)\RC.$$
\label{theo:triblockmid1}
\end{theo}

\noindent\textit{Proof:}
We construct three strategy lower bounds corresponding to the upper bounds 
in lemma \ref{lemm:triblockmidupper1}.  The first lower bound comes from 
considering three concatenated unfolded loops, labelled A, B and A, as in the 
proof of theorem \ref{theo:triblockend1}.  This gives the bound
$$\liminf_{n\to \infty} \Sfrac{1}{3n} \log T_{3n}^{(m)}(a,b,y) 
\geq \Sfrac{1}{3} (2\,\kappa(a)+\kappa(b)).$$
To obtain the second lower bound we consider three concatenated subwalks.  
The first and third are each an unfolded loop labelled A, each contributing 
$\sfrac{1}{3}\kappa(a)$ to the free energy.  The second is an unfolded loop 
with only the vertices of degree 1 in the surface and pulled at its mid-point.
This contributes $\sfrac{1}{3}\lambda(y^{1/2})$ to the free energy 
\cite{Rensburg2017}.  The total contribution of these three subwalks gives 
the lower bound
$$\liminf_{n\to \infty} \Sfrac{1}{3n} \log T_{3n}^{(m)}(a,b,y) 
\geq \Sfrac{1}{3}(2\,\kappa(a)+\lambda(y^{1/2})).$$
The third lower bound comes from concatenating a bridge (in the $x_3$-direction)
unfolded in the $x_1$-direction with $\lfloor \sfrac{3n}{2}\rfloor$ edges and 
pulled at its last vertex, with a walk unfolded in the $x_1$-direction and 
restricted to have the $x_3$-coordinates of its vertices at least as large as 
the top vertex of the bridge.  Since this unfolding \cite{HammersleyWelsh} 
and the confinement \cite{HammersleyWhittington} do not change the 
free energy, these walks give the lower bound 
$$\liminf_{n\to \infty} \Sfrac{1}{3n} \log T_{3n}^{(m)}(a,b,y) 
\geq  \frac{1}{2} (\lambda(y)+\log \mu_3).$$
These three bounds, together with the upper bounds from 
lemma \ref{lemm:triblockmidupper1}, prove the theorem. \hfill\hqed

This leaves the case $\kappa(a) <  \kappa(b)$ ($a <  b$). This is 
the situation where $B$-vertices adsorb more strongly than $A$-vertices.

\begin{lemm}
When $y \ge 1$ and $\kappa(b) > \kappa(a)$ ($a\leq b$),
$$\limsup_{n\to \infty} \Sfrac{1}{3n} \log T_{3n}^{(m)}(a,b,y) 
\leq \max\LC \Sfrac{1}{3} (2\,\kappa(a)+\kappa(b)),
\Sfrac{1}{2}(\lambda(y)+\log \mu_3) \RC.$$
\label{lemm:triblockmidupper2}
\end{lemm}

\noindent\textit{Proof:}
The arguments in lemma \ref{lemm:triblockmidupper1} also work when 
$\kappa(b) > \kappa(a)$, but we can improve that result here as follows:  
Note that
\begin{equation}
\Sfrac{1}{3}(2\,\kappa(a) + \lambda(y^{1/2})) 
   > \Sfrac{1}{3}(2\,\kappa(a) + \kappa(b))
\end{equation}
if and only if $\lambda(y^{1/2}) > \kappa(b)$.  But if 
$\lambda(y^{1/2}) > \kappa(b)$ then $\lambda(y^{1/2}) > \kappa(a)$.  
Therefore
\begin{equation}
\Sfrac{1}{2} \lambda(y) + \Sfrac{1}{2} \log \mu_3 
\geq \lambda(y^{1/2}) > \Sfrac{1}{3} (2\,\kappa(a) + \lambda(y^{1/2})).
\end{equation}
This completes the proof. \hfill\hqed 

Lemma \ref{lemm:triblockmidupper2} together with some strategy lower 
bounds determines the free energy, as stated in the following theorem.
\begin{theo}
When $y \geq 1$ and $\kappa(b) > \kappa(a)$ ($a\leq b)$, 
$$\lim_{n\to \infty} \Sfrac{1}{3n} \log T_{3n}^{(m)}(a,b,y) 
= \max\LC \Sfrac{1}{3}(2\,\kappa(a)+\kappa(b)), 
        \Sfrac{1}{2}(\lambda(y)+\log \mu_3)\RC.$$
\label{theo:triblockmid2}
\end{theo}

\noindent\textit{Proof:}
The two required lower bounds come from exactly the same arguments 
as those used in the proof of theorem \ref{theo:triblockmid1}.  Together 
with the upper bounds in lemma \ref{lemm:triblockmidupper2} these 
lower bounds establish the required result. \hfill \hqed

The results in theorems \ref{theo:triblockmid1} and \ref{theo:triblockmid2} 
give the complete phase diagram of this model.  The free energy is given by
\begin{equation*}
\tau_m(a,b,y) =
\begin{cases}
\sfrac{1}{3} (2\log \mu_3+\kappa(b)), 
& \hbox{if $a<a_c$ \& $2\,\kappa(b)+\log\mu_3>3\,\lambda(y)$}; \\
\sfrac{1}{3} (2\,\kappa(a)+\lambda(y^{1/2})), 
& \hbox{if $\kappa(b)<\lambda(y^{1/2})$ 
    \& $4\kappa(a)-3\log\mu_3>3\,\lambda(y)-2\,\lambda(y^{1/2})$}; \\
\sfrac{1}{3} (2\,\kappa(a)+\kappa(b)), 
& \hbox{if $a>a_c$, $\kappa(b)>\lambda(y^{1/2})$ 
    \& $4\kappa(a)+2\,\kappa(b)>3\,\lambda(y)+3\log\mu_3$}; \\
\sfrac{1}{2} (\lambda(y)+\log \mu_3), 
& \hbox{otherwise}.
\end{cases}
\end{equation*} 

\begin{figure}[t]
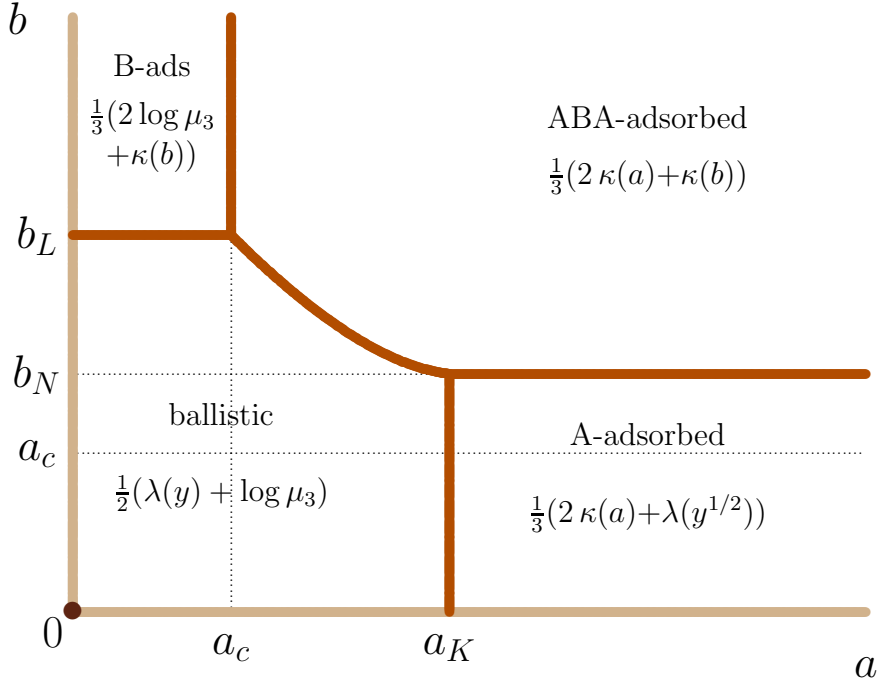

\beginpicture
\setcoordinatesystem units <1.5pt,1.5pt>
\setplotarea x from -70 to 300, y from -10 to 150
\setplotarea x from 0 to 300, y from 0 to 150

\setdots <2pt>
\plot 0 40 200 40 /
\plot 40 0 40 95 /
\plot 0 60 200 60 /
\setsolid

\setplotsymbol ({\footnotesize$\bullet$})

\color{Tan}
\plot 0 150 0 0 200 0 /
\color{Brown}
\plot 0 95 40 95 / 
\plot 95 0 95 60 /
\plot 95 60 200 60 /
\plot 40 95 40 150 /
\setquadratic
\plot 40 95 70 70 95 60 /
\setlinear

\color{Sepia}
\put {\LARGE$\bullet$} at 0 0 

\color{black}
\normalcolor

\put {\LARGE$a_c$} at 40 -9  \put {\LARGE$a$} at 200 -14
\put {\LARGE$a_c$} at -9 40  \put {\LARGE$b$} at -14 150
\put {\LARGE$0$} at -5 -5 
\put {\LARGE$b_N$} at -9 60 
\put {\LARGE$a_K$} at 95 -9
\put {\LARGE$b_L$} at -9 95

\put {\large ballistic} at 37.5 50  
\put {\large$\sfrac{1}{2}(\lambda(y)+\log \mu_3)$} at 37.5 30
\put {\large A-adsorbed} at 145 45  
\put {\large$\sfrac{1}{3}(2\, \kappa(a) {+} \lambda(y^{1/2}))$} at 145 25
\put {\large ABA-adsorbed} at 145 125  
\put {\large$\sfrac{1}{3}(2\, \kappa(a) {+} \kappa(b))$} at 145 110

\put {\large B-ads} at 20 138
\put {\large$\sfrac{1}{3}(2\log \mu_3$} at 20 125
\put {\large$+\kappa(b))$} at 20 115

\endpicture
\caption{The phase diagram of triblock $ABA$-copolymers pulled at the middle 
vertex.  There are four phases, including phases which are mixed.  The 
$ABA$-adsorbed phase is separated from the $A$-adsorbed phase at $b=b_N$ 
where $b_N$ is the solution of $\kappa(b)= \lambda(y^{1/2})$.  The $A$-adsorbed 
phase is seen when both $b<b_N$ and $a>a_K$, where $a_K$ is the solution of 
$4\,\kappa(a)-3\,\log\mu_3 = 3\,\lambda(y)-2\,\lambda(y^{1/2})$.
The  $ABA$-adsorbed phase is separated from the $B$-adsorbed phase at
$a=a_c$.  The $B$-adsorbed phase is seen when both $a<a_c$ and
$b>b_L$, where $b_L$ is the solution of $2\,\kappa(b)=3\,\lambda(y)-\log \mu_3$.
The ballistic phase is separated by the solution of $4\,\kappa(a)+2\,\kappa(b) 
= 3\,\lambda(y)+3\,\log\mu_3$.  This curved phase boundary is asymptotic 
to $a^4\,b^2=y^3\, \mu_3^3 / \mu_2^6$ and passes through the points 
$(a_c,b_L)$ and $(a_K,b_N)$, as shown.}
\label{figure13}   
\end{figure}

The phase diagram can be determined from the above and is shown in
figure \ref{figure13}.

\section{Discussion}
\label{sec:discussion}

Linear block copolymers are an interesting class of polymers because of their 
application as steric stabilizers of colloidal dispersions.   If one kind of 
block adsorbs strongly on the colloidal particle it anchors
the polymer while the other block extends into the dispersing medium. 
This paper has been
concerned with self-avoiding walk models of diblock and triblock 
copolymers adsorbing at a surface and being desorbed by the application of a force.
We have established the form of the phase diagram for several cases and we showed
that it depends on the nature of the copolymer and on where the force is applied.

The methods developed in this paper could be extended to handle copolymers with alternating blocks $A_mB_mA_mB_m \ldots$ where we have a total of $k$ blocks, all of length $m$, with $k$ fixed and $m \to \infty$.  Other models that are interesting are when $m$ is fixed and $k \to \infty$.  These include the strictly alternating case $ABABAB \ldots$ and would require a different approach.  See for instance \cite{Whittington1998} for the strictly alternating case without a force.

\section*{Acknowledgement}
EJJvR  and CES acknowledge financial support from NSERC (Canada) 
in the form of Discovery Grants RGPIN-2019-06303 and RGPIN-2020-06339, 
respectively.  CES is grateful to the Department of Mathematics and Statistics 
at York University for hosting her sabbatical visit while parts of this 
research were done.





\begin{thebibliography}{30}




\bibitem{Beaton2015} Beaton N R 2015  \textit{J. Phys. A: Math. Theor.} \textbf{48} 16FT03
\bibitem{GuttmannLawler} Beaton N R, Guttmann A J, Jensen I and Lawler G F I 2015 \textit{J.
Phys. A: Math. Theor.}  \textbf{48} 454001
\bibitem{Bradly2019} Bradly C, Janse van Rensburg E J, Owczarek A and Whittington S G
2019 \textit{J. Phys. A: Math. Theor.} \textbf{52} 405001
\bibitem{Fleer} Fleer G J, Cohen Stuart M A, Scheutjens J M H M, Cosgrove T and Vincent B 1993
\textit{Polymers at Interfaces} Chapman Hall, London
\bibitem{Guttmann2014} Guttmann A J, Jensen I and Whittington S G 2014
\textit{J. Phys. A: Math. Theor.} \textbf{47} 015004
\bibitem{Hammersley1957} Hammersley J M 1957
\textit{Proc. Camb. Phil. Soc.} \textbf{53} 642-645
\bibitem{HTW} Hammersley J M, Torrie G and Whittington S G 1982
\textit{J. Phys. A: Math. Gen.} \textbf{15} 539-571
\bibitem{HammersleyWelsh} Hammersley J M and Welsh D J A 1962 \textit{Quart.
J. Math. Oxford}  \textbf{13} 108-110
\bibitem{HammersleyWhittington} Hammersley J M and Whittington S G 1985 
\textit{J. Phys. A: Math. Gen.} \textbf{18} 101-111
\bibitem{Haupt1999}
Haupt B J, Ennis J and Sevick E M 1999 \textit{Langmuir} \textbf{15} 3886-3892
\bibitem{Iliev}
Iliev G and Janse van Rensburg E J 2012 \textit{J. Stat. Mech.} P01019
\bibitem{IoffeVelenik}
Ioffe D and Velenik Y 2008  \textit{Ballistic phase of self-interacting random walks}, Analysis and
Stochastics of Growth Processes and Interface Models (P. Morters, R. Moser, M. Penrose,
H. Schwetlick, and J. Zimmer, eds.), Oxford University Press, pp. 55-79.
\bibitem{IoffeVelenik2010} 
Ioffe D and Velenik Y 2010 \textit{Braz. J. Prob. Stat.} \textbf{24} 279-299
\bibitem{Rensburg1998} Janse van Rensburg E J 1998 \textit{J. Phys. A: Math. Gen.}
\textbf{31} 8295-8306
\bibitem{Rensburg2015} Janse van Rensburg E J 2015
\textit{The Statistical Mechanics of Interacting Walks, Polygons, Animals and Vesicles 2ed},
Oxford University Press, Oxford
\bibitem{Rensburg2009} Janse van Rensburg E J, Orlandini E, Tesi M C and
Whittington S G  2009 \textit{J. Stat. Mech.} P07014
\bibitem{Rensburg2013} Janse van Rensburg E J and Whittington S G 2013
\textit{J. Phys. A: Math. Theor.} \textbf{46} 435003
\bibitem{Rensburg2016a} Janse van Rensburg E J and Whittington S G 2016
\textit{J. Phys. A: Math. Theor.} \textbf{49} 11LT01
\bibitem{Rensburg2016b} Janse van Rensburg E J and Whittington S G 2016
\textit{J. Phys. A: Math. Theor.} \textbf{49} 244001
\bibitem{Rensburg2017} Janse van Rensburg E J and Whittington S G 2017
\textit{J. Phys. A: Math. Theor.} \textbf{50} 055001
\bibitem{Rensburg2019} Janse van Rensburg E J and Whittington S G 2019
\textit{J. Phys. A: Math. Theor.} \textbf{52} 115001
\bibitem{Krawczyk2005} Krawczyk J, Owczarek A L, Prellberg T and Rechnitzer A
2005 \textit{J. Stat. Mech.} P05008
\bibitem{Krawczyk2004} Krawczyk J, Prellberg T, Owczarek A L and Rechnitzer A
2004 \textit{J. Stat. Mech.} P10004
\bibitem{Madras} Madras N 2017 \textit{J. Phys. A: Math. Theor.} \textbf{50} 064003
\bibitem{MadrasSlade} Madras N and Slade G 1993 \textit{The Self-Avoiding Walk} 
Birkh\"{a}user, Boston
\bibitem{Madras88} Madras N, Soteros C E and Whittington S G 1988
\textit{J. Phys. A: Math. Gen.} \textbf{21} 4617-4635
\bibitem{Mishra2005} Mishra  P K, Kumar S and Singh Y 2005 \textit{Europhys. Lett.} \textbf{69} 102-108
\bibitem{Napper} Napper D 1983 \textit{Polymeric Stabilisation of Colloidal Dispersions} Academic Press, London
\bibitem{Orlandini} Orlandini E and Whittington S G 2016 \textit{J. Phys. A: Math. Theor.} \textbf{49} 343001
\bibitem{SoterosWhittington}
Soteros C E and Whittington S G 2004  \textit{J. Phys. A: Math. Gen.} \textbf{37} R279-R325
\bibitem{Whittington1975} Whittington S G 1975  \textit{J. Chem. Phys.} \textbf{63} 779-785
\bibitem{Whittington1998} Whittington S G 1998 \textit{J. Phys. A: Math. Gen.} \textbf{31} 3769-3775
\bibitem{Zhang2003} Zhang W and Zhang X 2003 \textit{Prog. Polym. Sci.} \textbf{28} 1271-1295



\end{thebibliography}
\end{document}